\documentclass[useAMS,usenatbib]{mn2e}
\pdfoutput=1

\usepackage{graphicx}
\graphicspath{{/home/nina/Dropbox/MacAir_analysis/sparcs/pro/}{/home/nina/Dropbox/MacAir_analysis/sparcs/pro/polletta_seds/}{/home/nina/Dropbox/Mac_office/SWIRE_CATALOGS/}{/home/nina/Dropbox/Mac_office/HERMES_DETECTIONS/}{/home/nina/Dropbox/swire_deep_catalogs/}{/home/nina/Dropbox/Mac_office/STACKED_IMAGES/FITS/}}

\usepackage{titlesec}
\usepackage{url}
\usepackage{float}

\renewcommand\tabcolsep{3pt}

\title[{Unveiling the Star-forming Far-IR SED of SpARCS Brightest Cluster Galaxies at $0 < z < 1.8$}]{{\it Red but not dead} : Unveiling the Star-forming Far-infrared Spectral Energy Distribution of SpARCS Brightest Cluster Galaxies at $0 < z < 1.8$ } 
\author[{N. R. Bonaventura et al.}]{N. R. Bonaventura$^{1}$, T.M.A. Webb$^{1}$, A. Muzzin$^{2}$, A. Noble$^{3}$, C. Lidman$^{4}$, G. Wilson$^{5}$, \and H.K.C. Yee$^{3}$ , J. Geach$^{6}$, Y. Hezaveh$^{7}$, D. Shupe$^{8}$, J. Surace$^{8}$ \\ \\ \\
$^{1}$McGill University Space Institute, 3550 Rue University, Montreal, QC, H3A 2A7, Canada \\ \\
$^{2}$York University, 4700 Keele Street, Toronto, ON, M3J 1P3, Canada \\ \\
$^{3}$Department of Astronomy \& Astrophysics, University of Toronto, 50 St. George St., Toronto, ON, M5S 3H4, Canada \\ \\
$^{4}$Australian Astronomical Observatory, PO Box 915, North Ryde, NSW 1670, Australia \\ \\
$^{5}$Department of Physics \& Astronomy, University of California, Riverside, 900 University Avenue, Riverside, CA, 92521, USA \\ \\
$^{6}$Centre for Astrophysics Research, University of Hertfordshire, Hatfield, Hertfordshire, AL10 9AB, United Kingdom \\ \\
$^{7}$Physics Astrophysics Building, Stanford University, 452 Lomita Mall, Stanford, CA, 94305-4085, USA \\ \\
$^{8}$California Institute of Technology, Pasadena CA, 91125, USA} 

\begin{document}

\date{Accepted 2017 March 22. Received 2017 March 21; in original form 2016 July 18.}

\pagerange{\pageref{firstpage}--\pageref{lastpage}} \pubyear{2016}

\maketitle

\label{firstpage}

\begin{abstract}
We present the results of a {\it Spitzer}/{\it Herschel} infrared photometric analysis of the largest (716) and highest-redshift ($z=1.8$) sample of Brightest Cluster Galaxies (BCGs), those from the Spitzer Adaptation of the Red-Sequence Cluster Survey (SpARCS). Given the tension that exists between model predictions and recent observations of BCGs at $z<2$, we aim to uncover the dominant physical mechanism(s) guiding the stellar mass buildup of this special class of galaxies, the most massive in the Universe and uniquely residing at the centres of galaxy clusters. Through a comparison of their stacked, broadband, infrared spectral energy distributions (SEDs) to a variety of SED model templates in the literature, we identify the major sources of their infrared energy output, in multiple redshift bins between $0 < z < 1.8$. We derive estimates of various BCG physical parameters from the stacked ${\nu}L_{\nu}$ SEDs, from which we infer a star-forming, as opposed to a `red and dead' population of galaxies, producing tens to hundreds of solar masses per year down to $z=0.5$. This discovery challenges the accepted belief that BCGs should only passively evolve through a series of gas-poor, minor mergers since $z\sim4$ (De Lucia \& Blaizot 2007), but agrees with the improved semi-analytic model of hierarchical structure formation of Tonini et al. (2012), which predicts star-forming BCGs throughout the epoch considered. We attribute the star formation inferred from the stacked infrared SEDs to both major and minor 'wet' (gas-rich) mergers, based on a lack of key signatures (to date) of cooling-flow-induced star-formation, as well as a number of observational and simulation-based studies that support this scenario.
\end{abstract}


\section{Introduction}
A brightest cluster galaxy, or BCG, refers to the most massive and optically luminous galaxy in a cluster of galaxies, located at or near the gravitational centre of the cluster. Due to their unique location in the Universe, BCGs exhibit a very uniform set of characteristics that distinguishes them from other bright galaxies, frequently casting them outside the scope of many studies of galaxy formation and evolution. They are typically found to be a few times brighter than the second and third brightest galaxies in a cluster, and as such do not appear to represent the high-mass end of the luminosity function of galaxies, but exist in a galaxy class of their own (Dressler et al. 1984; DeLucia \& Blaizot 2007, hereafter referred to as 'DB07'). While BCGs are formally classified as either elliptical or cD galaxies based on their optical morphology and the 'red' optical/near-infrared colours of their stellar populations (e.g., Dubinski 1998; Tonini et al. 2012; Bai et al. 2014; Zhao et al. 2015), they distinguish themselves from the typical cluster elliptical galaxy with their relatively large masses and luminosities, and presentation of a shallower surface brightness profile and higher central velocity dispersion (Dubinski 1998). 

BCGs have been studied across a wide range of wavelengths, from X-ray to radio, and, prior to the current study, out to a redshift of $z\sim1.6$ (e.g., Burke et al. 2000; Katayama et al. 2003, Whiley et al. 2008; Stott et al. 2008, 2010, \& 2012; Hicks et al. 2010; Donahue et al. 2010; McDonald et al. 2011 \& 2016; Rawle et al. 2012; Lidman et al. 2012; Hashimoto et al. 2014; Dutson et al. 2014; Webb et al. 2015); the sample used in this study pushes the redshift limit out to $z = 1.8$. BCGs that are detected in the mid- and far-infrared are likely hosts to star formation as well as Active Galactic Nuclei (AGN), as the UV light from young stars, or nuclear emission from an AGN, is absorbed and re-radiated by dust at these infrared wavelengths. Those hosting sufficiently unobscured and radiatively efficient AGN are also detected in the X-ray regime, typically co-located with the peak of the X-ray emission of the hot intra-cluster medium (ICM) used to identify galaxy clusters in X-ray imaging surveys (Jones \& Forman 1982). Finally, detectable radio emission may be induced by both star-formation and AGN activity in BCGs, with nearly all known cool-core clusters found to host ''radio-loud'' AGN at their centres (Fabian et al. 2001; McNamara et al. 2001; Blanton et al. 2004).

Despite the numerous, multi-wavelength observations of BCGs which have been carried out to date, a coherent story of BCG formation and evolution remains elusive. The `dissipationless merger' hypothesis (White 1976; Ostriker \& Hausman 1977; Dubinski 1998; DB07) has been the generally accepted scenario by which BCGs accumulate their large observed stellar masses, in which they only passively evolve as ''red and dead'' elliptical galaxies through a series of gas-poor minor mergers since $z\sim2.5$. While a number of observational studies appear to confirm these predictions (e.g., Bower et al. 1992; Aragon-Salamanca et al. 1993; van Dokkum et al. 1998), several more-recent, simulation-based and observational studies show evidence for significant star formation in BCGs down to much lower redshifts than predicted, in both optically and X-ray-selected clusters (Bildfell et al. 2008; Donahue et al. 2009 \& 2010; Hicks et al. 2010; Tonini et al. 2012; Liu et al. 2012; McDonald et al. 2016; Webb et al. (2015) and the current work).  Complicating matters further, there is the theoretical expectation that a BCG may not even exist as a single object through a redshift as late as 0.5, but as multiple merging galaxies (Dubinksi et al. 1998, DB07).

Due to their location at the bottom of the gravitational potential wells of the largest collapsed structures in the Universe, we expect BCGs to be enmeshed in a mix of galaxy merging, cluster cooling flows, AGN feedback, and star formation processes, with an evolving interplay between these various processes as a function of redshift. The combination of these various processes may be unique to this central cluster environment, and are not fully understood individually, let alone simultaneously or synergistically. Such complexity inherent to the study of BCGs demands a statistically significant sample size across a wide range in redshift in order to disentangle the effects of these individual processes on BCG development through cosmic time. In conjunction with Webb et al. (2015) (hereafter referred to as 'W15'), we conduct a comprehensive study of the mid- and far-infrared properties of the largest and highest-redshift sample of optically and near-infrared-selected BCGs to date, with the aim of identifying the dominant physical process(es) responsible for their infrared energy output, as a function redshift. We stack the fluxes of 675 {\it SpARCS} BCGs in multiple redshift and flux bins, in eleven {\it Spitzer} IRAC and MIPS, and {\it Herschel} SPIRE and PACS wavebands, creating broadband infrared SEDs to which we compare a variety of infrared SED templates in the literature. From the stacked rest-frame SEDs we derive estimates of the BCG total infrared luminosity, star formation rate and efficiency, stellar mass, and effective dust temperature. 

In section 2, we describe the infrared data used in this study, in addition to our BCG selection technique. In sections 3 \& 4, we explain the methodology used to create stacked data images of BCGs from which we derive photometric flux measurements and create broadband SEDs. In section 5, we step through the SED analysis procedure and discuss the implications of the results; and finally, in section 6, we present our conclusions of this study. 
Throughout the analysis we assume a cold dark matter cosmology with $H_{0} = 70$ km/s/Mpc, $\Omega_{M} = 0.3$, and $\Omega_{vac} = 0.7$, and a Salpeter Initial Mass Function (IMF) in the calculation of BCG stellar masses, presented in Section 5.

\section{DATA AND OBSERVATIONS}
\subsection{{\it Spitzer} Adaptation of the Red-sequence Cluster Survey}

We gather our sample of BCGs from the {\it Spitzer} Adaption of the Red-Sequence Cluster Survey ({\it SpARCS}) (Muzzin et al. 2009; Wilson et al. 2009), a deep, {\it z'}-band imaging survey spanning forty-two of the fifty square degrees of the six {\it Spitzer} Wide-area InfraRed Extragalactic (SWIRE) Legacy fields (Lonsdale et al. 2003). {\it SpARCS} galaxy clusters are identified using the 'red sequence' method of cluster detection (Gladders \& Yee, 2000), which exploits the tight correlation between the optical colour and magnitude, e.g., B-R and R, of elliptical galaxies in colour-magnitude space. When the 'blue-red' colour is chosen to span the so-called '4000-Angstrom break', a well-known age indicator, the older and redder cluster galaxies trace out a track or 'red sequence' in this space because they all lie within a narrow 'red' colour range compared to the younger and 'bluer' field galaxies.  {\it SpARCS} replaces the optical/optical colour with the optical/infrared [z]--[3.6] colour in order to detect cluster galaxies at high redshift, where the 4000-Angstrom break shifts into the longer-wavelength, infrared regime. Spectroscopic redshifts are available for 14.8\% of the SpARCS cluster sample, shown plotted against the corresponding photometric redshifts in Figure 2 of Webb et al. (2015). Furthermore, the SpARCS cluster sample has been filtered of false-positive cluster detections, as described in Muzzin et al. (2008).

\subsubsection{BCG Identification}

Our sample of 716 BCG candidates are selected from all {\it SpARCS} clusters with a ''flux'' signal-to-noise ratio (S/N) of at least four in the over-density detection map, and a richness value ($N_{gal}$) greater than twelve, as the galaxy having the highest {\it Spitzer} IRAC 3.6$\umu$m flux, and whose {\it z'}-3.6$\umu$m colour lies within $\pm 0.5$ of the red sequence predicted according to Muzzin et al. (2009) (noting that the colour restriction used in our BCG identification method is inherently biased against BCGs with exceptionally 'blue' or 'red' colours which are inconsistent with the red-sequence, resulting in a potential rejection of additional star-forming BCGs from the sample). In addition, the Sloan Digital Sky Survey (SDSS) optical images of low-redshift ($z < 1$) BCGs in all but the southern CDFS and ELAIS-S1 fields (which lack SDSS coverage) were visually inspected to confirm that they exhibit the morphology consistent with that expected of a low-redshift BCG (Hashimoto et al. 2014) -- i.e., a red and bright round or ellipsoid shape lacking the morphological features of obvious contaminating foreground galaxies, such as a resolved spiral structure and disk shape. This process resulted in the rejection of 22 BCGs as contaminants from the SED analysis.

The reader is referred to our companion paper, W15, for a detailed explanation of the complete BCG identification procedure utilised in our study, as well as the determination of the BCG redshifts.

\subsection{{\it Spitzer} Infrared Data}

\begin{table}
 \centering
  \begin{tabular}{@{}lcccc@{}}
   \hline
    Waveband   & Pixel scale     &  FWHM      & $N_{coadds}$ & Sensitivity limit \tabularnewline 
    ($\umu$m)  & (arcsec/pixel)  &  (pixels)  &            & (mJy)  \tabularnewline
   \hline  
   {\it Spitzer}  &       &     &       &          \tabularnewline 
                  &       &     &       &          \tabularnewline 
   IRAC 3.6       &  0.6  & 3.2 &  4.   & .0037    \tabularnewline 
   IRAC 4.5       &  0.6  & 3.3 &  4.   & .0054    \tabularnewline  
   IRAC 5.8       &  0.6  & 3.5 &  4.   & .048     \tabularnewline 
   IRAC 8.0         &  0.6  & 4.7 &  4.   & .0378  \tabularnewline 
   MIPS 24        &  1.2  & 5   &  40.  & .23      \tabularnewline 
   MIPS 70        &  4.   & 5   &  20.  & 18.      \tabularnewline 
   MIPS 160       &  8.   & 5   &  4.   & 150.     \tabularnewline 
   \hline 
                  &      &      &       &          \tabularnewline 
  {\it Herschel}  &      &      &       &          \tabularnewline 
                  &      &      &       &          \tabularnewline 
   SPIRE 250      &  6.  & 3.   & n/a   & 24.0     \tabularnewline 
   SPIRE 350      &  10. & 2.5  & n/a   & 27.5      \tabularnewline 
   SPIRE 500      &  14. & 2.6  & n/a   & 30.5      \tabularnewline 
   PACS 100       &  1.6 & 4.2  & 1-6   & 81.93     \tabularnewline 
   PACS 160       &  3.2 & 3.4  & 1-6   & 163.82    \tabularnewline 
                  &      &      &       &           \tabularnewline
   \hline
  \end{tabular}
 \caption{Summary of the {\it Spitzer} and {\it Herschel} SWIRE observations used in the stacking and photometric analysis. The sensitivity limits quoted for {\it Spitzer} wavebands refer to the SWIRE estimated $5\sigma$ limiting sensitivities, and those for the {\it Herschel} SPIRE bands the confusion noise estimated by Nguyen et al. (2010), after a $5\sigma$ cut to the data. For the {\it Herschel} PACS bands, we calculate the average $5\sigma$ instrument sensitivities over the six fields stacked in our analysis, using the corresponding limits in units of $mJy \sqrt N_{scans}$ presented in Table 3 of Oliver et al. (2012) for the fast parallel mode, plus the number of scans contained in each field image (between 1 and 6 scans per field for both the PACS 100-${\umu}m$ 160-${\umu}m$ maps).}

\end{table}

As {\it SpARCS} was carried out over the SWIRE Legacy fields, we utilise the publicly available Data Release 5 {\it Spitzer} infrared images of all six fields (ELAIS-N1, ELAIS-N2, XMM-LSS, Lockman, CDFS, ELAIS-S1) to characterise the mid-to-far-infrared emission of the BCGs in our sample. We exploit the NASA/IPAC Infrared Science Archive Image Cutout Service to generate cutout images around each BCG position in the {\it Spitzer} Infrared Array Camera (IRAC) 3.6, 4.5, 5.8, and 8.0 $\umu$m bands; and download from the main SWIRE Data Access web page the full-field mosaic images in each of the {\it Spitzer} Multiband Imaging Photometer (MIPS) 24, 70, and 160 $\umu$m bands. A full summary of the {\it Spitzer} SWIRE data is contained in Surace et al. (2005); in Table 1, we list the details relevant to our analysis, namely the estimated $5\sigma$ SWIRE sensitivity limits per channel, the Full Width at Half Maximum (FWHM) of the PSF, the image pixel scale, as well as the number of coadds contributing to each data image.

\begin{table*}
 \centering
 \begin{minipage}{140mm}
  \label{tab_one}
  \begin{tabular}{lccccccc}

    \hline                    
              & $z_{avg} = 0.185$ & $z_{avg} = 0.479$ & $z_{avg} = 0.772$ & $z_{avg} = 1.066$ & $z_{avg} = 1.360$ & $z_{avg} = 1.653$  \\
    \hline
    $24\umu$m-bright BCGs          & 17  & 24 & 48  & 54 & 14 & 8  \\
    $24\umu$m-faint  BCGs          & 50  & 103 & 190  & 125 & 28 & 14   \\
    All BCGs                        & 67  & 127 & 238  & 179 & 42 & 22  \\ 
    \hline
  \end{tabular}
 \caption{The number of BCGs contributing to the stacked flux in each data bin considered in the current analysis.}
 \end{minipage}
\end{table*}

We also utilise the {\it Spitzer} IRAC point-source fluxes associated with the BCG positions in the public SWIRE ELAIS N1, ELAIS N2, Lockman, and XMM-LSS Region Spring 2005 Spitzer Catalog, and the SWIRE ELAIS S1 and CDFS Region Fall 2005 Spitzer Catalog. While these catalogs also contain MIPS 24$\umu$m fluxes, we use the corresponding MIPS 24$\umu$m fluxes from a private catalog provided by the SWIRE team, with a fainter detection threshold than the $5\sigma$ detection threshold used in the public catalogs, specifically 100 $\umu$Jy ($S/N\sim3.3$) versus 230 $\umu$Jy for the latter.

\subsection{{\it Herschel} Infrared Data}

To constrain the far-infrared component of the BCG SED, we incorporate into our analysis all available {\it Herschel} data overlapping the SWIRE fields, in the SPIRE 250, 350, 500 $\umu$m and PACS 100 $\umu$m wavebands (noting that we also analysed PACS 160 $\umu$m data but did not include it in our final analysis, due to the higher signal-to-noise photometry achievable with {\it Spitzer} MIPS 160 $\umu$m data). We use the science-ready data maps for the SWIRE XMM, Lockman, CDFS, and ELAIS-S1 fields available in the {\it Herschel HerMES} database of SPIRE observations (Oliver et al. 2012); and manually combine all observations available in the Herschel Data Archive (HSA) for each of the two northern ELAIS fields, as the {\it HerMES} ELAIS-N1 and ELAIS-N2 coverage is significantly smaller than the corresponding SWIRE coverage. Likewise, the PACS data maps used in our analysis are the result of the manual coaddition of the separate PACS observations available for each SWIRE field in the HSA. While the {\it HerMES} survey limiting sensitivities are computed as the $5\sigma$ instrument sensitivities, ignoring confusion noise (see Oliver at al. 2012, Table 3), in Table 1 we provide instead the {\it Herschel} confusion limits contained in Nguyen et al. (2010), as confusion noise is the dominant source of noise in these far-infrared wavebands.

\subsection{20 cm Radio Data}

Finally, as part of the effort to disentangle the contribution from AGN versus star formation activity to the infrared emission of the BCGs in our sample, we gather all available 20 cm radio fluxes associated with their positions in the major radio surveys, to which we compare the corresponding 24$\umu$m {\it Spitzer} flux using the Mid-infrared Radio Correlation (MRC). We cross-match the {\it z'}-band positions of the BCGs with the radio source catalogues whose coverage includes the area of the SWIRE fields, namely the Faint Images of the Radio Sky at Twenty cm (FIRST) Survey (Becker et al. 1995), 1.4 GHz NRAO VLA Sky Survey (NVSS, Condon et al. 1998), and the Australia Telescope Large Area Survey (ATLAS) at 1.4 GHz (Hales et al. 2014), via the HEASARC Master Radio Catalog online search interface. We use a search radius equal to the instrumental resolution respective to each survey, which is comparable to the astrometric uncertainty determined for each survey, with the exception of the NVSS, for which the instrument beam FWHM size is significantly larger: 45'' arcseconds versus 5'' and 10'' for FIRST and ATLAS, respectively. The detection limits for the surveys are $\sim1$ mJy for FIRST, $\sim2.5$ mJy for NVSS, and 30 mJy per beam at approximately 12 arcseconds by 6 arcseconds for ATLAS.

\begin{figure}
\centering
\vspace{0.2cm}
 \includegraphics[scale=0.265,trim=0cm 0cm 0cm 0cm]{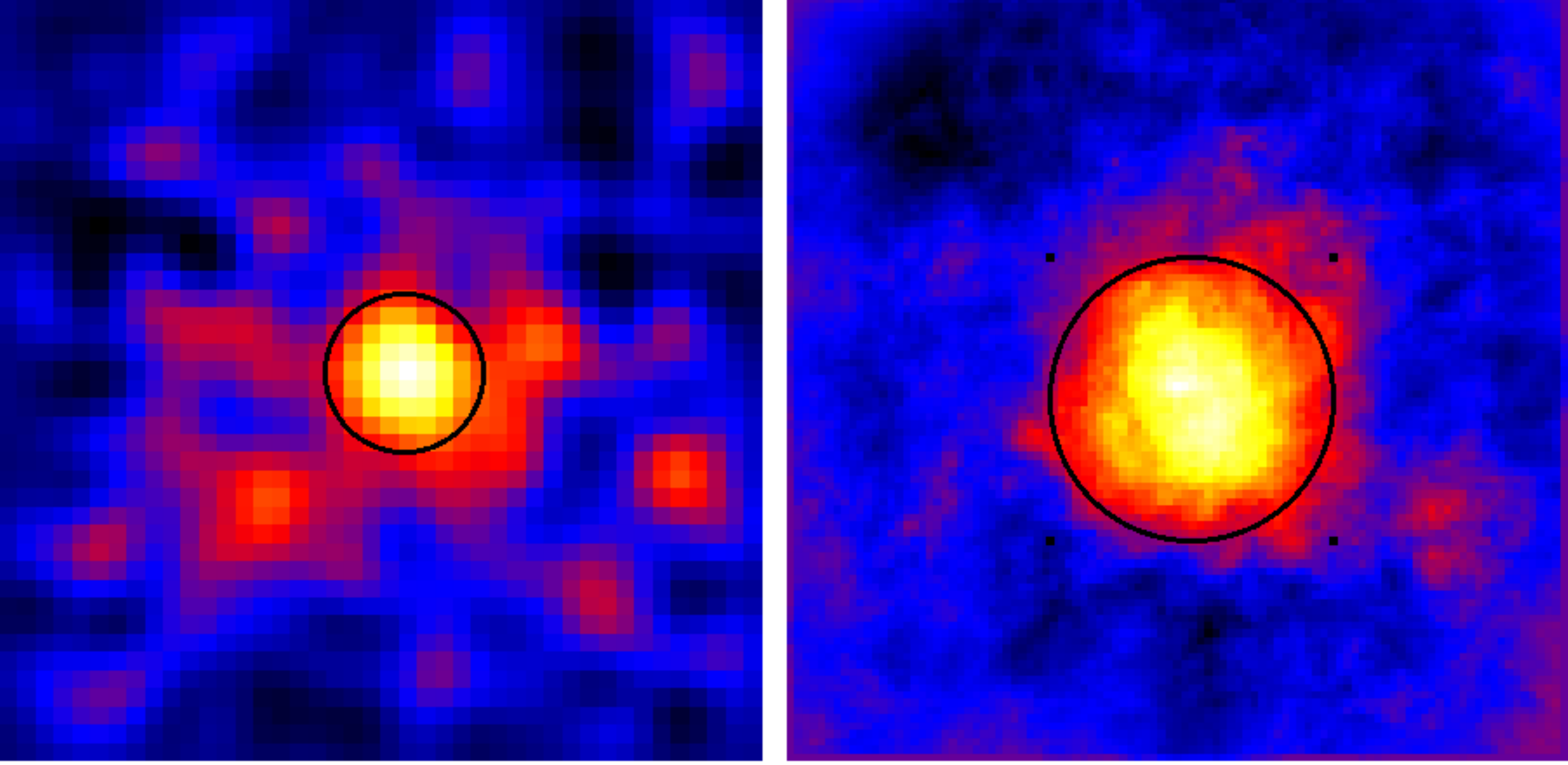}
\caption{An example of the procedure undertaken to minimise contaminating source flux and confusion noise in the far-infrared bands considered in our analysis (see section 3.1), whereby the stacked far-infrared fluxes are reduced by a factor equal to the ratio between the convolved MIPS 24$\umu$m flux obtained from the aperture shown on the right image, to the unconvolved MIPS 24$\umu$m flux from the smaller aperture shown on the left. These particular stacked images correspond to the 24$\umu$m-faint BCGs observed at $z_{avg} = 1.07$, with that on the right convolved with the SPIRE 250$\umu$m beam.}
\end{figure}

\begin{figure*}
\vspace{15pt}
\begin{tabular}{c@{\hskip 0.8cm}c@{\hskip 0.8cm}c@{\hskip 0.8cm}c@{\hskip 0.8cm}c@{\hskip 0.8cm}c@{\hskip 0.8cm}c@{\hskip 0.8cm}c@{\hskip 0.8cm}c@{\hskip 0.8cm}c@{\hskip 0.8cm}c}                     
 IRAC & IRAC & IRAC & IRAC & MIPS & MIPS & MIPS & PACS & SPIRE & SPIRE & SPIRE \\ 
 3.6$\umu$m & 4.5$\umu$m & 5.8$\umu$m & 8$\umu$m & 24$\umu$m & 70$\umu$m & 160$\umu$m & 100$\umu$m & 250$\umu$m & 350$\umu$m & 500$\umu$m  \\ 

19.8'' & 19.8'' & 19.8'' & 19.8'' & 15.75'' & 48'' & 96'' & 14.4'' & 65.8'' & 90'' & 126''  \\

\end{tabular}
\begin{tabular}{l}
 \includegraphics[scale=0.27,trim=0cm 0cm 0cm 0cm]{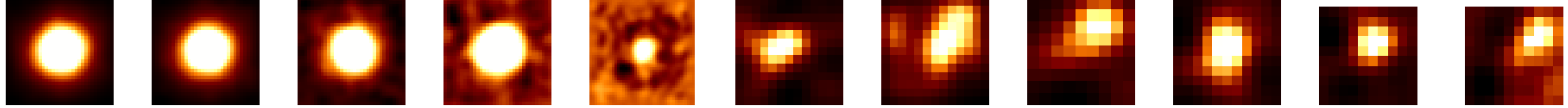}  \\
 $z_{avg} = 0.19$  \\
 \includegraphics[scale=0.27,trim=0cm 0cm 0cm 0cm]{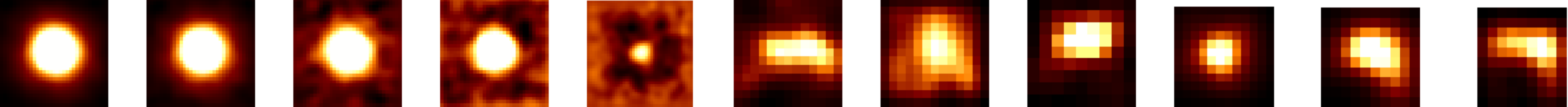}  \\
 $z_{avg} = 0.48$    \\
 \includegraphics[scale=0.27,trim=0cm 0cm 0cm 0cm]{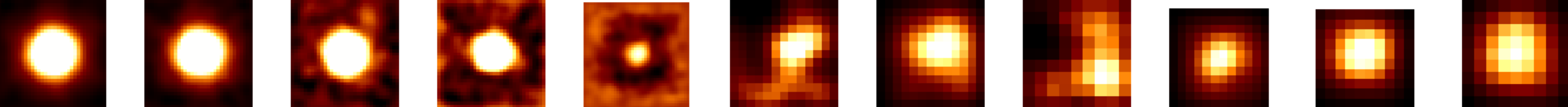}  \\
 $z_{avg} = 0.77$   \\
 \includegraphics[scale=0.27,trim=0cm 0cm 0cm 0cm]{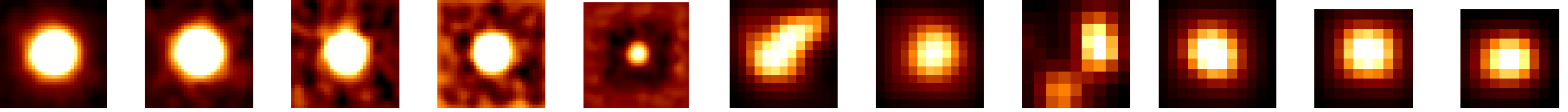}  \\
 $z_{avg} = 1.07$   \\
 \includegraphics[scale=0.27,trim=0cm 0cm 0cm 0cm]{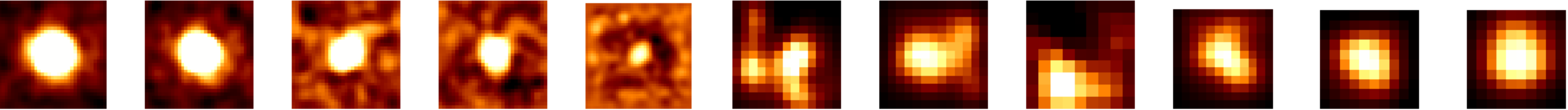}  \\
 $z_{avg} = 1.36$   \\
 \includegraphics[scale=0.27,trim=0cm 0cm 0cm 0cm]{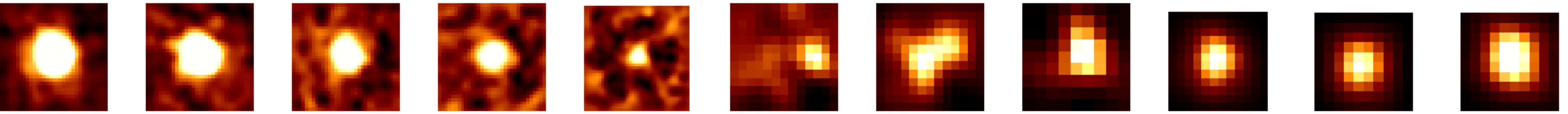}  \\
 $z_{avg} = 1.65$   \\
\end{tabular}
\vspace{0.3cm}
\caption{{\it Spitzer} and {\it Herschel} cutout images centred on the stacked BCG position within each of six redshift bins, shown in increasing order from top to bottom, in each of the {\it Spitzer} IRAC 3.6, 4.5, 5.8, 8 $\umu$m; {\it Spitzer} MIPS 24, 70, 160 $\umu$m; {\it Herschel} PACS 100 $\umu$m; and{\it Herschel} SPIRE 250, 350, and 500 $\umu$m wavebands, from left to right (we do not show the PACS 160$\umu$m images since they were not used in the SED analysis). The image cutout size in each band is three times larger than the largest flux aperture used in that band for photometry. All of the images appear smoothed with a $1.5\sigma$ Gaussian kernel (kernel radius = 3) using the SAO Image DS9 smoothing function (for presentation purposes, only). The IRAC and MIPS 24$\umu$m images are scaled with the DS9 {\it linear zscale} setting, and the remaining images with the {\it sinh zcale} setting. In several of the MIPS 24$\umu$m images, a dark, ring-like structure appears surrounding the target BCG, a result of the removal of contaminating point sources near the BCG position, described in section 3.1. The far-infrared images (longward of 24$\umu$m) appear {\it with} contaminating source flux, however, since the removal of this contamination was done differently than for the IRAC and MIPS 24$\umu$m bands. Therefore, where a single source appears significantly off-centre in these images (in most of the PACS 100-$\umu$m images and the MIPS 70-$\umu$m $z_{avg} = 1.65$ image), not enough flux overlaps the centred, target BCG aperture for these stacked 'detections' to be considered robust (i.e., $S/N < 3$).}
\end{figure*}

\section{STACKING ANALYSIS}

While each of the BCGs in our sample has an entry in the SWIRE IRAC-24$\umu$m Point Source Catalog within 3'' of its {\it z}-band RA and Dec position, only $50\%$ (358/716) are detected at all SWIRE IRAC 3.6/4.5/5.8/8.0$\umu$m infrared bands, with the remaining 50\% lacking 5.8 $\umu$m and/or 8 $\umu$m detections. Furthermore, only $23.3\%$ (167/716) are detected in the {\it Spitzer} MIPS 24 $\umu$m band at a flux of 100 $\umu$Jy or higher in the deeper SWIRE point-source catalogue made available through private communication with the SWIRE team. Finally, none of the BCGs are individually detected in the MIPS 70 and 160 $\umu$m bands within a search radius of 4'', and only 12\% are detected in one or more of the {\it Herschel} far-infrared SPIRE 250/350/500 bands, based on their inclusion in the HerMES point-source catalog.  Consequently, we perform a standard image stacking analysis in which we take the median of the individual fluxes of the BCGs contained in multiple flux and redshift bins, in each of the seven {\it Spitzer} and five {\it Herschel} infrared wavebands considered in the analysis.  This process reduces the image background by a factor of $1/\sqrt{N}$, with $N$ representing the number of sources stacked within a given data bin, with the aim of teasing out a significant stacked detection from which to extract a flux measurement (noting that confusion noise, which is the dominant contribution to the background noise in the far-infrared bands, is a time-invariant quantity and therefore does not decrease with an effective increase in exposure time resulting from the stacking procedure, unlike instrument noise (Nguyen et al. 2010)). This type of stacking analysis therefore turns a loss of information on many individual faint sources into an opportunity to obtain a robust measure of the global, average properties of the population of sources as a whole, where there is a statistically significant sample, as in the case of the current study. 

We choose to conduct a median photometric stacking analysis over a mean, given that the mean stacked fluxes proved to be sensitive to bright outliers in the flux distribution, which also increases source Poisson noise and reduces the signal-to-noise ratio. Furthermore, even where the underlying flux distribution may be highly asymmetric or skewed, e.g. which could occur when stacking over the fluxes of different galaxy types, the mean stacked flux will still be biased towards the dominating portion of the distribution in all cases, with the median always being preferred. Finally, we note that for the small fraction of stacked fluxes in our analysis which exhibit a low signal-to-noise ratio ($S/N < 3$), the median and mean stacked values approximate each other, assuring us that a median stack (and also the mean, in this case) is still a reasonable choice, probing the representative flux value we seek (see, e.g., White et al. 2007, who demonstrate that in the low signal-to-noise limit of a highly asymmetric distribution, the median can 'get stuck' converging towards a local mean value and not the global mean, as {\it rms} noise increases).

\subsection{Removal of Contaminating Sources}

Prior to performing the image stacking procedure in the mid-infrared {\it Spitzer} IRAC and MIPS 24$\umu$m bands, we utilise the PSF-fitting and subtraction functions contained in the AstroPy {\it photutils} package to subtract contaminating flux from sources near the target BCG position, centred outside a radius equal to the FWHM from the BCG position, and detected at a $3{\sigma}$ level.

In the remaining {\it Spitzer} and {\it Herschel} far-infrared bands, we account for both the contaminating flux from nearby sources and the confusion noise which dominates these bands through an analysis of the associated MIPS 24$\umu$m data, assuming that any emission at 24$\umu$m has a counterpart at the longer infrared wavelengths, e.g., along the lines of the Rosebloom et al. (2010) study. We do this according to the following procedure: First, we simulate the far-infrared stacked flux using the associated (uncleaned) MIPS 24$\umu$m data in the given data bin, by convolving $120''\times120''$ MIPS 24$\umu$m cutout images with the PSF of the far-infrared band, and then stacking those convolved MIPS 24$\umu$m images (we note that we use a larger MIPS 24$\umu$m cutout image size in this context than for the MIPS 24$\umu$m photometry, described in Section 4.1, in order to properly capture contaminating flux from convolved sources outside the flux aperture). Next, we measure the stacked MIPS 24$\umu$m flux within the far-infrared flux aperture (see Table 4) on the convolved and stacked MIPS 24$\umu$m image to capture the 'extra' flux hiding within the larger far-infrared beam area in comparison to the smaller MIPS 24$\umu$m flux aperture. Finally, we divide the 24$\umu$m flux value obtained from the relatively large far-infrared aperture on the convolved stacked image by the 24$\umu$m flux value obtained from the {\it MIPS 24$\umu$m flux aperture} on the corresponding {\it unconvolved} stacked image, for the same data bin; the far-infrared stacked fluxes are then reduced by this ratio. An example of this procedure is shown in Figure 1, where the unconvolved stacked MIPS 24$\umu$m flux is shown alongside the MIPS 24$\umu$m flux convolved with the SPIRE 250$\umu$m beam size.  Table 3 lists the range of correction factors applied to the 24$\umu$m-bright and 24$\umu$m-faint (hereafter, 'bright' and 'faint') far-infrared stacked fluxes in each redshift bin of our analysis.

We note that in the $z_{avg} = 1.36$ bin of our SED analysis, where a rest-frame 9.7-$\umu$m silicate absorption feature of a star-forming galaxy SED could coincide with the observed 24-$\umu$m flux, the correction factor would be overestimated -- i.e., our assumption of a one-to-one correspondence between a 24-$\umu$m and far-infrared detection being based on both fluxes lying on the galaxy infrared flux continuum, would break down where the 24-$\umu$m flux lies significantly below the continuum towards the minimum of a spectral absorption feature. Similarly, where the observed 24-$\umu$m flux coincides with a rest-frame PAH emission feature, the correction factors would be underestimated. However, if the bulk of the contaminating flux in the 24-$\umu$m flux aperture used to estimate the far-infrared correction factors comes from cluster galaxies in the vicinity of the BCG which have similar star-forming features, then the correction factor would not be severely over- or underestimated (i.e., where the contamination would only be a function of {\it N} number of contaminating sources in the flux aperture, and not also of the the height/depth of an emission/absorption feature with respect to the continuum). While we are not yet in a position to make this claim at this stage of our analysis of SpARCS BCGs, we do find that altering the original correction factors to account for a 24-$\umu$m flux lying at or near the peak of a PAH emission line leads to an apparent over-correction (i.e., over-reduction) of the far-infrared fluxes and therefore an unphysical far-infrared SED fit, whereas the original correction factors lead to very good fits. On the other hand, modifying the correction factors to account for silicate absorption leads to overall improved \footnote{The far-infrared fluxes of the $z_{avg} = 1.36$ faint SED were apparently being over-corrected by the original correction factors, due to the anomalously low $L_{FIR}$ of this SED relative to the SEDs in redshifts bins below and above it; the updated correction factor places the $L_{FIR}$ literally in-line with the increasing infrared detection threshold with redshift (see grey curve Figure 7a), as expected. As for the $z_{avg} = 1.65$ faint SED, the ${\chi}^{2}$ value of the young starburst fit improves by 48.8\% with a reduction to the original correction factors. Finally, the updated correction factors in both bins very closely match the correction factors for the bright SEDs in the corresponding redshift bins.} results for the SEDs likely to be affected, both the faint $z_{avg} = 1.36$ and $z_{avg} = 1.65$ SEDs (where, for the latter SED, only a young starburst template with a relatively wide silicate absorption feature simultaneously fits both the 24-$\umu$m and 160-$\umu$m fluxes). 

We calculate the refinement to the correction factors based on the average 24-$\umu$m flux difference between the height/depth of a spectral feature and the continuum, from the range of star-forming model SEDs which fit the BCG SEDs between 3 and 24 $\umu$m (not considering the far-infrared portion of the SED in this process, given that the correction factors which we are trying to refine have influenced this fit). As explained above, this process leads to updated far-infrared correction factors for only the $z_{avg} = 1.36$ and $z_{avg} = 1.65$ faint SEDs. For the remaining SEDs, the range of star-forming models which fit each SED place the observed 24-$\umu$m flux on or close to the flux continuum, and therefore we assume the original far-infrared correction factors to be robust.

\begin{table*}
 \centering
 \begin{minipage}{140mm}
  \label{tab_one}
  \begin{tabular}{lccccccc}

    \hline                    
               & $z_{avg} = 0.185$ & $z_{avg} = 0.479$ & $z_{avg} = 0.772$ & $z_{avg} = 1.066$ & $z_{avg} = 1.360$ & $z_{avg} = 1.653$  \\
    \hline
    MIPS  $70 \umu$m           & 1.00-1.89  & 1.55-2.55  & 1.84-3.53  & 1.62-3.4   & 1.52-2.19  & 1.65-1.9  \\
    MIPS  $160 \umu$m          & 2.92-3.14 & 2.19-9.99  & 2.97-15.57 & 2.97-12.50 & 3.22-7.00 & 2.70-3.60  \\
    SPIRE $250 \umu$m         & 1.00-2.02  & 1.51-2.04  & 1.69-2.21  & 1.62-4.35  & 1.60-2.21  & 1.0-2.20   \\ 
    SPIRE $350 \umu$m         & 1.00-2.12  & 1.29-1.55  & 1.71-4.93  & 1.95-4.44  & 1.42-2.9 & 1.30-2.82  \\
    SPIRE $500 \umu$m         & 2.39-3.75 & 1.48-8.43  & 2.55-12.99 & 2.54-13.95 & 3.10-9.57 & 1.86-2.28  \\
    PACS  $100 \umu$m          & 1.00-1.41  & 1.00-1.40   & 1.00-1.455  & 1.00-1.46   & 1.00        & 1.21-1.49   \\ 
    \hline
  \end{tabular}
  \caption{The range of correction factors by which the far-infrared stacked fluxes in the 'bright', 'faint', and 'all BCGs' flux bins (see Section 3.2) are reduced in each redshift bin, to mitigate the contribution from contaminating confused and resolved/detected sources to the far-infrared SEDs. A value of 1.0 indicates that no 'excess' convolved 24$\umu$m flux was detected within the associated far-infrared beam area (refer to Section 3.1 for details of the derivation of the far-infrared stacked flux correction).}
 \end{minipage}
\end{table*}

\subsection{Stacking Procedure}

Stacking the BCG fluxes involves creating a small image cutout exactly centred on each {\it z}'-band BCG position in all corresponding {\it Spitzer} and {\it Herschel} data maps, assembling the cutouts into six evenly spaced redshift bins between the limits of our sample, $0 < z < 1.8$, and calculating the median of the stack of cutouts within each data bin, pixel by pixel, to create a single stacked image per bin; this process is repeated in each of the eleven {\it Spitzer} and {\it Herschel} infrared wavebands utilised in our analysis. As we must stack on the same set of BCG positions in each waveband to produce a meaningful broadband SED, we reject those BCG positions that are not simultaneously covered by the {\it Spitzer} and {\it Herschel} maps in all eleven wavebands; this results in the rejection of 19 BCGs, in addition to the 22 rejected as contaminants, resulting in a total of 675/716 BCGs considered in the infrared SED analysis. The stacked, median-flux images from which we derive BCG flux measurements are shown in Figure 2. 

In order to make a direct comparison between the bright (24$\umu$m flux $>$ 100$\umu$Jy) and faint (24$\umu$m flux $<$ 100$\umu$Jy) subsets of the BCG sample, following on the analysis presented in W15, we separately stack the fluxes of the detected 'bright' sources alongside the 'faint' sources, splitting each of the six redshift bins into two flux bins.  In total, eleven stacked fluxes are calculated per SED for each of seven {\it Spitzer} and five {\it Herschel} wavebands, with a single SED produced for each redshift and bright/faint flux data bin. The resulting six 'bright' and six 'faint' SEDs, in addition to six SEDs stacked over all fluxes (both bright and faint), are shown in Figure 3.

\section{PHOTOMETRY}

Fluxes for the stacked detections are calculated using the standard techniques of circular aperture photometry for a point source, in which the flux values of all pixels lying within a circular aperture centred on the stacked source are summed to obtain the total stacked source flux, which is then corrected by a factor equal to the inverse of the energy fraction encircled by the chosen aperture; all aperture sizes and aperture corrections used are contained in Table 4. For the {\it Spitzer} IRAC and {\it Herschel} PACS data images in particular, which do not come background-subtracted, we first subtract the median background level per pixel in each cutout image prior to stacking. We consider a stacked source detection to be robust if it has a signal-to-noise ratio of at least three, where the aperture noise includes contributions from both source and background counts, calculated as described in the following sub-section.

\begin{table*}
 \centering
 \begin{minipage}{140mm}
  \label{tab_one}
   \begin{tabular*}{\textwidth}{c @{\extracolsep{\fill}} c @{\extracolsep{\fill}} c @{\extracolsep{\fill}}}

   \hline
    \\
    Waveband  &  Circular aperture radius  &  Aperture correction   \\
    ($\umu$m) &  (pixels)                  &  (pixels)             \\
    \\
    \hline
    & & \\
   {\it Spitzer}  & &   \\
             &          &          \\
   IRAC 3.6  & 11, 10, 9, 7.5, 6.83, 6.83  & 1.0198, 1.0277, 1.0368, 1.0532, 1.087, 1.087  \\
   IRAC 4.5  & 10, 9, 8, 7, 6, 6  & 1.0721, 1.0855, 1.1014, 1.1203, 1.12, 1.12  \\
   IRAC 5.8  & 9.66, 6, 6, 6, 4.83, 4.83  & 1.0416, 1.135, 1.135, 1.135, 1.25, 1.25  \\
   IRAC 8    & 7, 4.83, 4.83, 4, 4, 4  & 1.015, 1.429, 1.429, 1.568, 1.568, 1.568  \\
   MIPS 24   & 4.375  & 2.228       \\
   MIPS 70   & 4   &  2.07      \\
   MIPS 160  & 4   &  1.971   \\
             &    &        \\ 
\hline
              & & \\
   {\it Herschel} & &    \\
              &   &                 \\
   SPIRE 250  &  3.66    & 1.28    \\
   SPIRE 350  &  3     &  1.196  \\ 
   SPIRE 500  &  3     &  1.26    \\
   PACS 100   &  3     &  1.666   \\
             &    &        \\ 
   \hline
  \end{tabular*}
 \caption{Aperture photometry parameters used to compute BCG stacked fluxes. Where a range of aperture radii and associated aperture correction factors are listed, they represent the apertures customised to each redshift bin in a given band, in order of increasing redshift. The values for IRAC were taken from the IRAC Instrument Handbook and SWIRE Data Release 2 document (Surace et al. 2005), those for MIPS from the SWIRE Data Release 2 document, those for SPIRE from Pearson et al. (2014) , and those for PACS from Balogh et al. (2014)}
 \end{minipage}
\end{table*}

\subsection{Mid-infrared Stacked Flux \& Flux Uncertainty}

In performing circular aperture photometry on the stacked {\it Spitzer} IRAC images, we choose those reference aperture sizes and associated corrections that maximise the flux from the target BCG while minimising noise from the surrounding background, either from the IRAC Instrument Handbook or the {\it Spitzer} SWIRE Data Release 2 document (Surace et al. 2005). For each of the the MIPS bands, we adopt the recommended aperture size and correction, and additional calibration factor based on instrument performance, from the SWIRE Data Release 2 document. 

The complete error on each stacked flux is represented by the quadratic sum of the statistical error associated with the stacking procedure, the aperture background noise represented by the median background level of the corresponding $3\sigma$-clipped master {\it Spitzer} data map, and the Poisson noise from the source itself.  We compute the statistical error on each data bin using the Median Absolute Deviation (MAD) ($\sigma \sim 1.4826$ MAD) of the individual cutout images contributing to each median stacked image, on a pixel by pixel basis, utilising the multiplicative factor of 1.4826 based on the assumption of normally distributed values. The resulting statistical error on each pixel in the median stacked image is then propagated through the number of pixels contained within the source aperture from which the stacked source flux is measured.

The aperture background noise of each stacked detection is represented by the median value of 500 random measures of the median background level of the corresponding $3\sigma$-clipped master {\it Spitzer} data map, in apertures matching the size of apertures used in this analysis. As each stacked cutout image contains BCG contributions from multiple SWIRE Legacy fields, this procedure was carried out separately for each field map contributing to the stack, per {\it Spitzer} waveband, with the results being averaged according to the number of BCGs contributing from each respective field. We then combine this aperture noise with the Poisson or 'shot' noise from the source itself in the usual way: $\Delta F = \sqrt \Sigma(\sigma_{i}^{2} + f_{i}/g_{eff})$ , where $\Sigma$ indicates a summation operation, $f_{i}$ and $\sigma_{i}$ represent the stacked source flux and error of pixel {\it i}, respectively, and $g_{eff}$ the effective gain value, which incorporates the number of coadded images comprising the corresponding {\it Spitzer} map from which the cutouts were created (not to be confused with the number of individual BCGs contributing to the stack).

\begin{figure*}
\begin{tabular}{ccc}                  
\includegraphics[scale=0.09, width=6.1cm, trim=2cm 7.9cm 7cm 15.5cm]{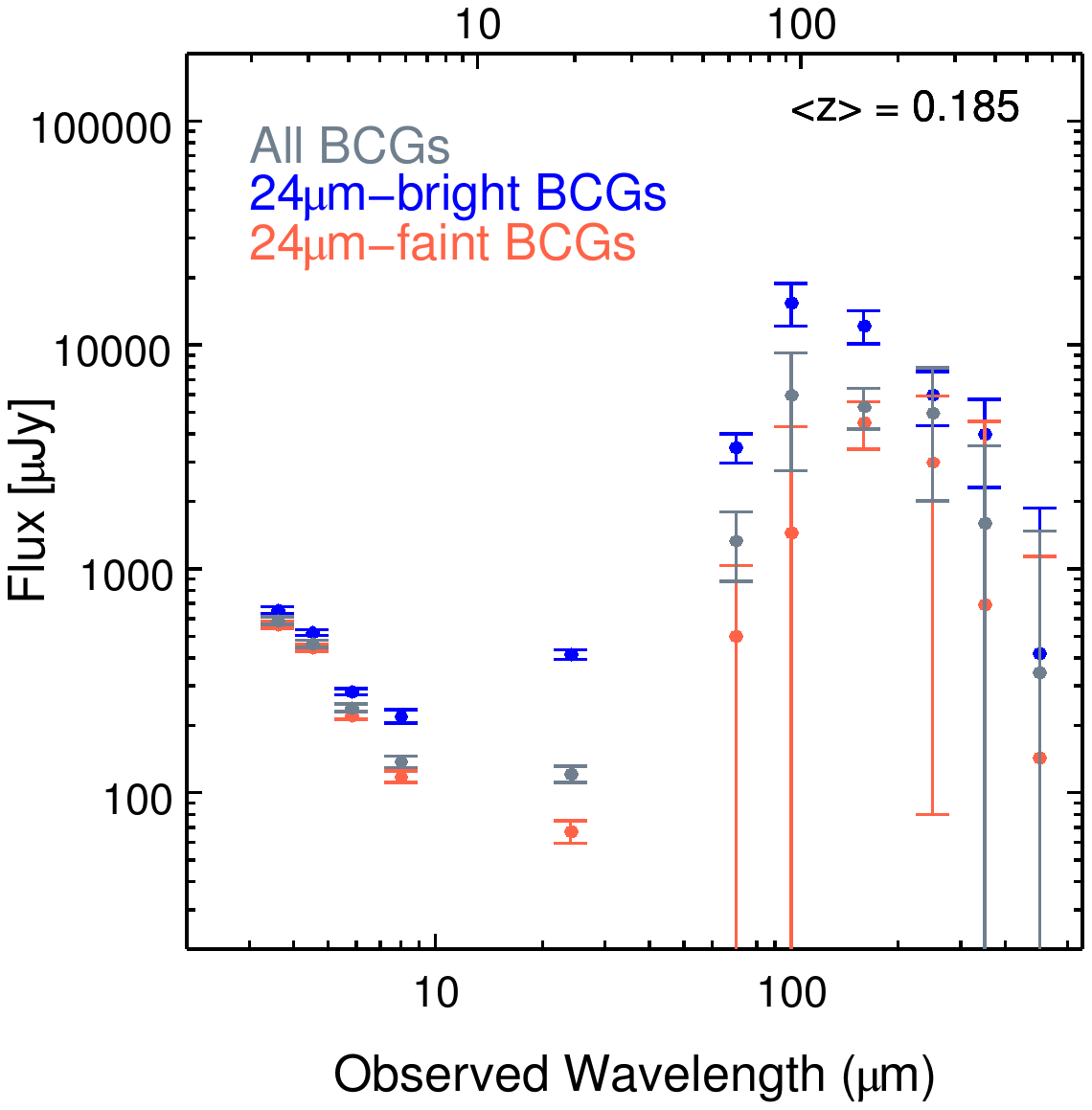} &
\includegraphics[scale=0.09, width=6.1cm, trim=2cm 7.9cm 7cm 15.5cm]{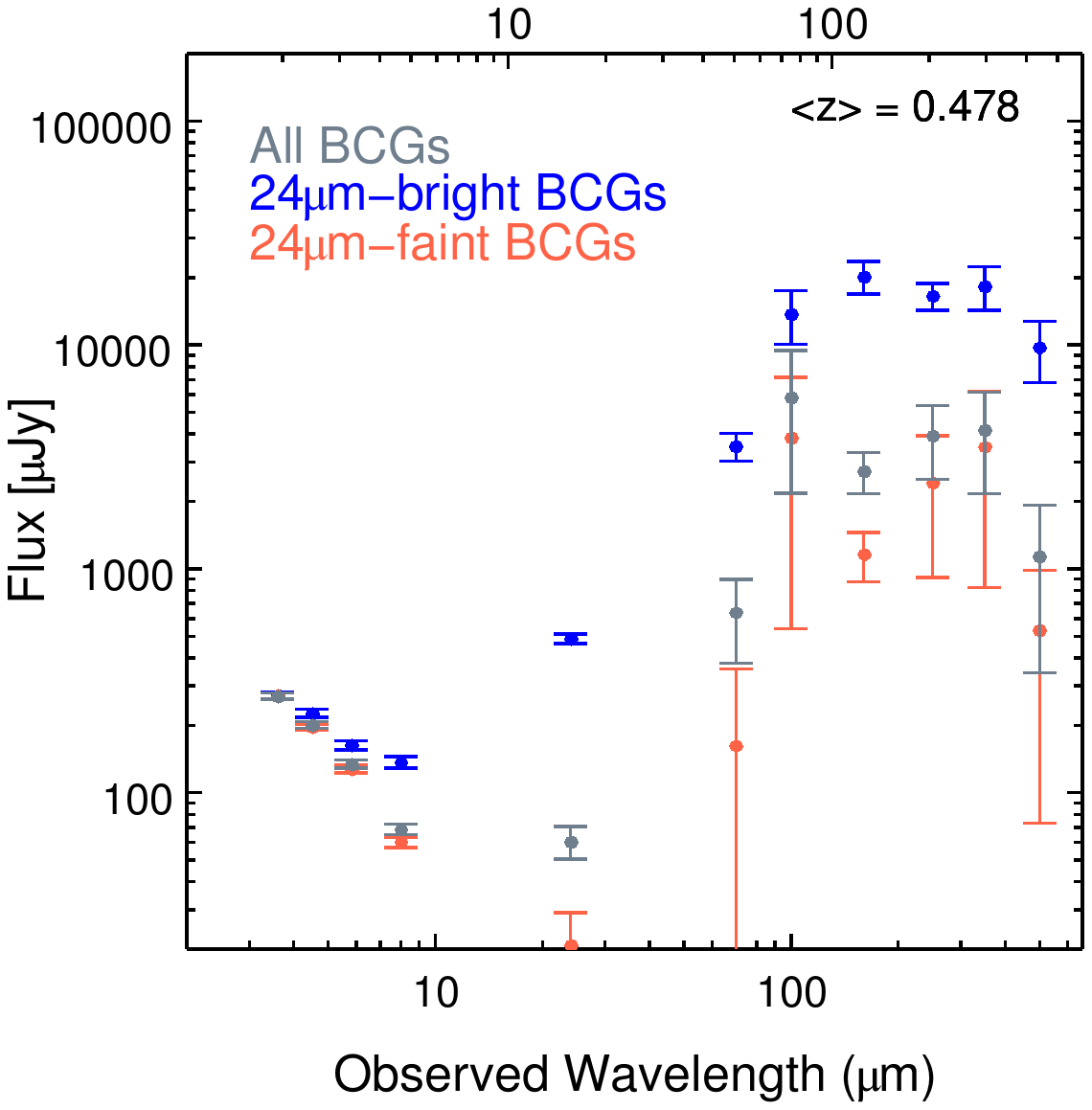} &
\includegraphics[scale=0.09, width=6.1cm, trim=2cm 7.9cm 7cm 15.5cm]{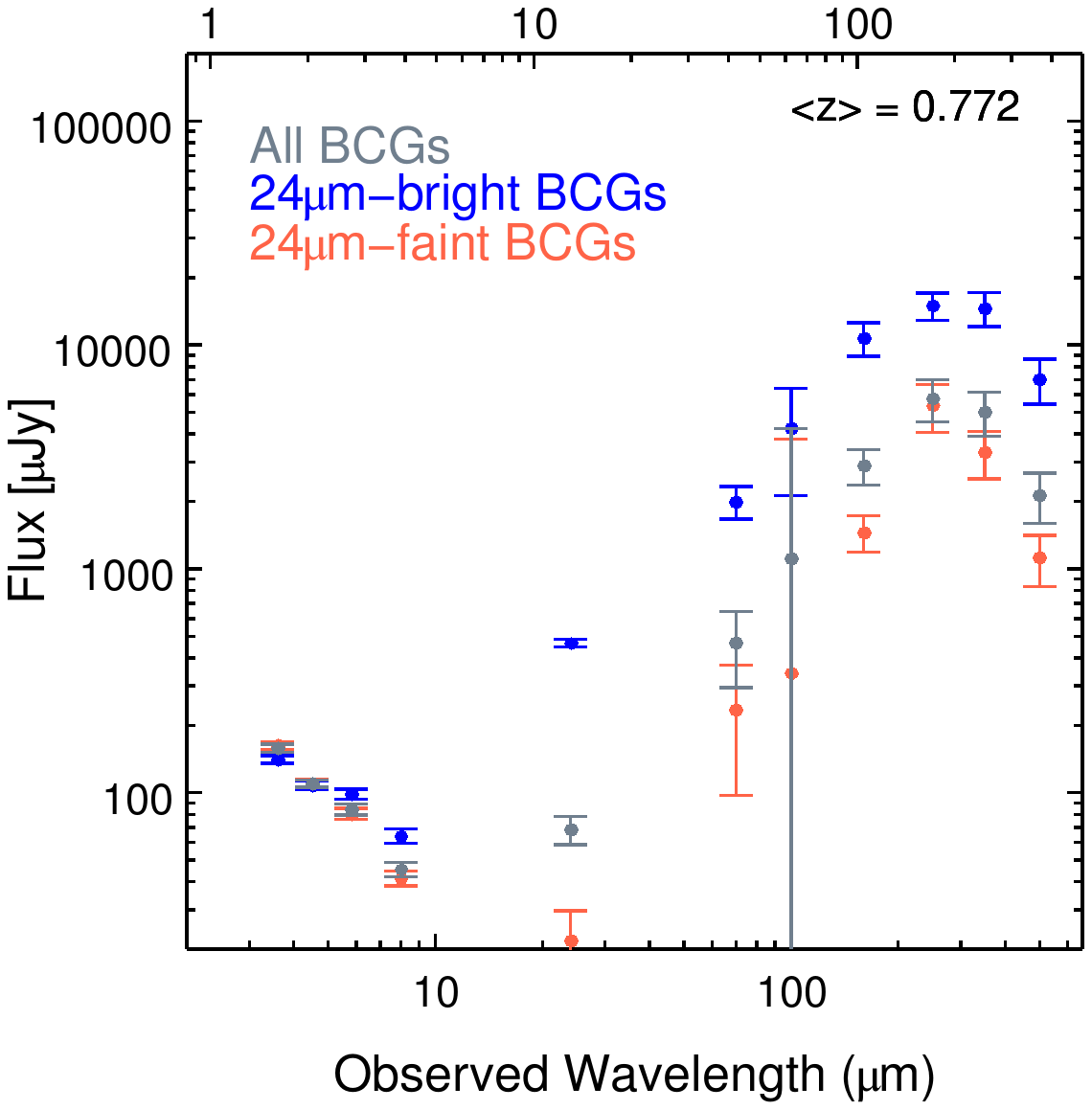}  \\

\includegraphics[scale=0.09, width=6.1cm, trim=2cm 7.7cm 7cm 8.3cm]{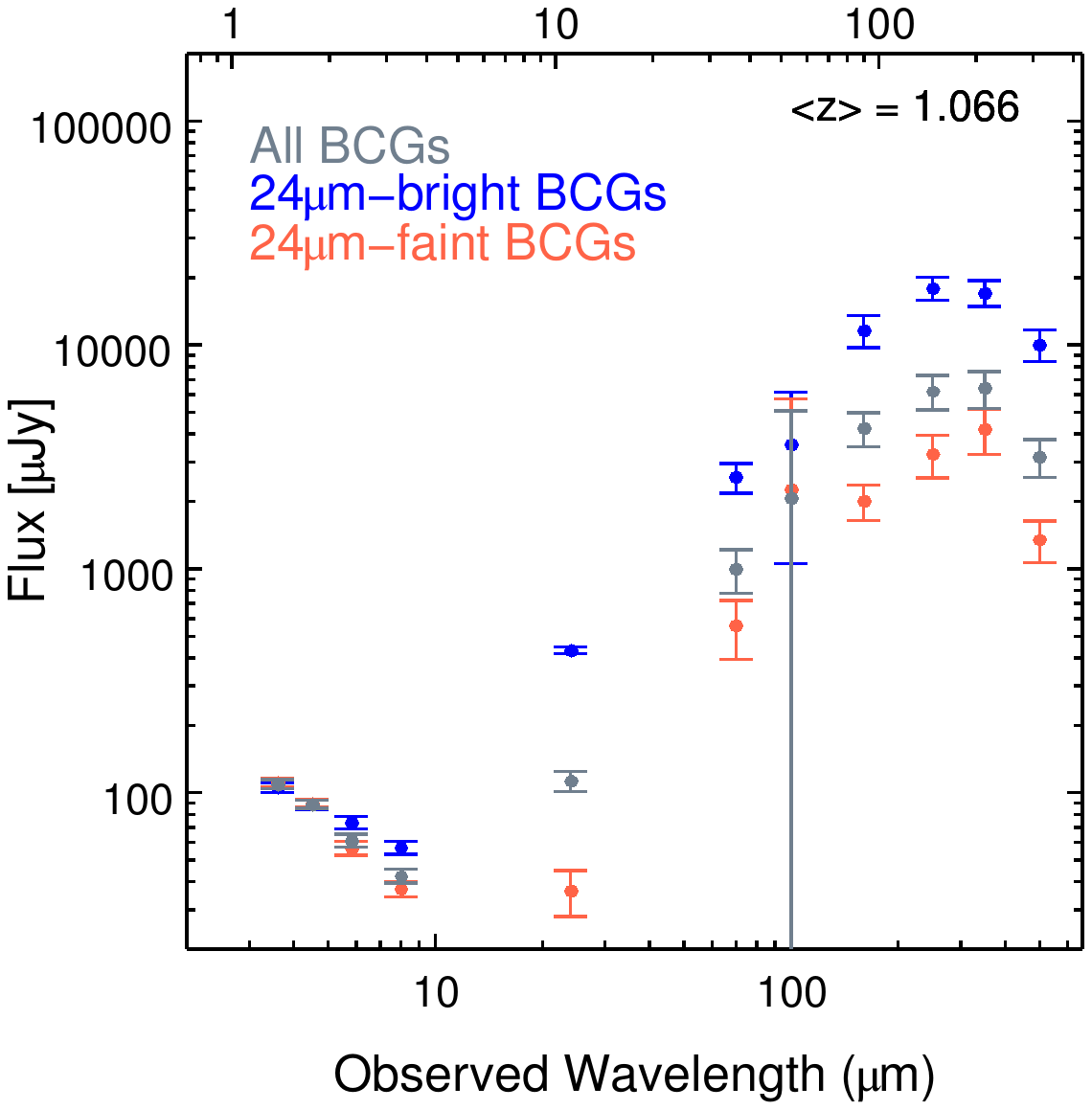} &
\includegraphics[scale=0.09, width=6.1cm, trim=2cm 7.7cm 7cm 8.3cm]{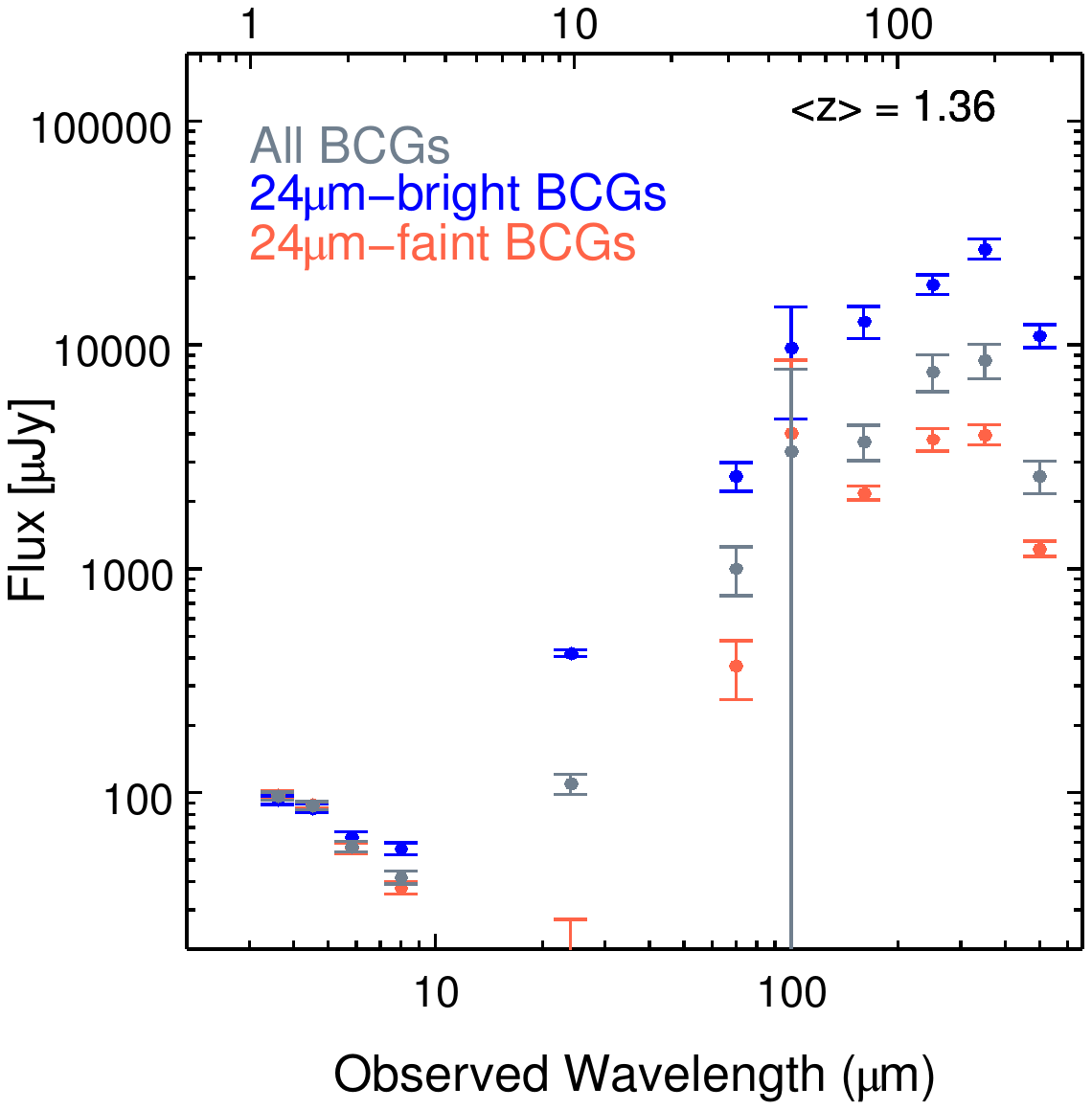} &
\includegraphics[scale=0.09, width=6.1cm, trim=2cm 7.7cm 7cm 8.3cm]{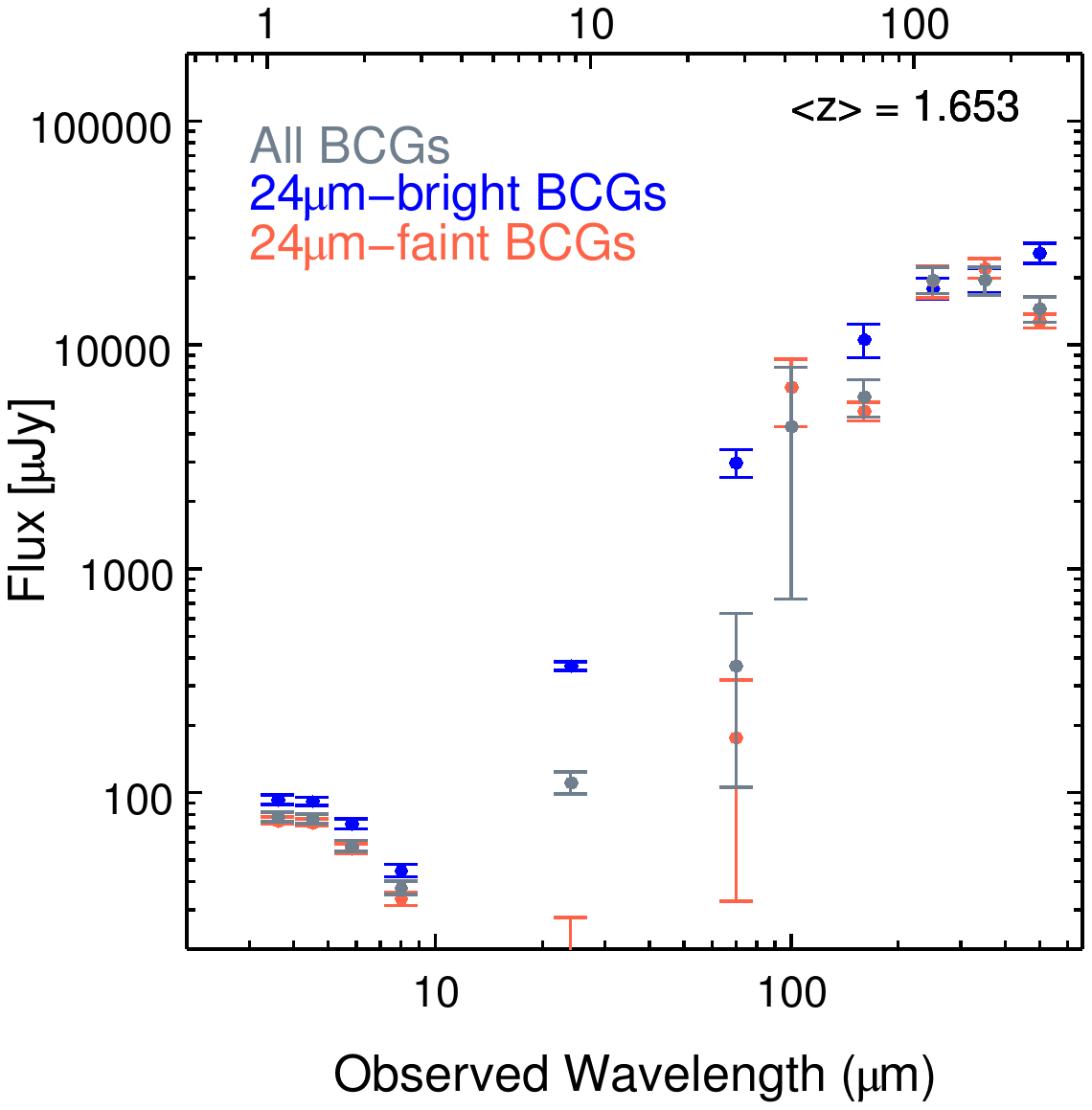} 
\end{tabular}
\vspace{3.5cm}  
\caption{(a) The stacked, broadband infrared flux SEDs of 675 {\it SpARCS} BCGs divided into six equally spaced redshift bins and two flux bins: those which are detected in the {\it Spitzer} 24-$\umu$m band, and those which are not (with the detection threshold set at 100 $\umu$Jy). The upper X-axis shows the rest-frame wavelength values corresponding to observed wavelength on the bottom X-axis.}
\end{figure*}

\subsection{Far-infrared Stacked Flux \& Flux Uncertainty}

To measure the stacked source flux in the {\it Spitzer} MIPS 70 $\umu$m and 160 $\umu$m, {\it Herschel} SPIRE 250/350/500 $\umu$m and PACS 100 $\umu$m bands, we adopt the reference aperture radii and associated aperture corrections from the SWIRE Data Release 2 document, Pearson et al. (2014), and Balogh et al. (2014), respectively.

To calculate the stacked flux uncertainty in the far-infrared bands, we employ the same procedure as that for the mid-infrared data, described in the previous section, with the following exceptions: 1) before performing aperture photometry on the {\it Herschel} SPIRE data maps in particular, which are provided in units of {\it Jy/beam}, we convert the data to {\it Jy/pixel}, heeding the advice of Nguyen et al. (2010); 2) we do not consider the Poisson noise from the source, as the effect is minimal in these wavebands where the target BCG is faint; and 3) we take an extra step to minimise the contribution from confusion noise which dominates these far-infrared wavebands, by reducing the final stacked fluxes by the factor by which the MIPS 24-$\umu$m flux increases after convolving the MIPS 24-$\umu$m flux with the far-infrared PSF, and then measuring the total MIPS 24-$\umu$m flux within the larger far-infrared flux aperture (see Section 3.1 for details). Finally, we apply the appropriate handbook calibration uncertainties to the stacked far-infrared fluxes of $5\%$ and $5.5\%$ for the {\it Herschel} SPIRE and PACS bands, respectively.

\begin{figure*}
 \includegraphics[scale=0.8,trim=1cm 0cm 0cm 13cm,clip]{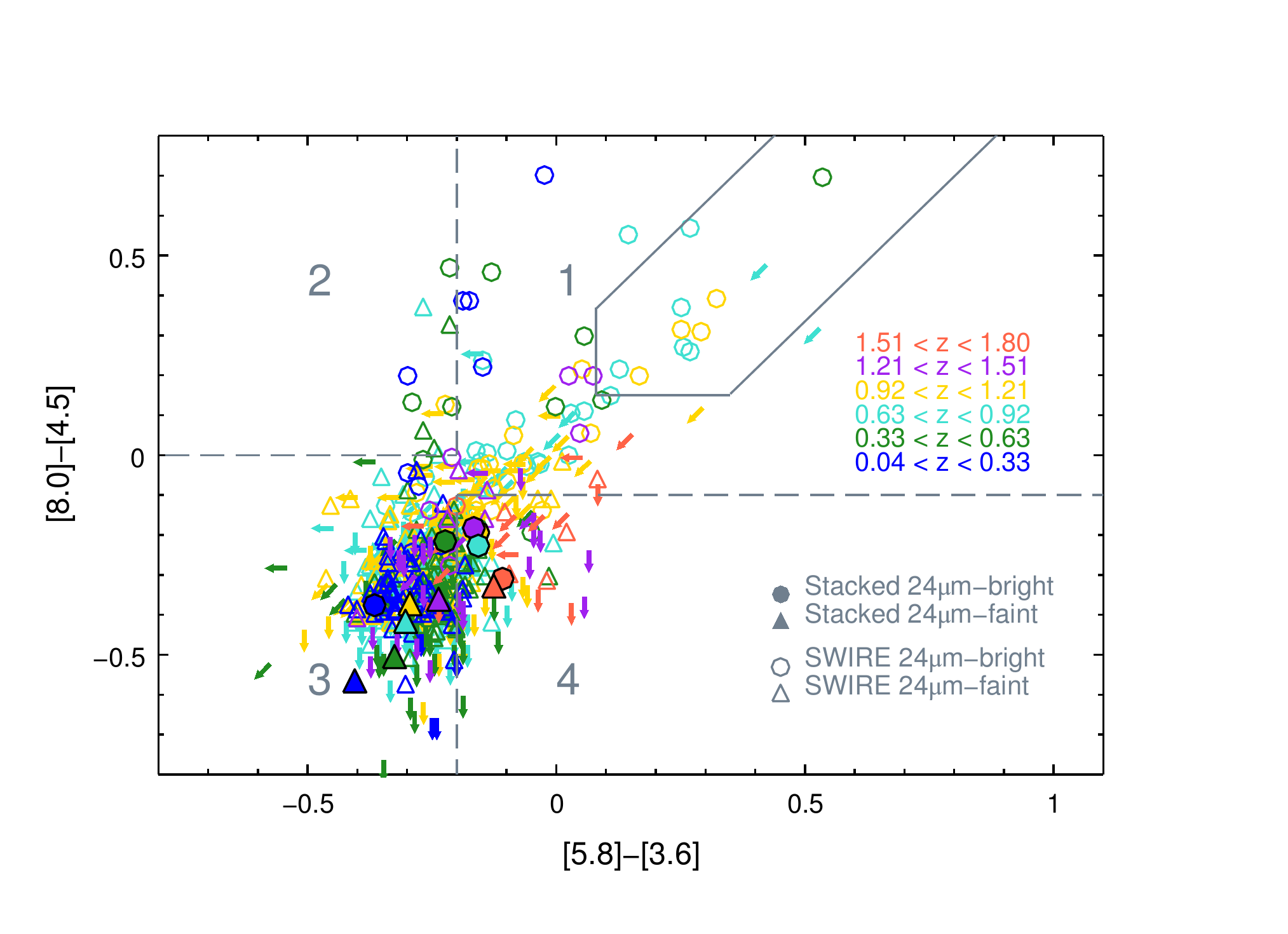}
\caption{{\it Spitzer} IRAC colour--colour diagnostic of Sajina et al. (2005) applied to the BCG stacked IRAC colours (filled symbols) resulting from the current analysis, in addition to the SWIRE IRAC colours of the 51\% of the sample which are individually detected in all four IRAC bands, and derived from the associated IRAC fluxes contained in the {\it Opt-IRAC-MIPS24} Spring 2005 {\it SWIRE} catalog. The SWIRE catalog sources missing 5.8 $\umu$m and/or 8 $\umu$m flux values are represented by arrow symbols, corresponding to the upper limits defined by the {\it Spitzer} $5-\sigma$ 5.8 $\umu$m and/or 8 $\umu$m limiting sensitivities. Regions 1-4 represent those of Figure 10 in Sajina et al. (2005), which they determine to preferentially select continuum dominated sources at $z \sim 0-2$, PAH-dominated sources at $z \sim 0.05-0.3$, stellar- and PAH-dominated sources at $z \sim 0.3-1.6$, and stellar- and PAH-dominated sources at $z \ga 1.6$, respectively. The 'AGN wedge' of Donley et al. (2012) is also shown. Both horizontal and vertical error bars associated with each data point are smaller than the symbol size (the enhanced sizes of the filled symbols representing stacked colours is for visual clarity, only, and does not reflect error sizes, which are much smaller.)}
\end{figure*}

\section{SED ANALYSIS}
\subsection{Mid-infrared Characterisation}

The stacked, median-flux, infrared {\it Spitzer}/{\it Herschel} SEDs resulting from this photometry are shown in Figure 3 in grey points, with the bright and faint contributions to each data bin displayed in coloured points for comparison. What is immediately noticeable is that the basic shape of the SED is consistent across all redshift and flux bins, suggesting a relatively uniform BCG population. 

In each of the six faint SEDs, we observe a downward-pointing slope through the 3.6-to-8.0$\umu$m {\it Spitzer} IRAC bands, a 24$\umu$m-to-8$\umu$m flux ratio smaller than unity, and a prominent far-infrared blackbody 'bump' longward of 40 $\umu$m.  Similar features are observed in the bright SEDs, with the exception of a flatter negative slope through the IRAC bands than the corresponding faint SED, as well as a 24$\umu$m-to-8$\umu$m flux ratio greater than unity in each of the bright SEDs.

\subsubsection{AGN versus Star-forming Contribution}

\underline{{\it Spitzer} IRAC Colour-Colour Diagnostic}

As in W15, we invoke the {\it Spitzer} IRAC $\log (S_{8.0}/S_{4.5})$ versus $\log (S_{5.8}/S_{3.6})$  colour--colour diagnostic, displayed in Figure 4 -- now including an additional 231 BCGs from the SWIRE ELAIS-S1 and CDFS fields which were not included in that study -- to quantify these trends and make a preliminary determination of the dominant source of infrared energy output of the BCGs as implied by their mid-infrared stacked fluxes, whether from stellar emission, AGN emission, or polycyclic aromatic hydrocarbon (PAH) molecular line emission resulting from star formation (e.g., Lacy et al. 2004 \& 2007; Stern et al. 2005; Sajina et al. 2005; O'Dea et al. 2008). According to this diagnostic, a downward-pointing slope through the 3.6-to-8.0$\umu$m range in the SED is the hallmark of a galaxy dominated by emission from a stellar population consisting of old, 'red' stars; as opposed to an AGN, which would manifest itself as an upward-pointing slope through this wavelength range due to hot emission from Very Small Grains (VSGs) (at a minimum between 5.8 and 8 $\umu$m, if not completely through the full 3.6 to 8 $\umu$m range, depending on dust geometry and galaxy viewing-angle (O'Dea et al. 2008)). The other alternative, a negative slope between 3.6 and 5.8 $\umu$m succeeded by an upward jump at 8 $\umu$m, indicates a galaxy dominated by star formation, due to the PAH emission which occurs in the rest-frame 7-12 $\umu$m wavelength range when PAH dust absorbs and re-emits UV photons from newly formed stars.

We apply the IRAC colour-colour diagnostic of Sajina et al. (2005) and the refined 'AGN wedge' of Donley et al. (2012) to both the stacked and individual IRAC colours of the BCGs in Figure 4, the former resulting from our SED analysis and the latter derived from the associated SWIRE catalog fluxes. The dashed lines in Figure 4 demarcate the four different regions shown by Sajina et al. (2005) to preferentially select different types of sources, taking into account the redshift evolution of their colours (see Figure 10 of the same work).  We find BCGs with IRAC colours spanning all four regions of the plot, but mostly concentrated in Regions 3 and 4, where mainly starlight-dominated sources reside at redshifts between 0.2 and 0.4, and a mix of starlight- and PAH-dominated sources intermingle in the 0.5-2.0 redshift range; the stacked IRAC colours of both the average bright and faint BCGs, marked by filled circular and triangular symbols, respectively, lie in these regions. Region 1, which corresponds to AGN or continuum-dominated sources through  $z = 2$, contains a significant fraction of the 24${\umu}$m-detected BCG population, while Region 2, where only low-redshift PAH-dominated galaxies are typically found, is sparse.
  
The AGN selection 'wedge' of Donley et al. (2012), marked by solid lines within Region 1 of Figure 4, has been shown to effectively remove contamination from high-redshift star-forming galaxies and recover X-ray-detected luminous AGN to a high level of completeness and reliability (Donley et al. 2012). This selection criterion yields a contribution of 10 luminous AGN to our sub-sample of 167 24-$\umu$m-bright BCGs, with all but one lying within the $0.78 < z < 1.36$ redshift range. We note, however, that these numbers reflect only {\it luminous} AGN, as the Donley et al. (2012) criterion misses 39\% of spectroscopically confirmed mid-infrared-bright AGN (Kirkpatrick et al. 2013), i.e. low-luminosity and/or host-dominated AGN, and those with a strong 9.7-$\umu$m silicate absorption feature. However, we expect that the fraction of AGN missed by the Donley wedge on account of dust attenuation to be small, and that those which are, are likely to reside in host-dominated, star-forming systems, based on the findings of Goulding et al. (2012): less than half of all nearby, hard X-ray-selected, Compton-thick AGN exhibit strong silicate absorption, with the dust attenuation appearing to originate in extra-nuclear star-forming regions within the host galaxy, as opposed to the AGN dust torus, as expected for these systems.

Finally, if we assume that all of the BCGs in Region 1 of the IRAC colour-colour plot which lie outside the Donley wedge are likely to represent AGN/star-forming composite systems (i.e., without interloping high-redshift, purely star-forming galaxies), the fraction of bright BCGs in our sample harbouring AGN activity would rise to 50\%, the majority still lying at $0.78 < z < 1.36$ (though we note that X-ray-detected AGN have been discovered 'hiding' in all four areas of the IRAC colour space, according to, e.g., Mendez et al. 2013).  As for the remaining half of the bright BCG subset, their IRAC colours indicate a mix of starlight- and PAH-dominated emission. Therefore, while the exact number of AGN in the SpARCS BCG sample cannot be constrained through mid-infrared colour selection due to the potential for contamination in various regions, we {\it can} learn that, if they exist in larger numbers than implied by this diagnostic, very few of them dominate over the emission from their host galaxy.

\begin{table}
 \centering
 \begin{minipage}{140mm}
  \label{tab_one}
  \begin{tabular}{@{}lcc@{}}
   \hline
                           & Radio-loud            &      Radio-quiet            \\
    \hline
    $24\umu$m-bright &     20/148 (4 mid-IR AGN) &      6/148 (2 mid-IR AGN)     \\
    $24\umu$m-faint  &     0/148               &      122/148 (3 mid-IR AGN)     \\
   \hline
  \end{tabular}
 \caption{The fraction of the radio-detected BCGs in our sample \newline as a function of radio-loudness and 24-${\umu}$m flux based on the \newline Mid-infrared Radio Correlation. The number of BCGs in each \newline category which are classified as AGN by the IRAC colour-colour \newline diagnostic are noted, yielding a total of only 9 mid-IR AGN \newline candidates with a radio detection.}
 \end{minipage}
\end{table}

\begin{figure}
 \includegraphics[scale=0.64,trim=1cm 0cm 1cm 15cm,clip]{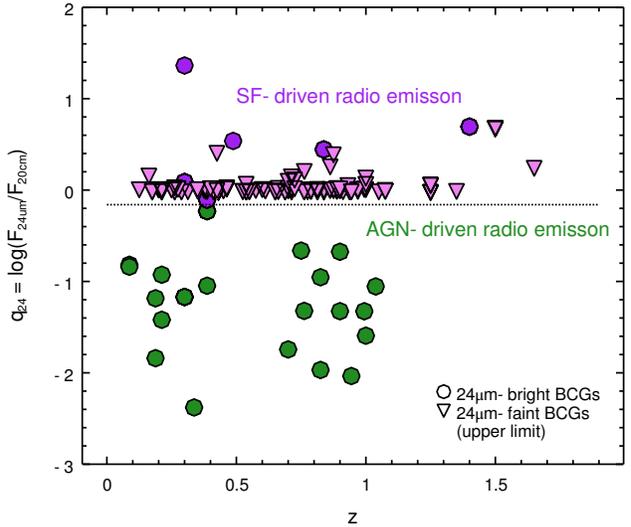}
\vspace{-.5cm}
\caption{The Mid-infrared Radio Correlation ($q_{24}$ parameter) of the radio-detected BCGs in our sample, as a function of redshift. (The upper-limit values for the radio-detected 24$\umu$m-faint BCGs, corresponding to the 100 ${\umu}$Jy detection threshold, are shown for completeness.)}
\end{figure}


\underline{Mid-infrared Radio Correlation}

In Figure 5 we plot the Mid-Infrared Radio Correlation (MRC), or $q_{24}$ ($= log(F_{24um}/F_{20cm})$) parameter, as a function of redshift for the 24$\umu$m-detected BCGs in our sample that have an associated 20 cm (1.4 GHz) radio detection. This simple parameter has been shown to robustly identify AGN-driven radio activity as being characterised by a $q_{24}$ value of less than $-0.16$ (e.g., Norris et al. 2006; Middelberg et al. 2008; Mao et al. 2012), and therefore relative radio-loudness, largely independent of the redshift of the emission or the host galaxy contribution (Appleton et al. 2004; Wilkes et al. 2009; Sargent et al. 2010; Mao et al. 2011; Hales et al. 2014). While a single-band radio analysis does not afford a complete characterisation of the radio properties of the BCGs in our sample -- e.g., to determine the number, character, physical size, or age of the distinct radio components which may be associated with the BCGs (Hogan et al. 2012) -- the $q_{24}$ parameter offers an additional, reliable means by which to determine the fraction of BCGs in our sample with a dominant AGN.

We find that 21.3\% (148/716) of the the full BCG sample are associated with radio sources within the beam size of each radio survey searched, noting that an AGN with a low/moderate radio luminosity of $L_{1.4GHz} = 10^{45}$ $W/Hz$ would fall below the flux density detection limit of the most sensitive of the radio surveys we employ (1 mJy) beyond a redshift of $z\sim0.3$. Furthermore, we did not attempt to determine the reliability of the matches, neither through a visual comparison of corresponding radio and infrared data images of each BCG, nor by calculating a false-source detection rate. We show the breakdown of the detections into the four permutations of bright versus faint and radio-loud versus radio-quiet categories in Table 5. While the MRC places the upper-limit $q_{24}$ values for the radio-detected 24$\umu$m-faint BCGs in the 'star-forming' region of Figure 5 (assuming the 24-$\umu$m flux threshold value of 100 ${\umu}$Jy), these points are only depicted for the sake of completeness, as it is clear that this diagnostic is only meaningful for those galaxies with a 24$\umu$m-flux detection.

The application of the MRC to our BCG sample diagnoses $20/148$ of the radio-detected BCGs as radio-loud AGN, all of which are 24$\umu$m-bright, noting that only four of these BCGs are also classified as AGN based on the IRAC colour-colour diagnostic discussed previously. For the remaining sixteen radio-loud sources, we consider three possible explanations for their lack of a matching mid-infrared AGN classification, given that only high-redshift ($z > 1$) radio AGN with extremely outlying radio-to-mid-infrared flux ratios, and also lacking a 3.6-$\umu$m detection (see discussion of infrared-faint radio sources, IFRS, in Zinn et al. 2011 and references therein), would be expected to exhibit the radio luminosity of an AGN, but no mid-infrared AGN signature: 1) they are candidates for radio mismatches to the corresponding BCG optical position and therefore are physically unassociated with the BCG, but rather are associated with a neighbouring galaxy or a mini-halo resulting from the 'sloshing' of the surrounding intracluster medium centred on the BCG position (Hogan et al. 2014 and references therein); 2) the radio emission arises from the lobes of the BCG, as opposed to the core, and therefore may be temporally separated from the infrared emission stemming from the core (i.e., that the radio lobe emission traces past activity of an inactive central engine, Hogan et al. 2014); or 3) if their radio emission is the result of non-thermal synchrotron emission emanating from the jets associated with a black-hole accretion disk at the core of the BCG, the thermal mid-infrared signature of the disk must be too weakly radiating to dominate over the emission from the host galaxy.

Regardless, for the purposes of our statistical study, we consider only that a minority, 10.3\% of SpARCS BCGs at $z < 1$ (with only high radio luminosities detectable at $z > ~0.3$), are classified as radio-loud AGN by the MRC, with 80\% of these systems lacking a corroborating mid-infrared AGN signature. This leads us to conclude that these radio-loud AGN are low-luminosity AGN (i.e., their accretion disks are too weakly emitting to dominate the host galaxy mid-infrared emission) --  or, if their radio emission has been misidentified, as offered as a possibility above, then they lack non-thermal synchrotron emission emanating from jets associated with a black-hole accretion disk. While we cannot comment on the fraction of low-to-moderate radio AGN residing amongst the high-redshift BCG population, the MRC shows no high-luminosity AGN above $z=1$.

Furthermore, while the $q_{24}$ parameter itself cannot readily distinguish between the radio emission from a radio-weak AGN and that from a galaxy dominated by star formation processes (the radio emission in this case being due mainly to synchrotron emission from cosmic ray electrons which are accelerated by supernovae shocks, Mao et al. 2011), it is more likely that the 128 radio-quiet detections in our sample are star-forming galaxies as a) they are not classified as AGN by the IRAC colour-colour diagnostic; and b) radio-quiet AGN are not likely to be found in elliptical galaxies (Wilson et al. 1994). Furthermore, while only 10\% of AGN from the SDSS survey are radio-loud (Netzer 2013), implying that they are rare, every known BCG in the centre of an X-ray-detected, cool-core cluster hosts a radio-loud AGN, widely accepted to be the direct result of a cooling flow (e.g., Sun et al. 2009); and Chiaberge et al. (2015) show that radio-loud AGN in the field unambiguously reside in major mergers. Therefore, with BCGs being subjected to higher merger rates than field galaxies (e.g., Mihos et al. 2003), and also uniquely interacting with cluster cooling flows, we might expect a larger fraction of AGN in BCGs to exhibit radio loudness. With cool-core clusters comprising 50-70\% of all low-redshift, X-ray-detected clusters (Santos et al. 2008), we can conclude that cool-core clusters do not occupy a large fraction of our low-redshift sample based on the low number of candidate radio-loud AGN detected. Furthermore, the results of the MRC applied to our sample do not contradict the observations that gas-rich major mergers are rare events, and that not every major merger ignites an AGN.

Therefore, the MRC merely adds to a growing body of evidence that the SpARCS BCG sample is likely to be dominated by star-forming and quiescent systems, adding only 16 potential low-luminosity AGN candidates which may be hidden in Regions 1, 3, or 4 of the IRAC colour-colour plot.

\begin{figure*}
\begin{tabular}{lll}
\includegraphics[scale=0.1, width=6.1cm, trim=2cm 7.9cm 7cm 15.5cm]{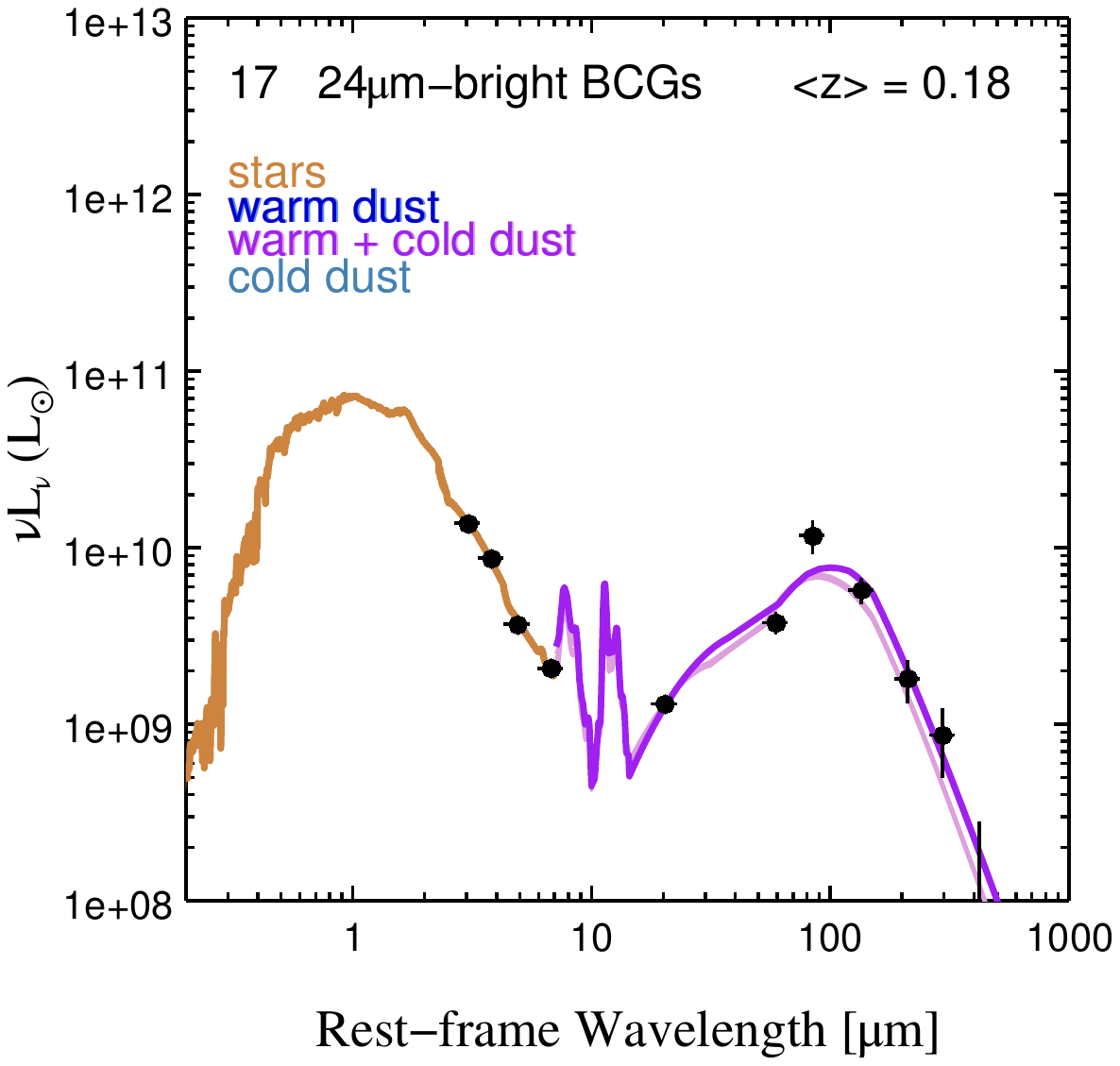}  &
\includegraphics[scale=0.1, width=6.1cm, trim=2cm 7.9cm 7cm 15.5cm]{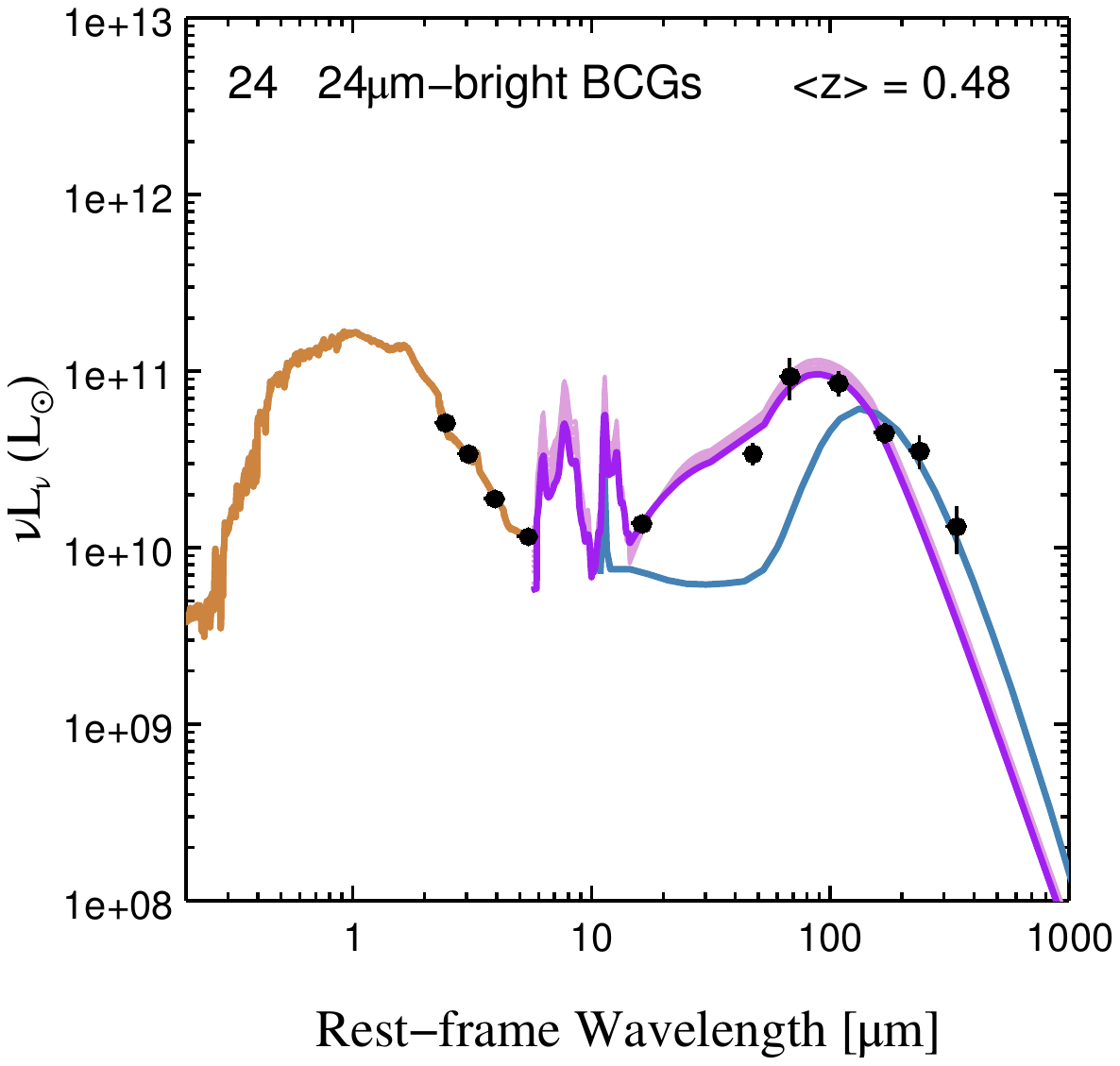}  &
\includegraphics[scale=0.1, width=6.1cm, trim=2cm 7.9cm 7cm 15.5cm]{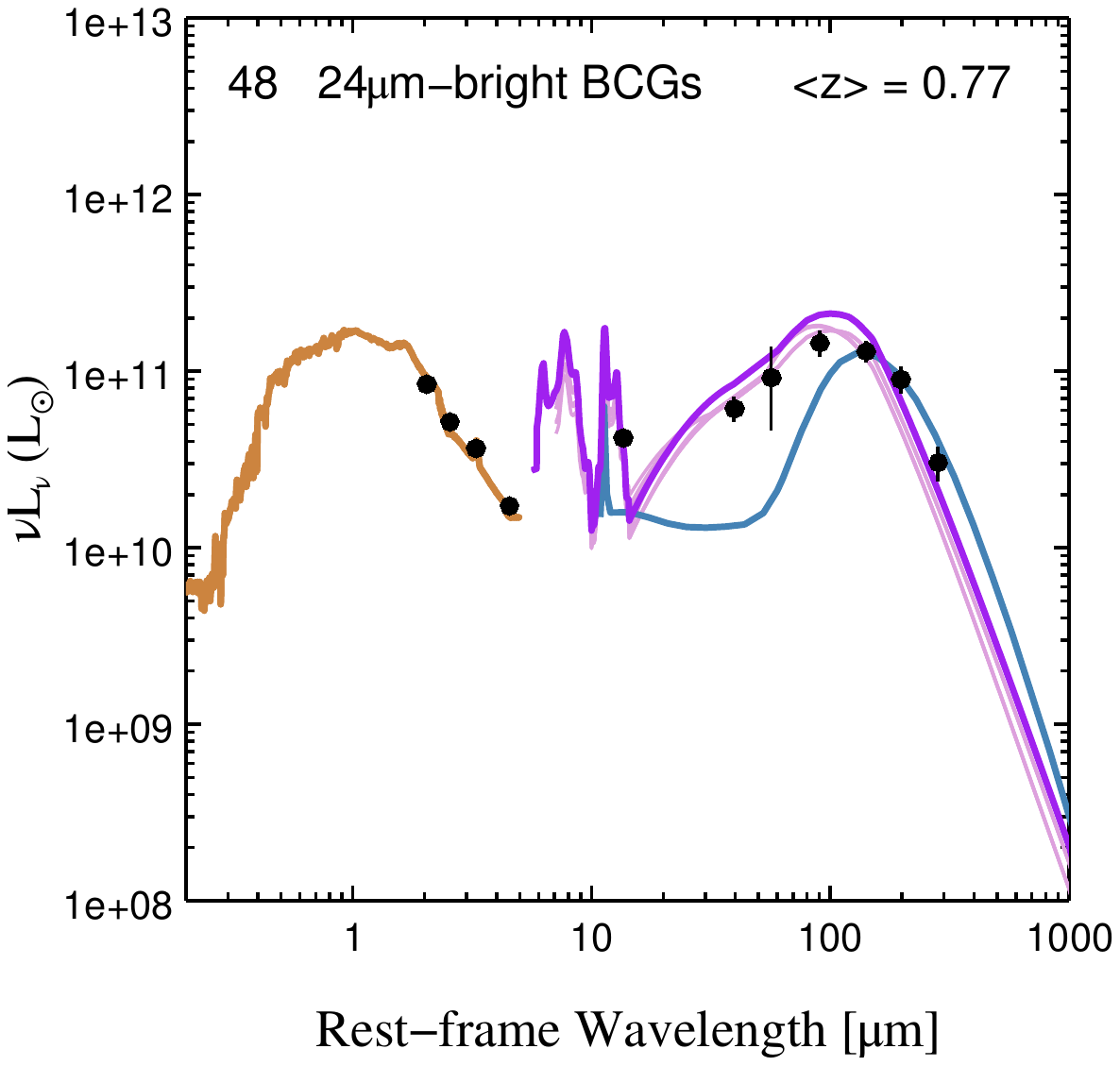}  \\
\includegraphics[scale=0.1, width=6.1cm, trim=2cm 7.9cm 7cm 8.3cm]{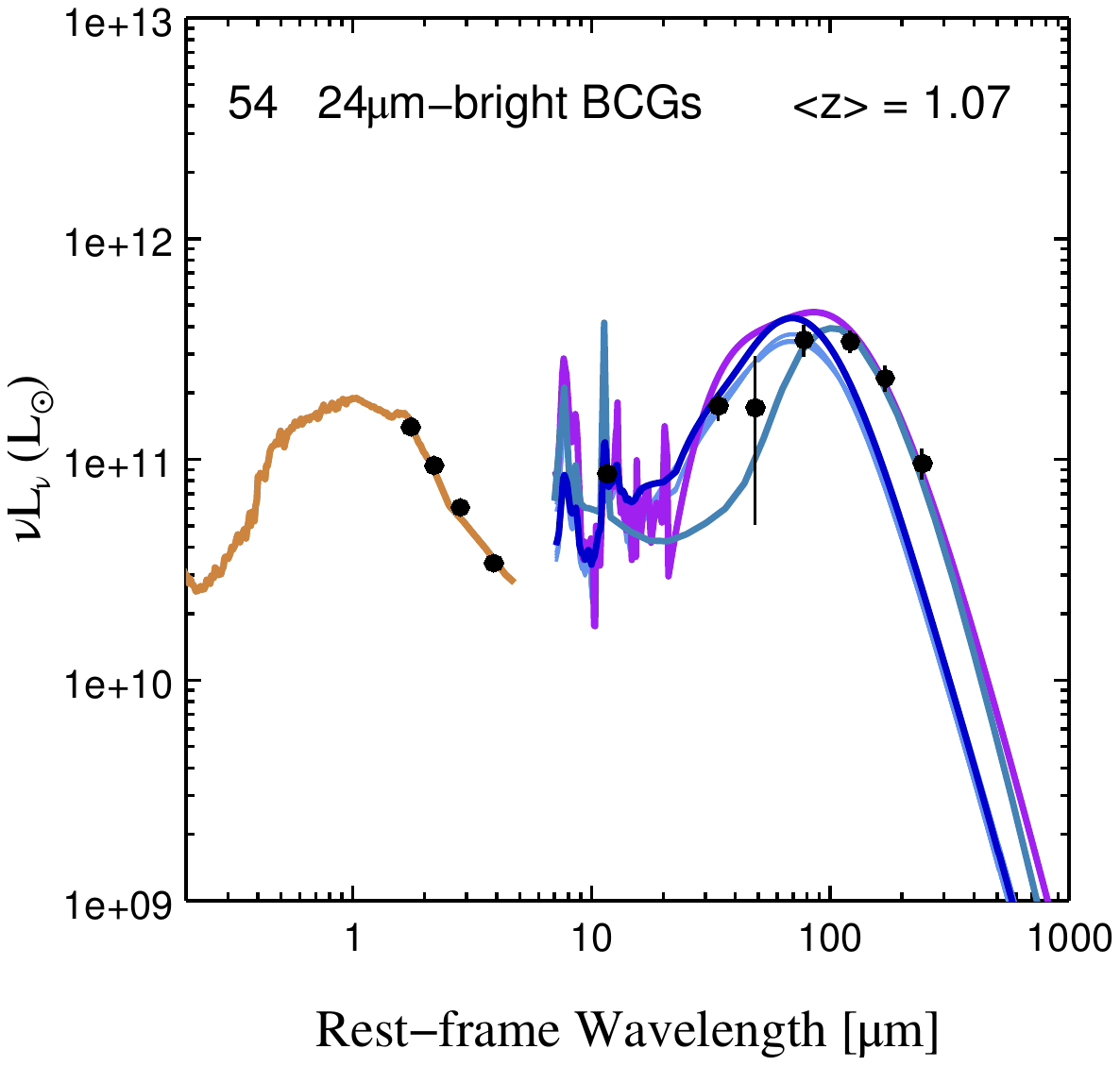}  &
\includegraphics[scale=0.1, width=6.1cm, trim=2cm 7.9cm 7cm 8.3cm]{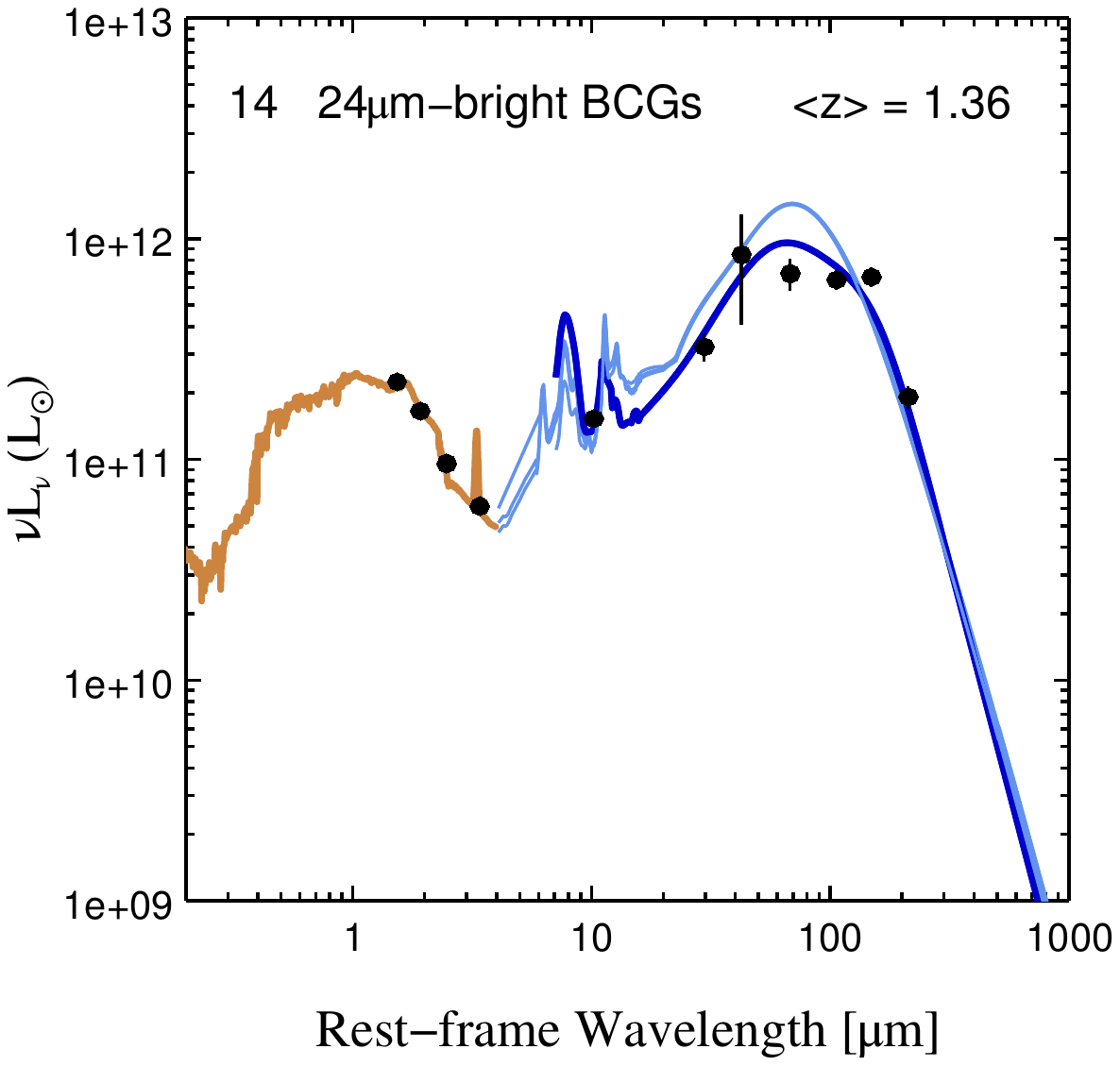}  &
\includegraphics[scale=0.1, width=6.1cm, trim=2cm 7.9cm 7cm 8.3cm]{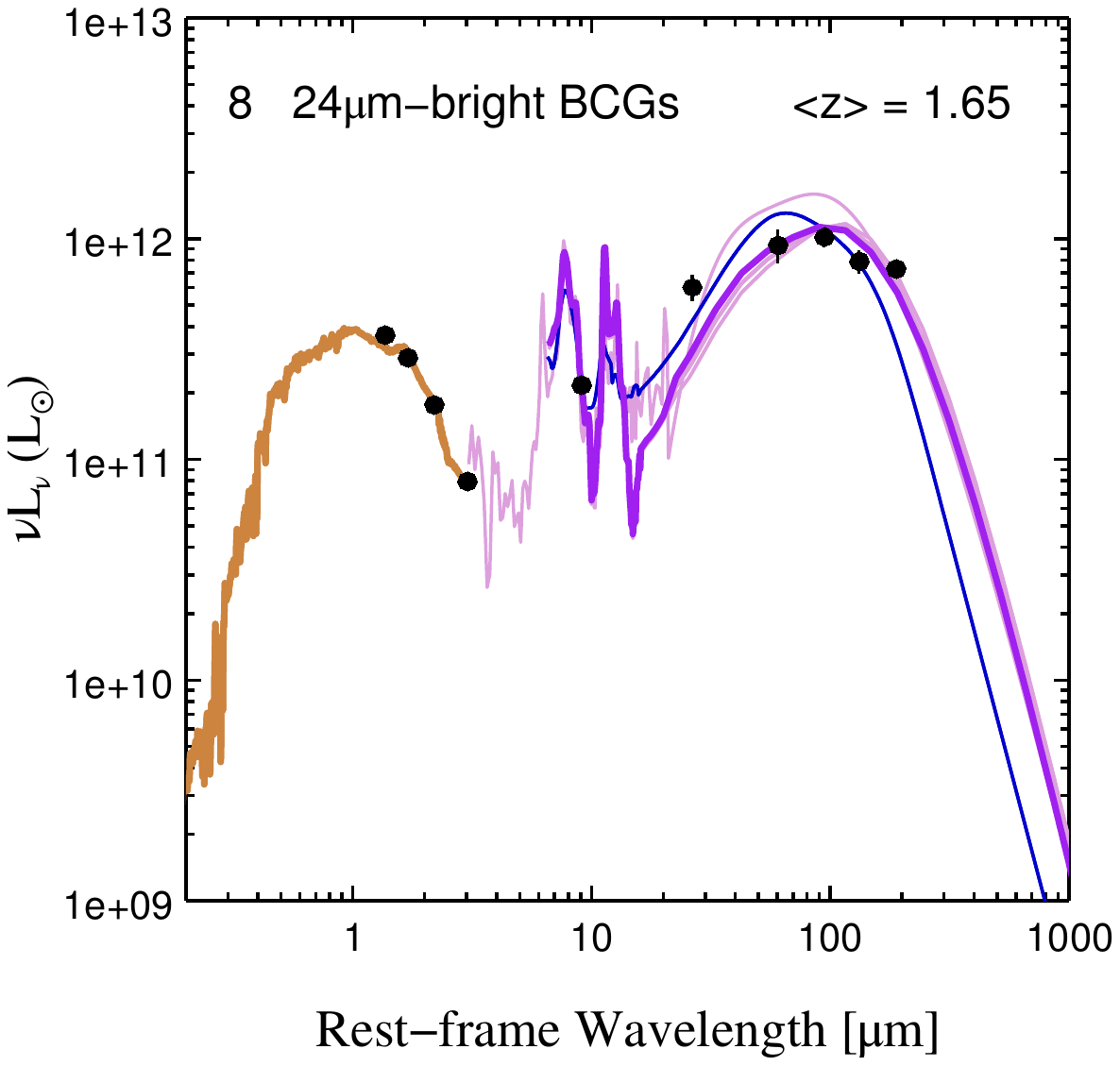}   \\
\includegraphics[scale=0.1, width=6.1cm, trim=2cm 7.9cm 7cm 8.3cm]{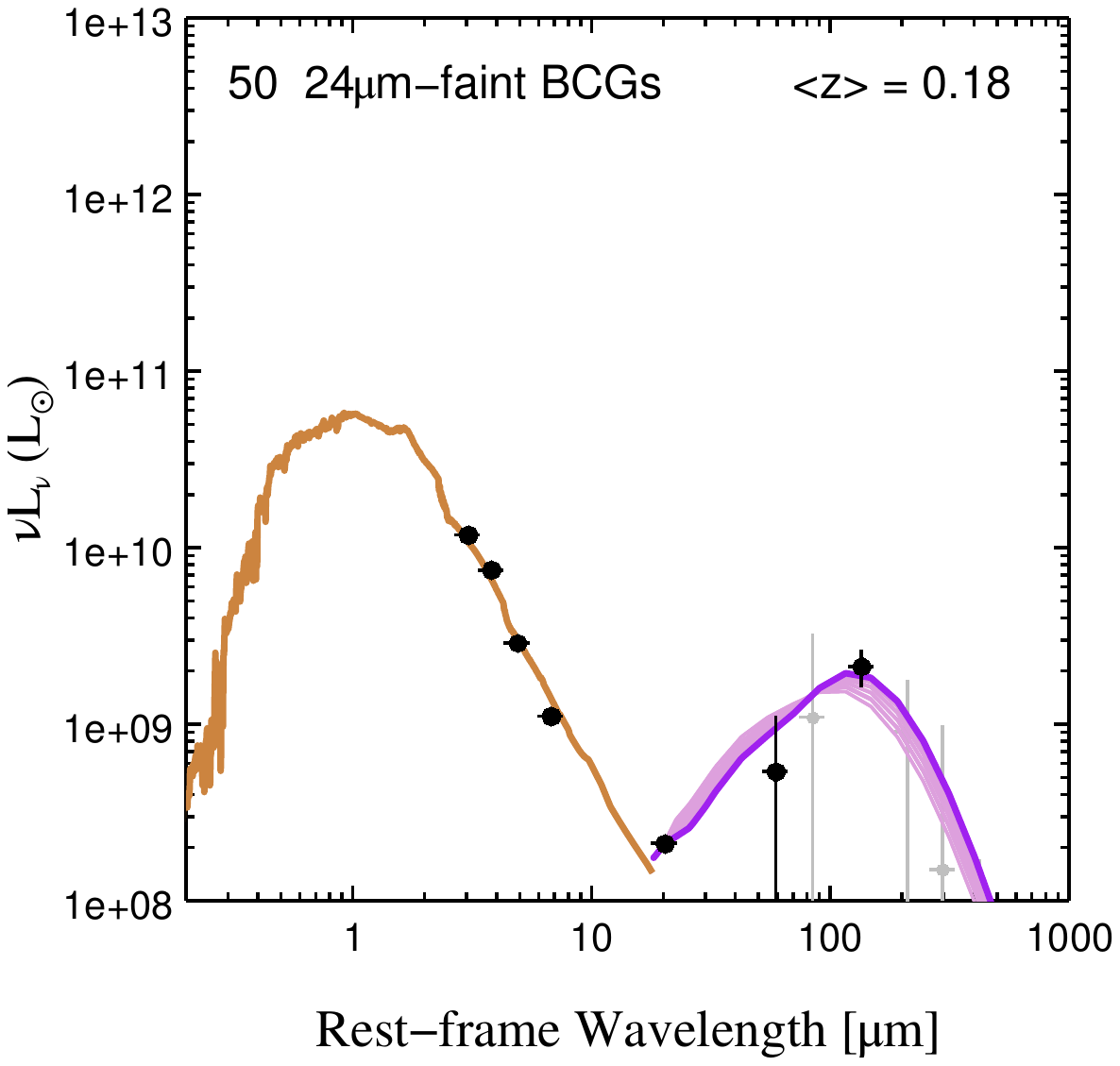} & 
\includegraphics[scale=0.1, width=6.1cm, trim=2cm 7.9cm 7cm 8.3cm]{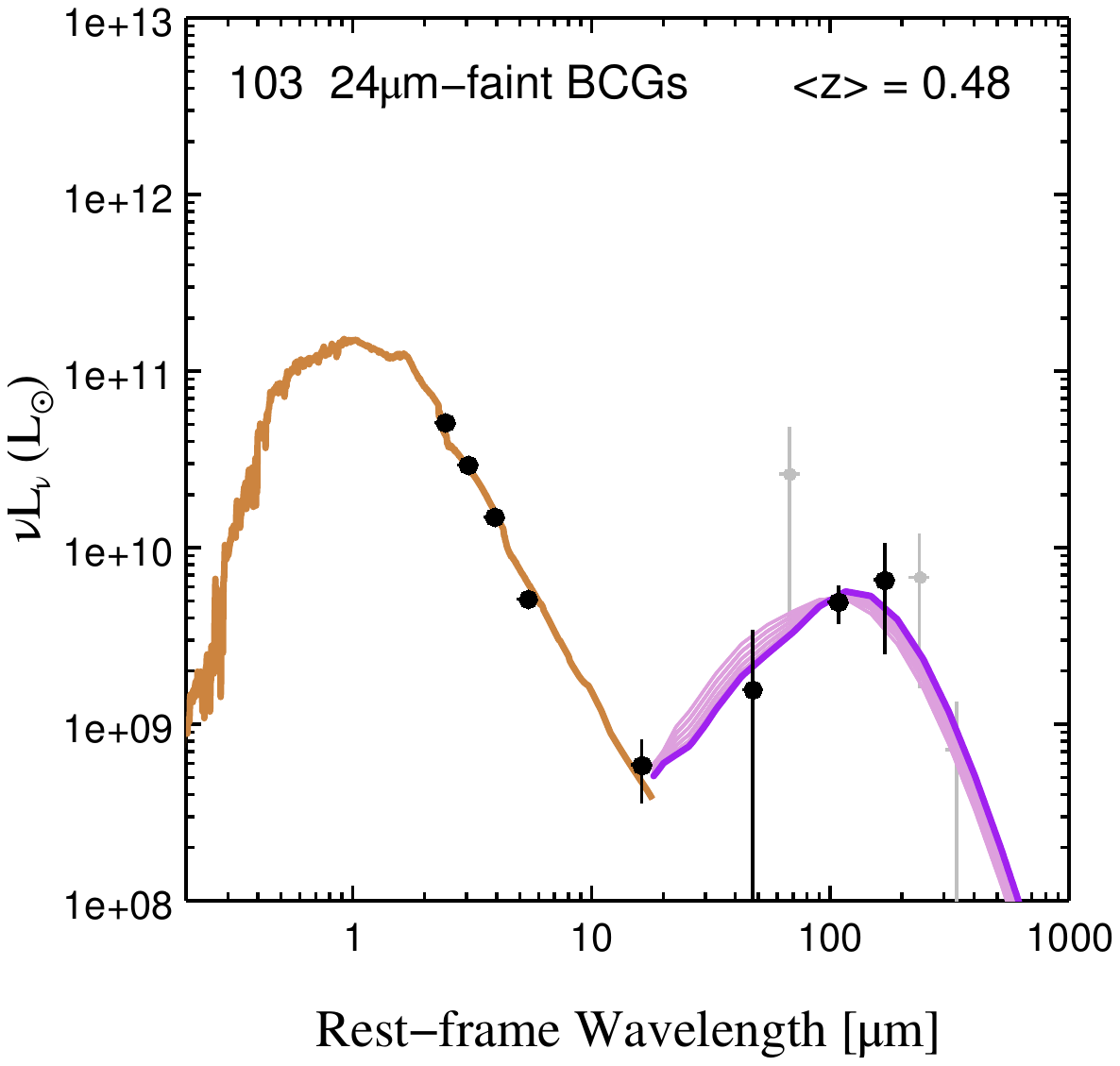} &
\includegraphics[scale=0.1, width=6.1cm, trim=2cm 7.9cm 7cm 8.3cm]{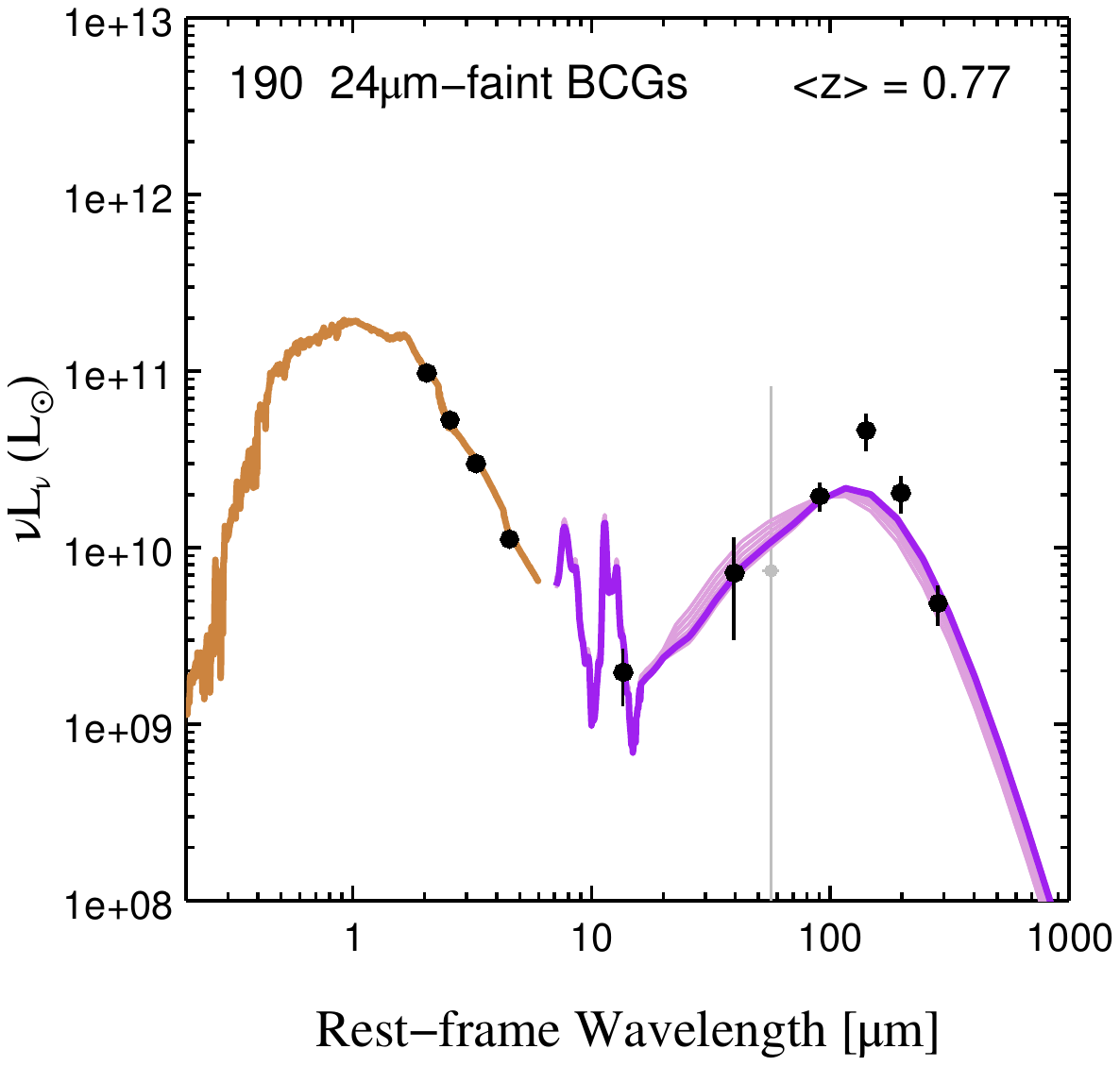}  \\
\includegraphics[scale=0.1, width=6.1cm, trim=2cm 7.7cm 7cm 8.3cm]{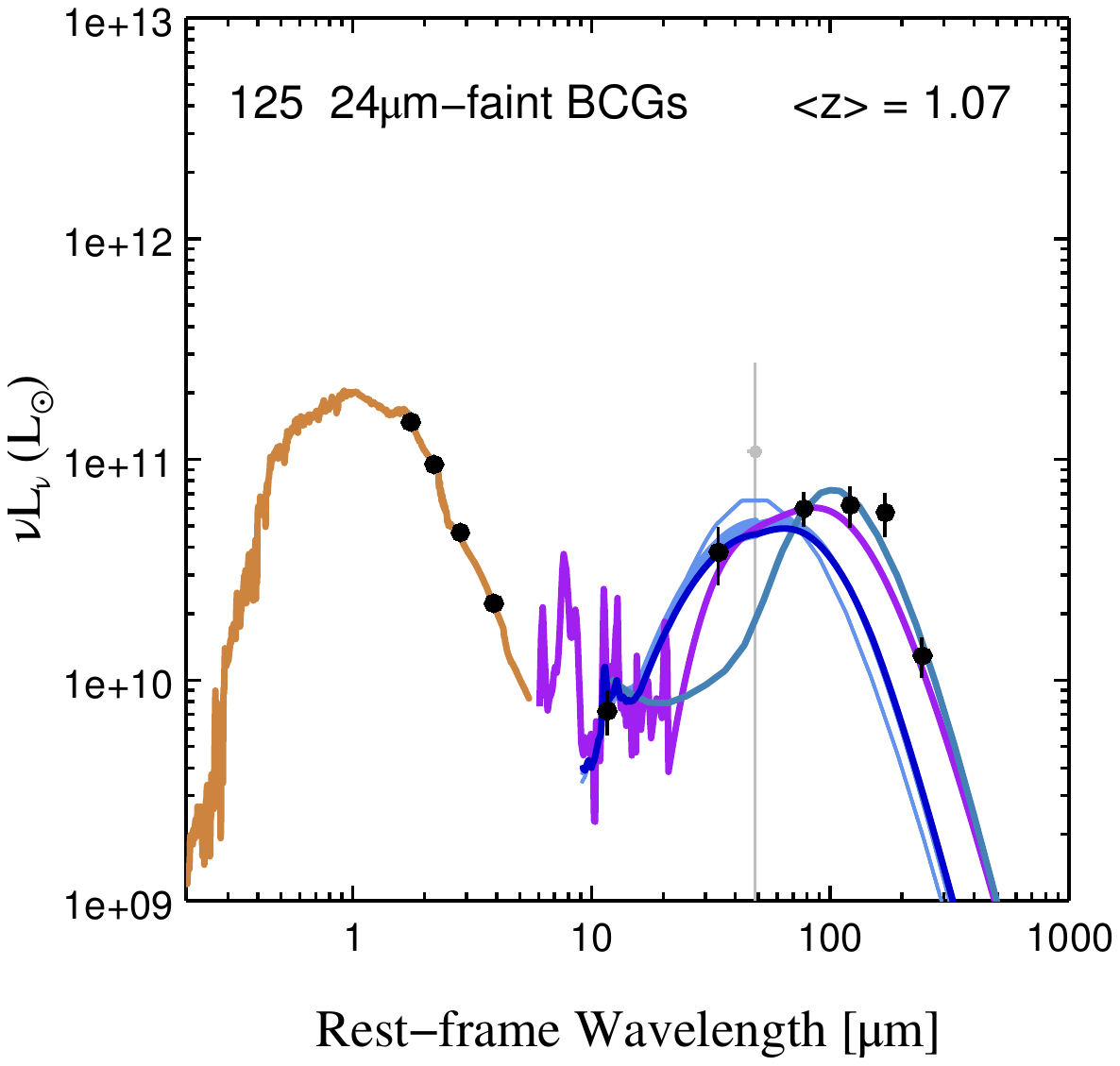} &
\includegraphics[scale=0.1, width=6.1cm, trim=2cm 7.7cm 7cm 8.3cm]{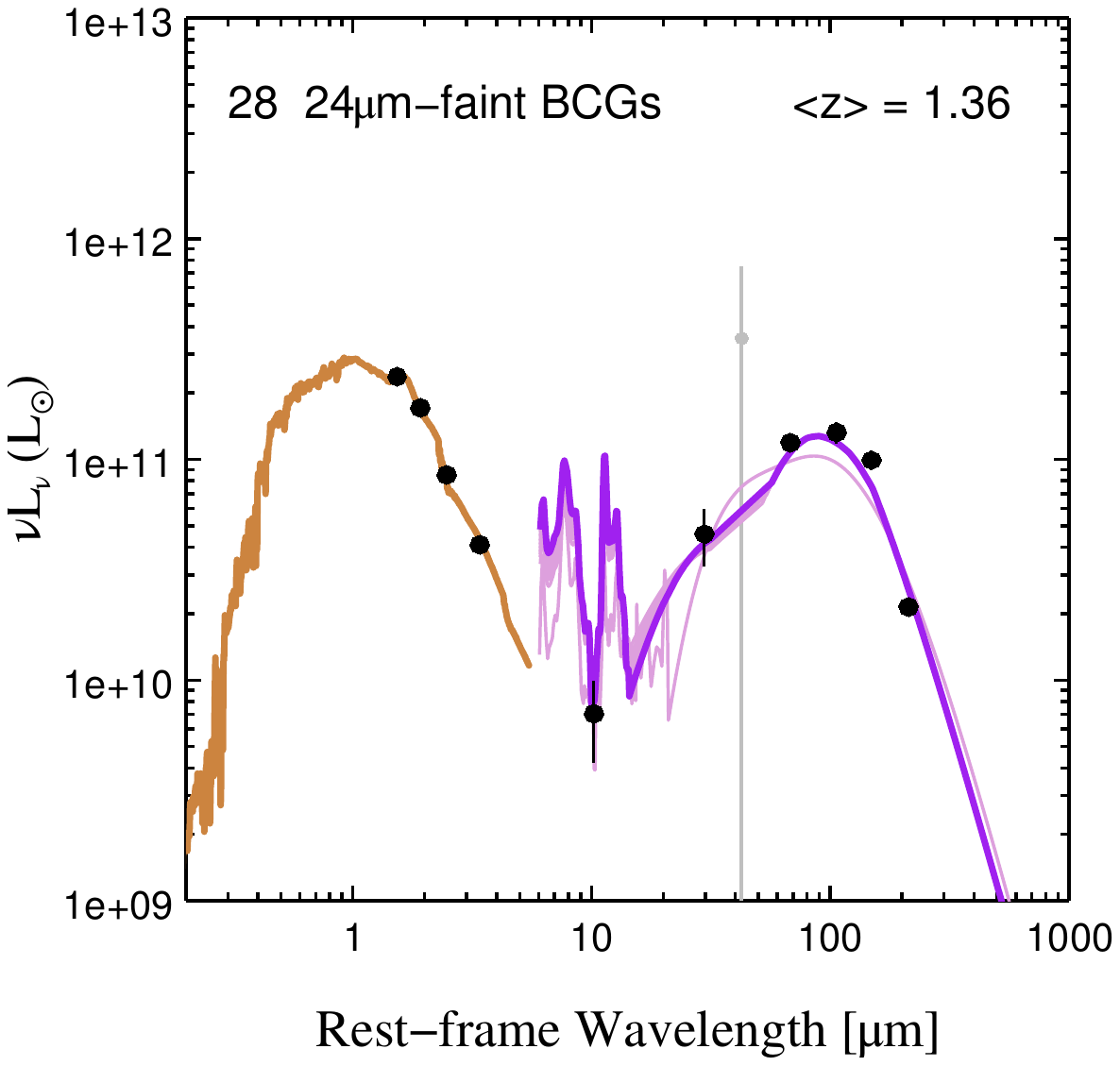} &
\includegraphics[scale=0.1, width=6.1cm, trim=2cm 7.7cm 7cm 8.3cm]{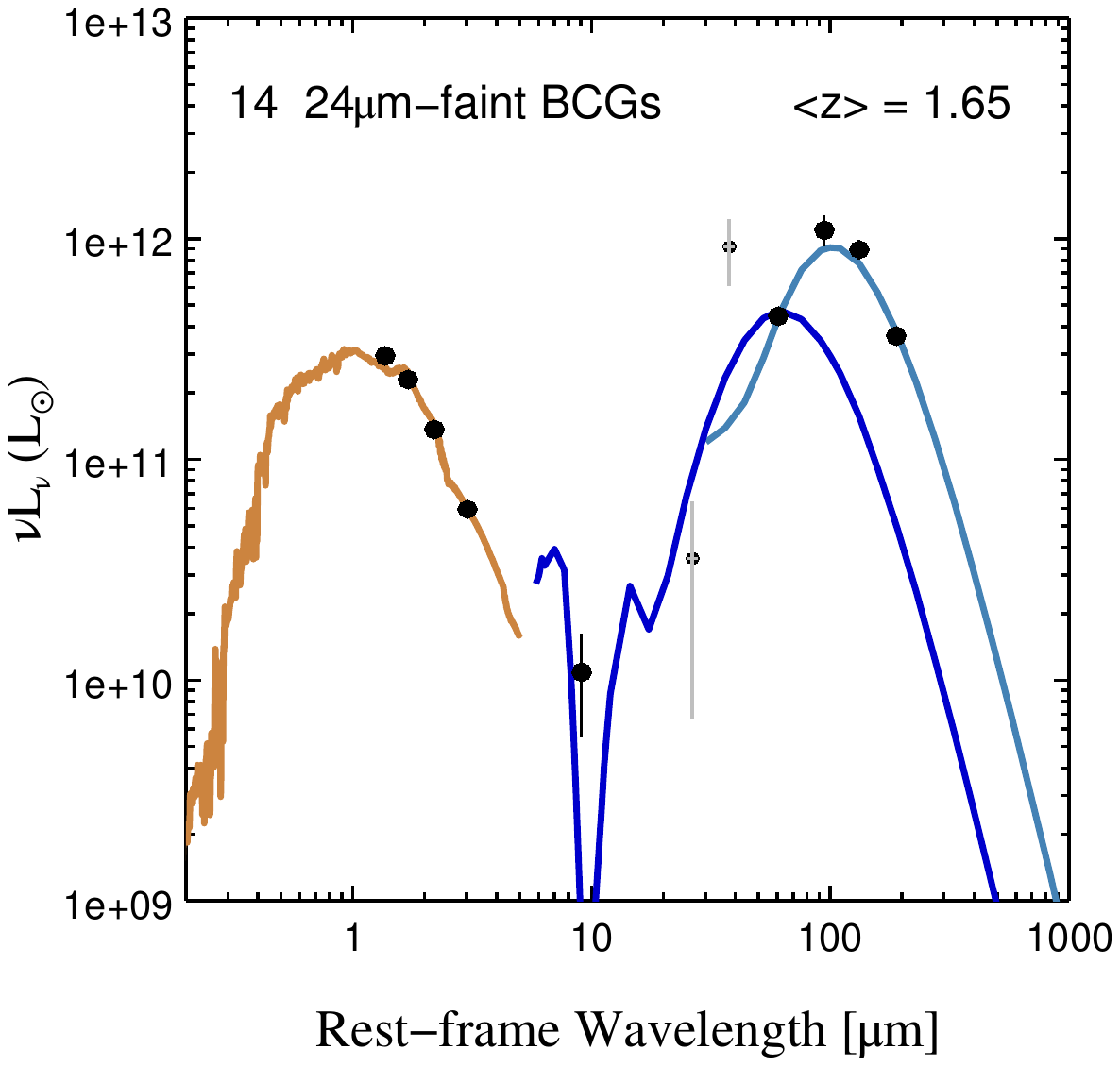}  \\
\end{tabular}
\vspace{3cm} 
\end{figure*}

\begin{figure*}
\begin{tabular}{lll}
\includegraphics[scale=0.1, width=6.1cm, trim=2cm 7.9cm 7cm 15.5cm]{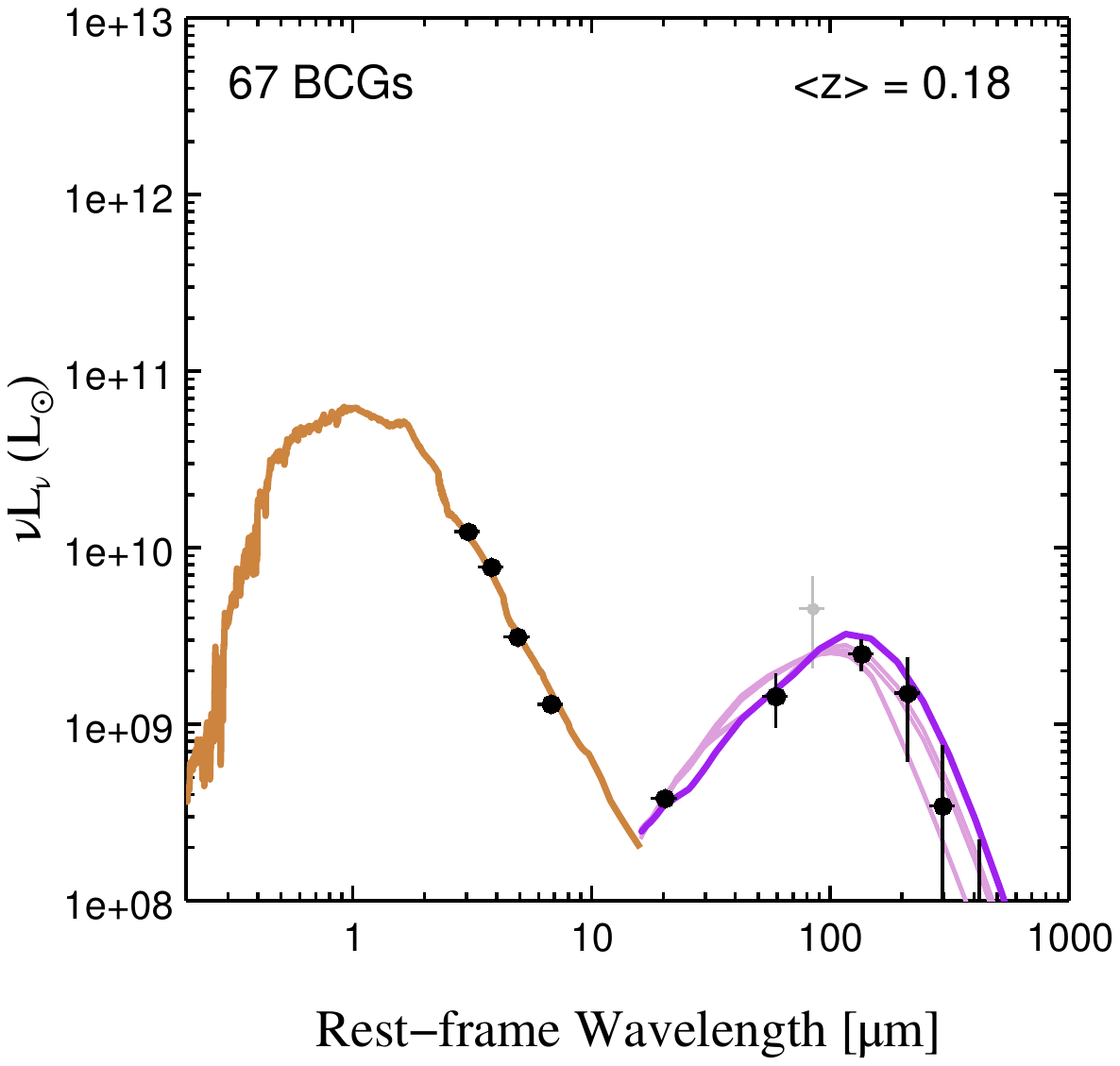} &
\includegraphics[scale=0.1, width=6.1cm, trim=2cm 7.9cm 7cm 15.5cm]{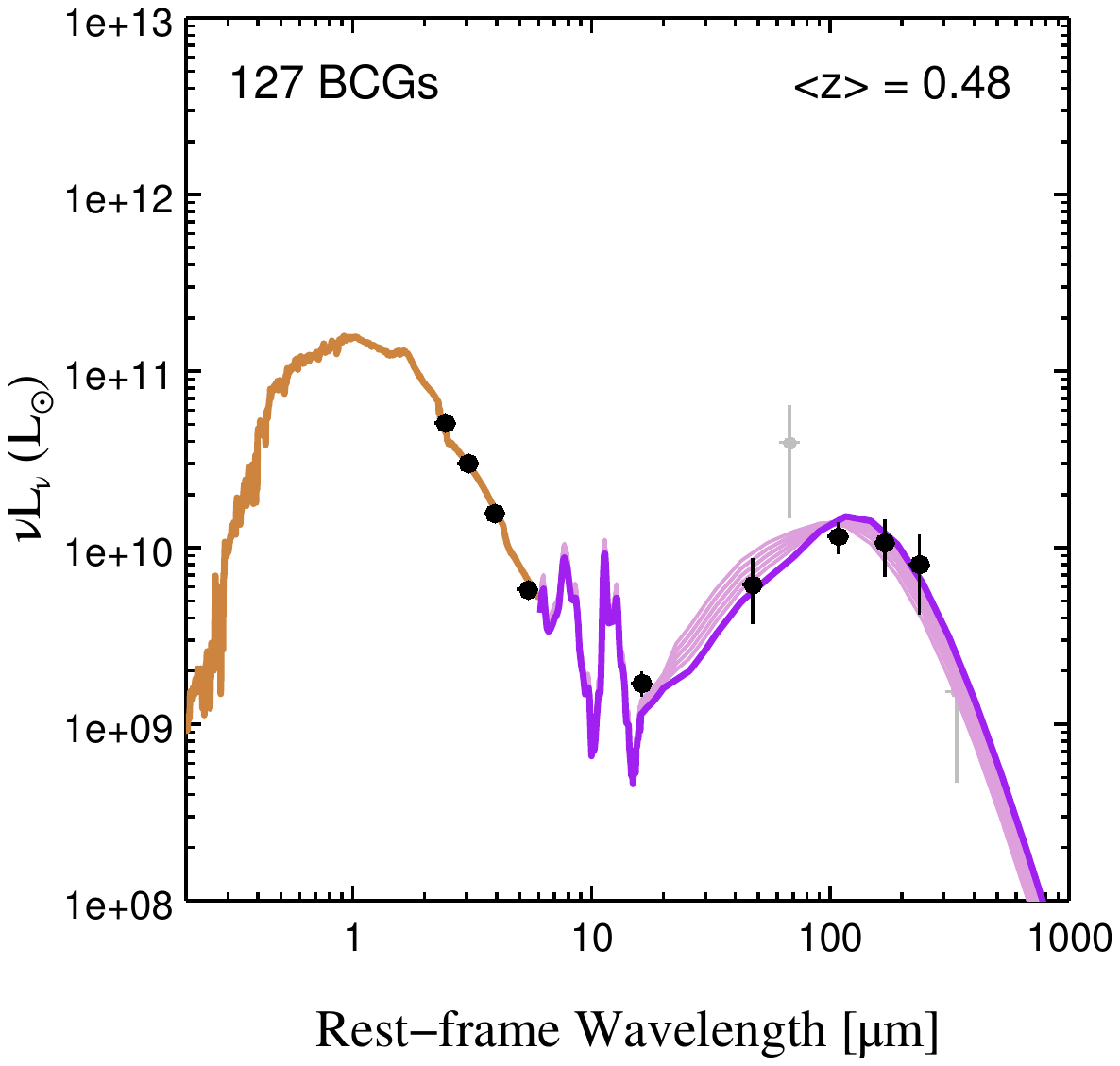} &
\includegraphics[scale=0.1, width=6.1cm, trim=2cm 7.9cm 7cm 15.5cm]{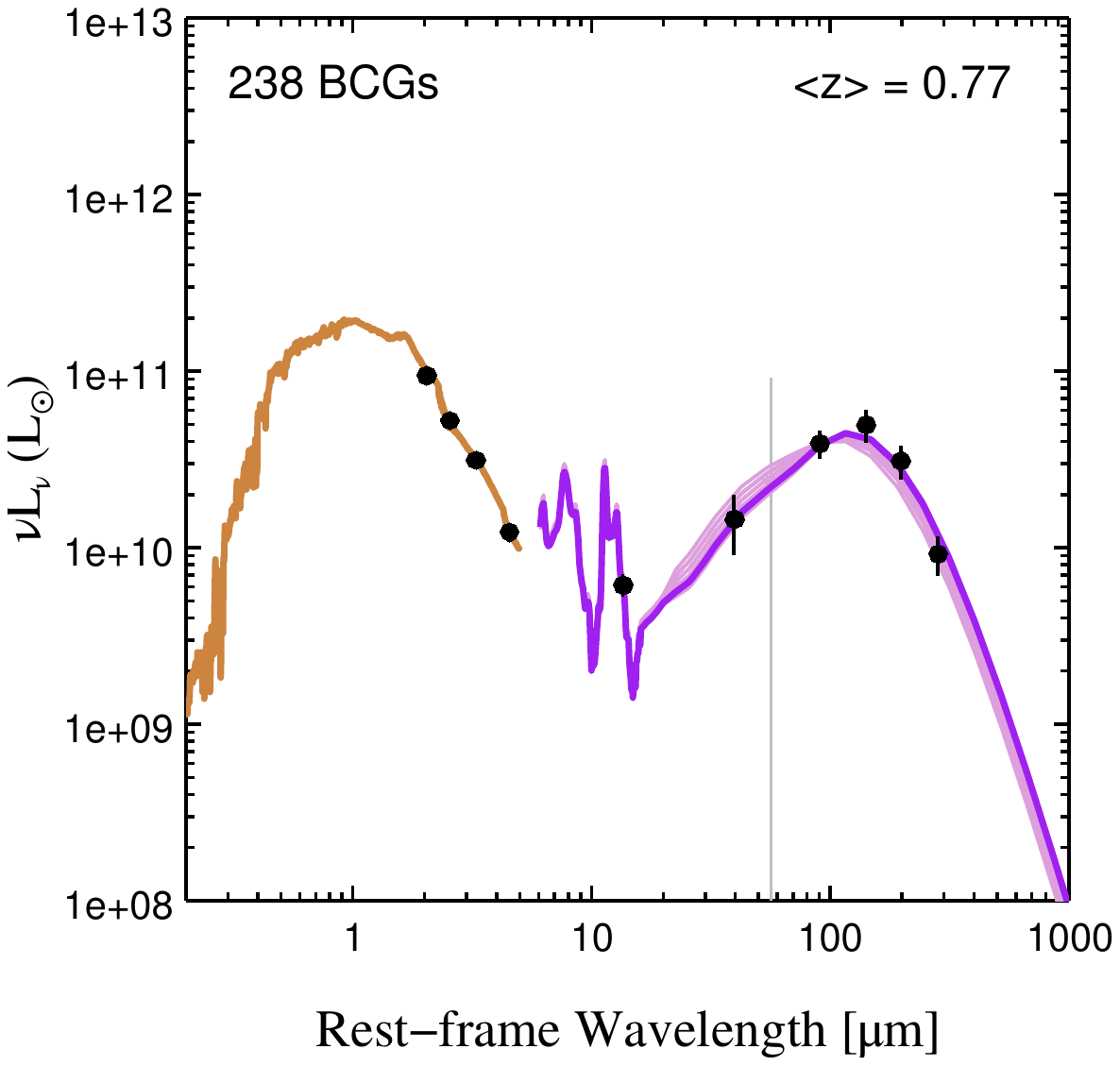}  \\
\includegraphics[scale=0.1, width=6.1cm, trim=2cm 7.7cm 7cm 8.3cm]{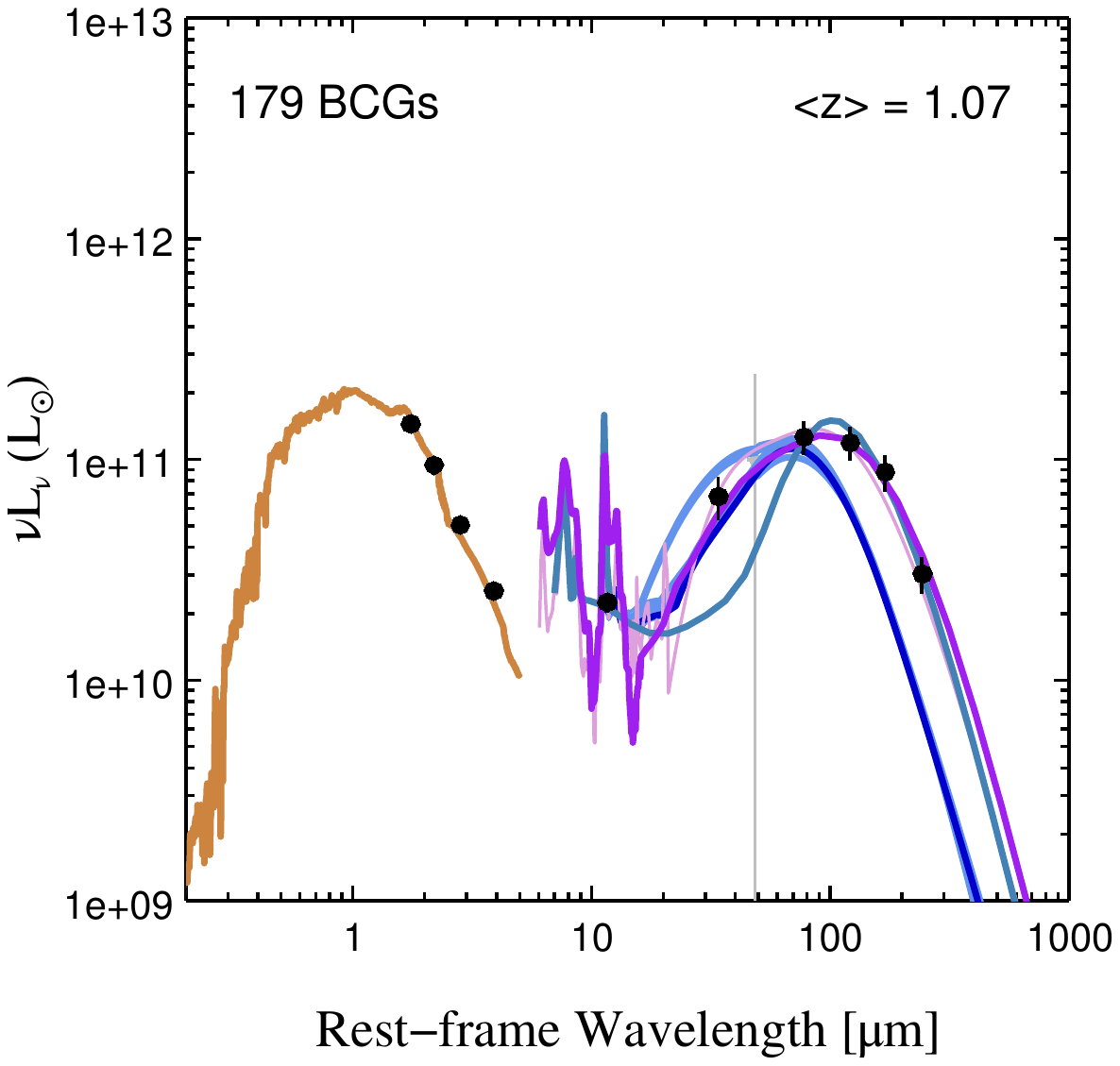} &
\includegraphics[scale=0.1, width=6.1cm, trim=2cm 7.7cm 7cm 8.3cm]{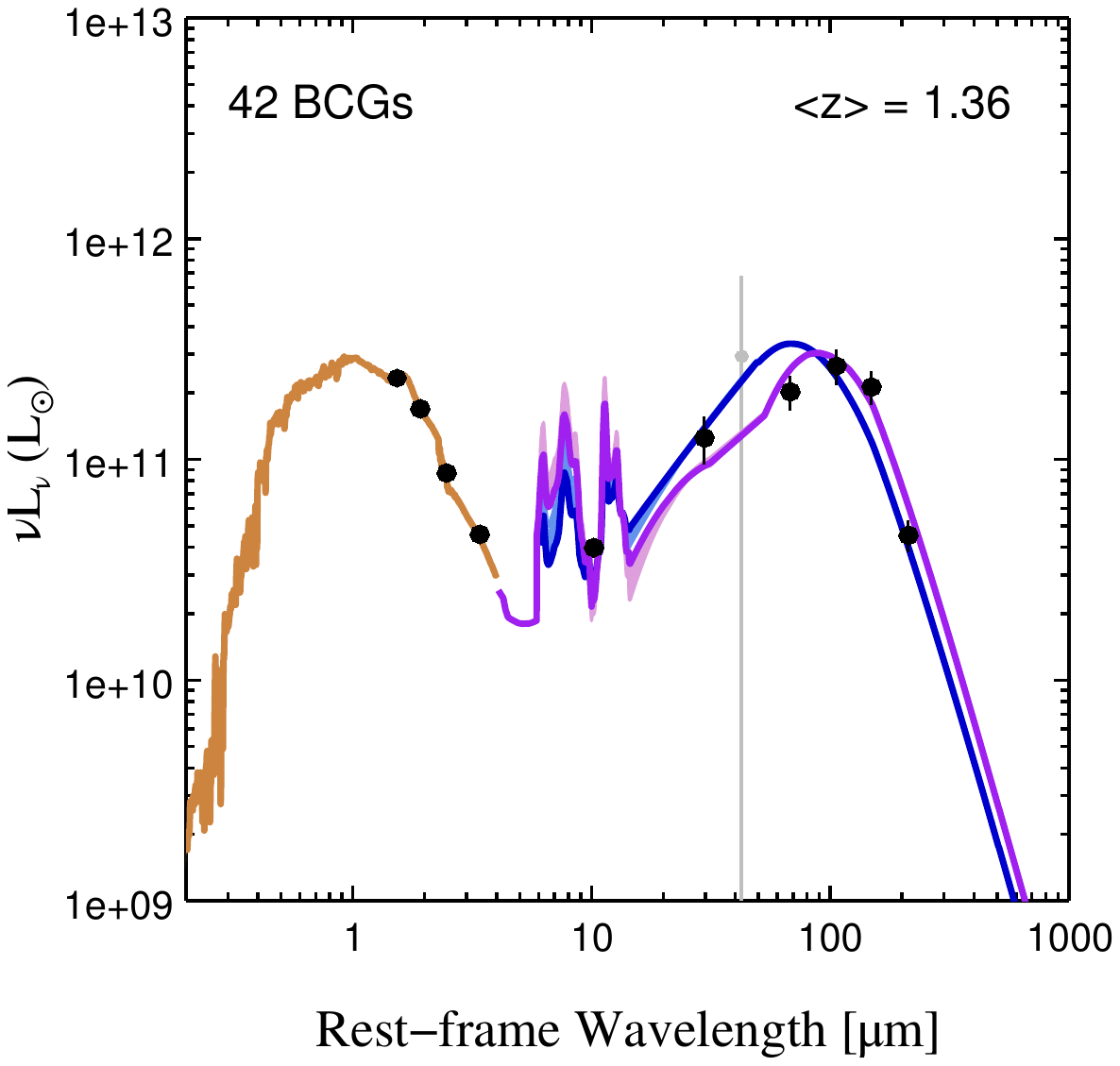} &
\includegraphics[scale=0.1, width=6.1cm, trim=2cm 7.7cm 7cm 8.3cm]{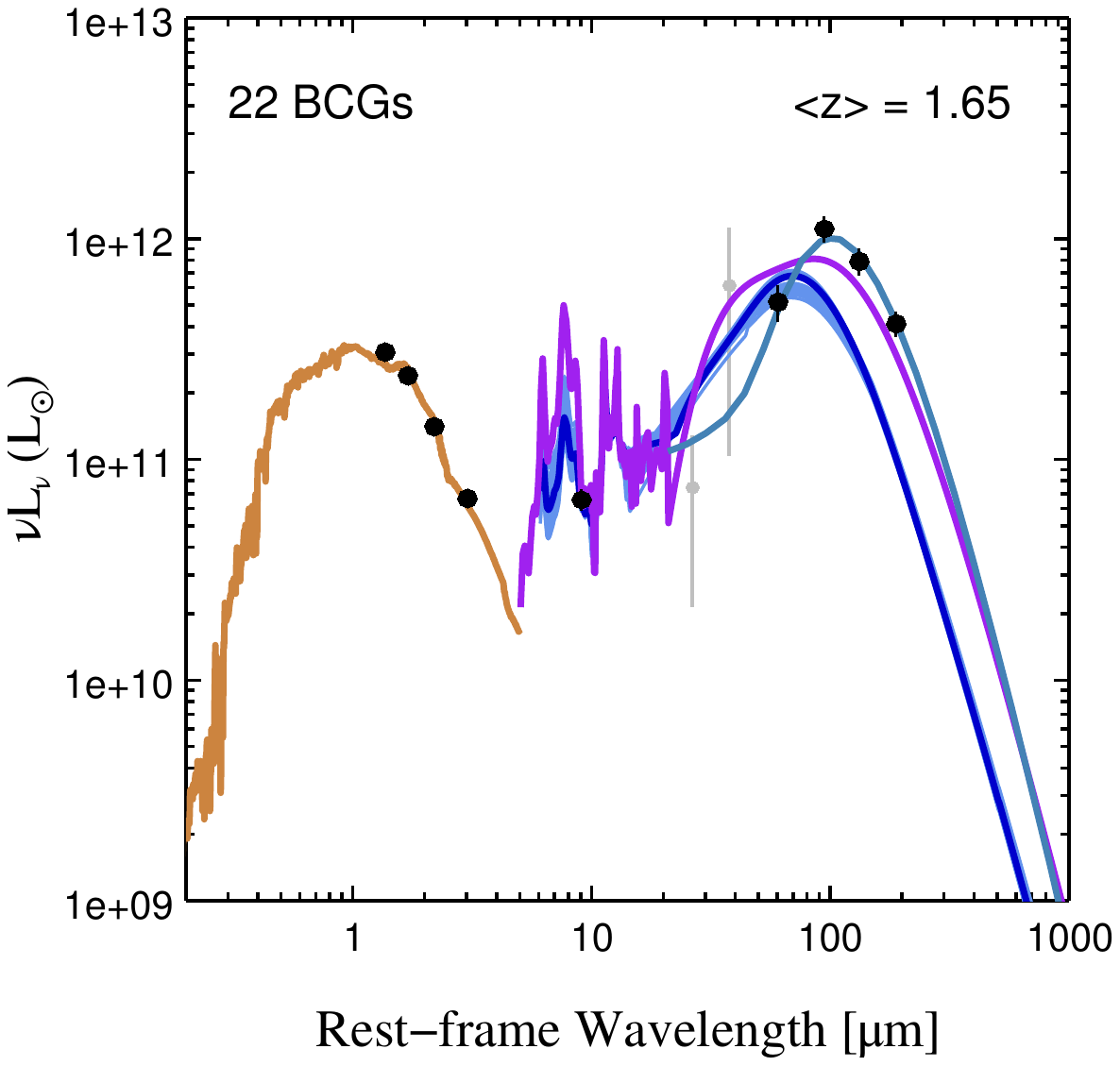} \\
\end{tabular}
\vspace{3.5cm} 
\caption{Stacked $\nu L_{\nu}$ BCG SEDs fitted with a stellar model component in gold, a 'warm' star-forming component in light and dark blue, 'cold' cirrus dust emission in teal (green/blue), and a 'warm plus cold' star-forming component in light and dark pink/purple, customised to each data bin according to the description in Section 5.2.3. The 'warm' and 'cold' blackbody components are considered to be those with a peak wavelength shortward and equal to or longward of 100 $\umu$m, respectively, while the broader 'warm plus cold' component includes clear contributions from both a warm and cold component, with an effective peak wavelength at $\sim100$ $\umu$m. Stacked fluxes with $S/N < 3$ appear as grey data points, and the non-detections with stacked fluxes $\sim 0$ $\umu$Jy do not appear. The top 5-20 CE01 star-forming models with the highest likelihoods of truly representing the data appear in light blue for the warm component and pink for the warm-plus-cool component, and the single star-forming model (either from the CE01 or Polletta template library, or Kirkpatrick et al. 2012) which best matches the significant ($S/N \ge 3$) data points is traced by a darker, thicker line.} 
\end{figure*}

\subsection{Far-infrared Characterisation}

Since each of the observer-frame flux SEDs presented in the previous section samples a different part of the rest-frame SED at each redshift, we convert the SEDs from flux density units to $\nu L_{\nu}$ in solar luminosity units in the emitted frame prior to fitting each with star-forming, AGN, and quiescent infrared galaxy templates from the literature, to facilitate a direct comparison of the SEDs at different redshifts. From the best-fitting templates we derive estimates of the BCG total far-infrared luminosity, star formation rate and efficiency, stellar mass, and effective dust temperature.

\subsubsection{SED Model Comparison}
\underline{Model Template Libraries}

The SED-fitting process involves a quantitative comparison of the BCG rest-frame SEDs to the star-forming templates contained in the Chary \& Elbaz (2001) (''CE01'', hereafter) SED template library; the elliptical, star-forming, AGN, and star-forming/AGN composite templates from the SWIRE template library (Polletta et al. 2007; Lonsdale et al. 200~4; hereafter referred to as the ''Polletta'' templates); the star-forming/AGN composite templates of Dale et al. (2014) (''D14'', hereafter); the $z=1$ star-forming, and 'smooth' and 'silicate power law' AGN templates of Kirkpatrick et al. (2012); the cold 'cirrus' dust and dusty AGN torus models of Rowan-Robinson et al. (2008) (''RR08'', hereafter); as well as the young starburst model from Rowan-Robinson et al. (2010) (''RR10'', hereafter).

The CE01 templates reproduce the mid-to-far-infrared ISO, IRAS, and SCUBA data of nearby star-forming galaxies, scaled to 105 different infrared luminosities based on redshift, and evolved to match the star formation history of the Universe (CE01).  The template bearing the closest match to the observed 24$\umu$m flux in a given infrared SED is considered by CE01 to be the best-matched template, based on the tight correlation they observe between 24$\umu$m flux and total infrared luminosity. However, due to the well-known 'mid-IR excess' problem (e.g., E. J. Murphy et al. 2009, Muzzin et al. 2010, Elbaz et al. 2011), whereby such local templates systematically overestimate the rest-frame far-infrared flux for sources lying beyond $z\sim1.5$ when extrapolating the far-infrared blackbody emission from the MIPS 24$\umu$m flux, we compare our stacked SEDs to each template in the full CE01 library in order to determine the best match, as well as employ the $z\sim1$ star-forming template from Kirkpatrick et al. (2012) for the high-redshift fits.

From the Polletta (SWIRE) template library we select the M82 and Arp220 'normal' and starburst/ULIRG star-forming templates, respectively, as well as the {\it Sey1.8} Seyfert 1.8 template, the {\it IRAS19254-7245 South} Seyfert 2+starburst/ULIRG model and {\it I22491} starburst/ULIRG templates, and the 13-Gyr-old {\it Elliptical 13 Gyr} template. These model SEDs accurately reproduce the observed SEDs of representative galaxies of each infrared galaxy type, and were generated by the GRASIL code as described in Silva et al. (1998), with the exception of the PAH region between 5-12 $\umu$m, which consists of observed infrared spectra from the PHT-S spectrometer on the Infrared Space Observatory and from IRS on {\it Spitzer}.  

The D14 templates consist of 100 star-forming templates which range from a purely star-forming template with a 0\% AGN contribution to a 100\% AGN-dominated template, where each SED in between gains a 1\% incremental increase in the AGN contribution.  These templates were designed to allow for statistical constraints on the star-forming versus AGN properties of a large sample of galaxies, as opposed to a detailed analysis of individual systems; and include a contribution from a single AGN model, the median unobscured Type 1 quasar model of Shi et al. (2013).

The templates from RR08 and RR10 were used in an attempt to simultaneously model the individual components contributing to the stacked BCG luminosity SEDs, specifically those due to an AGN torus, cold cirrus dust, and a young starburst. This approach to modelling the SED is also carried out in their 2010 study of {\it Herschel}-SPIRE-detected star-forming galaxies out to high redshift, in which both the young starburst and cirrus models are found to be required by the SED fits in addition to a normal star-forming template. The young starburst template is an Arp220-like template but with a higher optical depth  ($A_{v}=150$); and the three cirrus templates used reflect a dust grain temperature in the 9.8 K to 24.1 K range, determined by an interstellar radiation field value equal to 5, 1, and 0.1 times the interstellar radiation field in the vicinity of the sun.

\underline{SED Fitting Procedure}

The SED fitting is performed with {\it Sherpa} -- a multi-wavelength, multi-purpose spectral fitting package provided by the {\it Chandra X-ray Centre} -- using a $\chi^{2}$ minimisation method involving one free parameter, the SED model amplitude. Each of the SED template models we employ in this study is read into a {\it Sherpa} table model and then fit to the BCG stacked, rest-frame luminosity data using the {\it ${\chi}^{2}$ CHI DVAR} fit statistic\footnote{\url{http://cxc.cfa.harvard.edu/sherpa/statistics/}} and robust {\it NelderMead} optimisation method\footnote{\url{http://cxc.cfa.harvard.edu/sherpa/methods/index.html}}. 

The uncertainty on each SED fit is represented by the 99\% confidence interval on the single fit parameter, the SED model amplitude, which we derive assuming a Gaussian and uncorrelated probability distribution about the best-fitting parameter value. We generate this probability distribution by randomly sampling the amplitude fit parameter 10,000 times within the associated flux uncertainties, described in Sections 4.1 and 4.2, and then calculate the $2.58\sigma$ confidence interval of this parameter distribution.

\underline{Composite Fits}

While the {\it Spitzer} IRAC colours of the bright subset of the BCG sample provide evidence for AGN/star-forming composite systems, the lack of warm- and hot-dust AGN signatures in the broadband SEDs led us to reject the poorly fitting star-forming/AGN templates of the D14 library, based on a reduced $\chi^{2}$ analysis of the far-infrared fit between 8 and 1000 $\umu$m (all fits yielding reduced $\chi^{2}$ values in excess of 100).  Instead, the star-forming templates of the CE01 library provide a good match to the far-infrared portion of the majority of the SEDs, as well as the star-forming M82 and star-forming/AGN composite S18 templates from the Polletta library (where we only utilise the far-infrared portion of the S18 template which matches that of a purely star-forming SED), and finally the $z\sim1$ star-forming template from Kirkpatrick et al. (2012). The best-fitting reduced $\chi^{2}$ values for these star-forming fits are $<2$ for all SEDs except in the following bins, which have reduced $\chi^{2}$ values of 5 or less: redshift bins 2-6 of the bright SEDs, redshift bin 6 of the faint SEDs, and redshift bins 5 and 6 of the SEDs including both bright and faint BCGs.

In all of the SEDs, we fit the stellar emission shortward of $\sim10$ $\umu$m separately from the warm and cold dust emission at longer wavelengths, despite many of the star-forming models fitting the full SED. The motivation for this follows: unlike other classes of galaxies, e.g., non-BCG elliptical galaxies, main-sequence and starburst galaxies, and AGN of various types, we have little {\it a priori} knowledge of how the full infrared SED of a BCG ought to appear, as the current study represents the first attempt to rigorously model it (Rawle et al. 2012 fit the detected and stacked undetected {\it Herschel} far-infrared fluxes of a sample of 65 X-ray-bright BCGs at $0.5 < z< 1.0$ to derive $L_{TIR}$ and the associated dust-obscured star formation rate, dust temperature and mass, but neither attempt to mitigate confusion noise in these bands, nor match the observed far-infrared and 24-$\umu$m emission in their SED template fitting (which are not simultaneously fit for significant number of their sources).) Furthermore, the results of the template fitting show that a customised combination of components provides a better description of the BCG stacked fluxes in most instances than any single template from a pre-existing library, particularly at $z<1$. It is clear that an elliptical template alone is not sufficient to match any of the SEDs, as the far-infrared emission of the BCGs is 1-2 orders of magnitude brighter than that of a typical 'red and dead' elliptical galaxy template. As for the star-forming fit, based on the dust scaling relations empirically derived from observations of Herschel Reference Survey (HRS) galaxies of a large range of spectral type, environment, and stellar mass in the local Universe (Cortese et al. 2012) -- specifically the inverse proportionality between dust-to-stellar mass ratio and the stellar mass and sSFR in local galaxies -- we expect the BCG SED to exhibit a significantly lower dust-to-stellar mass ratio in comparison to a less-massive non-BCG star-forming template, which is exactly what we find. At the lowest redshifts we consider, $z_{avg} = 0.185$, a typical star-forming SED matches the BCG SED longward of 10 $\umu$m, but the BCG 'stellar bump' rises significantly above that of the star-forming template.  While these dust scaling relations were derived for local field and cluster galaxies (including those in cluster cores), we do not yet know if they apply to galaxies at higher redshifts (though we might expect them to if we assume an understanding of the physics underlying the trends). Whereas massive elliptical galaxies in the local Universe exhibit relatively low dust-to-stellar mass ratios on account of a diminished level of star formation activity with respect to local star-forming late-type galaxies, we would expect an increasing dust-to-stellar mass ratio for elliptical galaxies of a fixed stellar mass with increasing specific star formation rate towards higher redshift, as observed in our sample (reminding the reader that we select BCGs from clusters of the same mass at every redshift, therefore the BCGs are of the same mass at every redshift based on the linear scaling of BCG stellar mass and host-cluster mass, e.g. Lidman et al. 2012). In fact, as we move towards higher redshift, we observe the star-forming templates begin to match the full extent of the BCG SED better and better, as opposed to just the 'far-IR bump' at lower redshifts. 

Therefore, we fit the stellar emission shortward of $\sim10$ $\umu$m in the faint SEDs with only the Polletta {\it 13-Gyr Elliptical} model, but add to this model either of the Polletta {\it M82} or {\it S18} templates to fit the stellar emission in the bright SEDs. In the mid- and far-infrared regions of all of the SEDs, we fit each of the 105 star-forming CE01 templates and take the Bayesian weighted model average (Hoeting et al. 1999) of the 5-20 best-fitting templates to distill the subset of SEDs which have the highest likelihood of truly representing the data, and to properly account for the scatter in the best-fitting SEDs (where each template is weighted by its associated ${\chi}^{2}$ fit value). We also include in this average fit either of the Polletta {\it M82}, {\it Seyfert 1.8}, {\it A19254-7245 South}, and/or {\it IRAS 22491-1808} templates, as well as the RR10 young starburst model, where they also provide a good fit to the star-forming component of the SED. Finally, where it is clear that none of the star-forming templates from the CE01 or Polletta libraries contain a contribution from cold dust which is substantial enough to fit the full wavelength extent of the far-infrared emission, we add to the fit an additional cold dust component from the RR08 library -- the combination of which nicely matches the $z\sim1$ star-forming template from Kirkpatrick et al. (2012) in the $z_{avg}=1.07$ bin of all flux categories considered in our SED analysis. In these cases, the total far-infrared emission is represented by an average of the Kirkpatrick et al. (2012) model and the combination (sum) of the CE01 star-forming and cirrus components. 

The resulting customised, composite SED fits appear in Figure 7, with the stellar emission depicted in yellow and the 'warm', 'warm plus cold, and 'cold' dust components of the galaxy in blue, purple, and dark red, respectively, where a warm-dust blackbody component is considered to be one with a peak wavelength shortward of 100 $\umu$m, and the cold-dust model one with a peak wavelength longward of 100 $\umu$m. Non-detections with flux levels equal to the stacked image background ($\sim 0$ $\umu$Jy) do not appear in the plots, and stacked fluxes with $S/N < 3$ appear in grey. The top 5-20 CE01 star-forming models with the highest likelihoods of truly representing the data appear in light blue, and the single star-forming model (either from CE01, P07, or Kirkpatrick et al. 2012) which best matches only the significant ($S/N \ge 3$) data points is traced by a thicker, darker line.

These results clearly show that the SED of both the average bright and faint BCG throughout the redshift range of our sample is fit best by a star-forming model, with an obvious 'stellar bump' shortward of $\sim 4$ $\umu$m caused by the emission from old stars, and a relatively broad 'far-infrared bump' of blackbody emission at longer wavelengths resulting from the 'warm' heating of large dust grains by star formation processes and 'cold' heating from the ambient interstellar radiation field. This is in-line with the expectation of a significant contribution from cold dust in Herschel-detected galaxies, as found in RR08 (where our Herschel detections are {\it stacked} detections), as well as the significant heating from a massive old stellar population suggested by the elliptical model fit to the stellar bump in the SEDs.  We do not observe the warm dust emission indicative of an AGN contribution in the mid-infrared portion of the SED, or the 'warm' effective dust temperature attributed to AGN heating in the far-infrared blackbody emission.

\subsubsection{AGN contribution}

While the {\it Spitzer} IRAC colour diagnostic discussed previously would suggest that up to $50\%$ of the bright BCGs in our sample could be significantly dominated by AGN emission, ignoring possible contamination from high-redshift star-forming galaxies to Region 1 of the plot in Figure 4  -- and also considering that low-luminosity/host-dominated AGN could 'hide' anywhere in the colour space --  the stacked SED representing the average bright BCG in each redshift bin is clearly dominated by stellar emission at wavelengths shortward of $\sim 4$ $\umu$m and that from star formation processes at longer wavelengths. The R08 dusty torus AGN template is consistently rejected during SED fitting, contributing 0\% to the combined SED fit, and we also find that none of the AGN/SF composite SEDs in the D14 template library provide a good fit to the stacked bright SEDs based on a ${\chi}^{2}$ analysis. 

Given that the D14 templates include a contribution from only a single AGN model, the median unobscured Type 1 quasar model of Shi et al. (2013), we use the IDL routine DeCompIR (Mullaney et al. 2011) as a final check of the AGN contribution to the bright SEDs and the associated $L_{IR}$-inferred star formation rate, as this program includes an average AGN model based on many different manifestations of AGN in infrared SEDs. DeCompIR flexibly fits this AGN model to the input SED, along with five averaged ''host galaxy'' starburst templates with far-infrared blackbody curves spanning a wide range in peak temperature.  However, we find again that none of the star-forming/AGN composite models of this routine provide a meaningful fit to the SEDs, as all are fit with a reduced ${\chi}^{2}$ value in excess of 100, except in one case, where the AGN contribution is a negligible 0.4\%. 

While the addition of the far-infrared SED modelling analysis to the IRAC colour diagnostic does not help to uncover the exact fraction of low- or moderate-luminosity AGN which may be residing in completely host-dominated SpARCS BCGs, it should 'pick up' the mid-infrared emission of those AGN which are not classified as continuum-dominated in IRAC colour space, longward of the IRAC bands in the infrared SED (as also noted in, e.g., Alberts et al. 2016). Where potentially many AGN/star-forming BCGs could reside all over the IRAC colour space, they should not be able to 'hide' as easily in the SED between 8 and 40 $\umu$m (after which point AGN emission begins to steeply drop off), where we do not observe a combined AGN/star-forming signature, not even at the 1\% AGN level. With a lack of an AGN signature in both the IRAC colour diagram and in the mid-infrared portion of the SED, we are left with the choice of a low-luminosity or radiatively inefficient average AGN in our sample.  Furthermore, where an AGN is not energetic enough to heat a circumnuclear accretion disk and surrounding dust torus, as evidenced by the lack of a hallmark near-to-mid-infrared power law component to the stacked SED, one would not expect large dust grains farther out in the host galaxy to be significantly heated by the AGN and therefore contribute to the far-infrared emission and associated SFR (Hiner et al. 2009). Therefore, we assume a 0\% contribution to the far-infrared emission from AGN in the derivation of the physical BCG parameters discussed in the following sections.

\subsubsection{Star-Formation Diagnostics}
\underline{Total Infrared Luminosity \& Star Formation Rate}

With the availability of far-infrared and sub-mm data in the post-{\it Spitzer} era, it is now possible to directly estimate the total infrared luminosity of star-forming galaxies out to high redshift by integrating the full luminosity SED, ending the need to extrapolate the far-infrared blackbody emission from the MIPS 24$\umu$m flux of the best-matched local galaxy template. Integrating the Bayesian weighted average of the 'warm', or 'warm' plus 'cold' star-forming and cirrus models fitted to the far-infrared portion (8-1000 $\umu$m) of each stacked SED shown in Figure 7, we arrive at the average total infrared luminosities displayed in Figure 7a as a function of redshift, along with the associated star formation rates (SFR) determined via the Kennicutt et al. (1998) relation, SFR = $1.724 \times 10^{-10}$  $L_{FIR}$. 

The uncertainty on the total infrared luminosity (and associated SFR) is estimated from the posterior variance associated with the posterior mean luminosity value, with the uncertainty on each individual model contributing to the Bayesian model average derived according to the explanation in Section 5.2.1.

\begin{figure*}
 \vspace{90pt}
\begin{tabular}{cc}
\includegraphics[scale=0.62, trim=1.5cm 0cm 6cm 17.4cm]{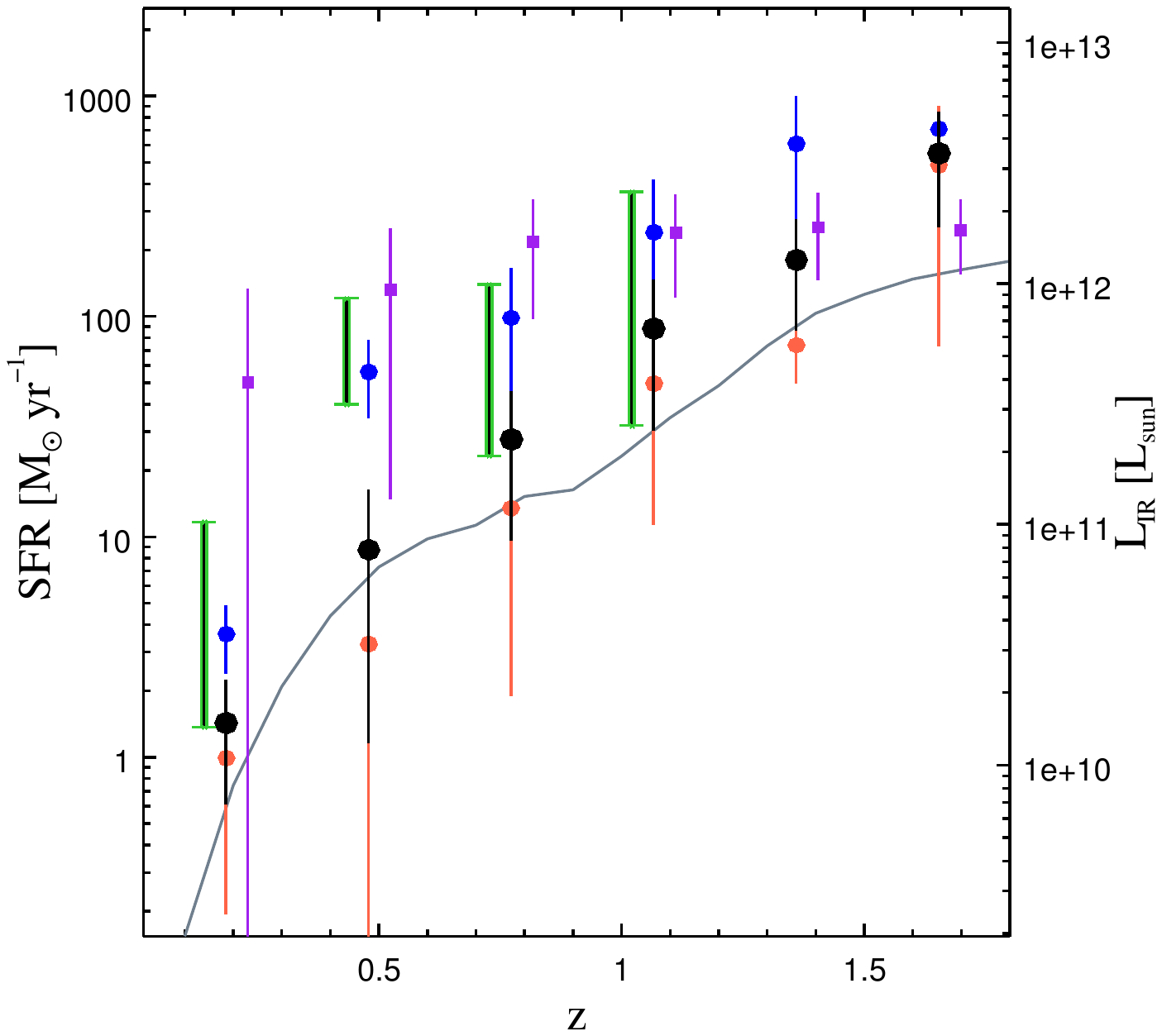} &
\includegraphics[scale=0.62, trim=1cm 0cm 6cm 17.4cm]{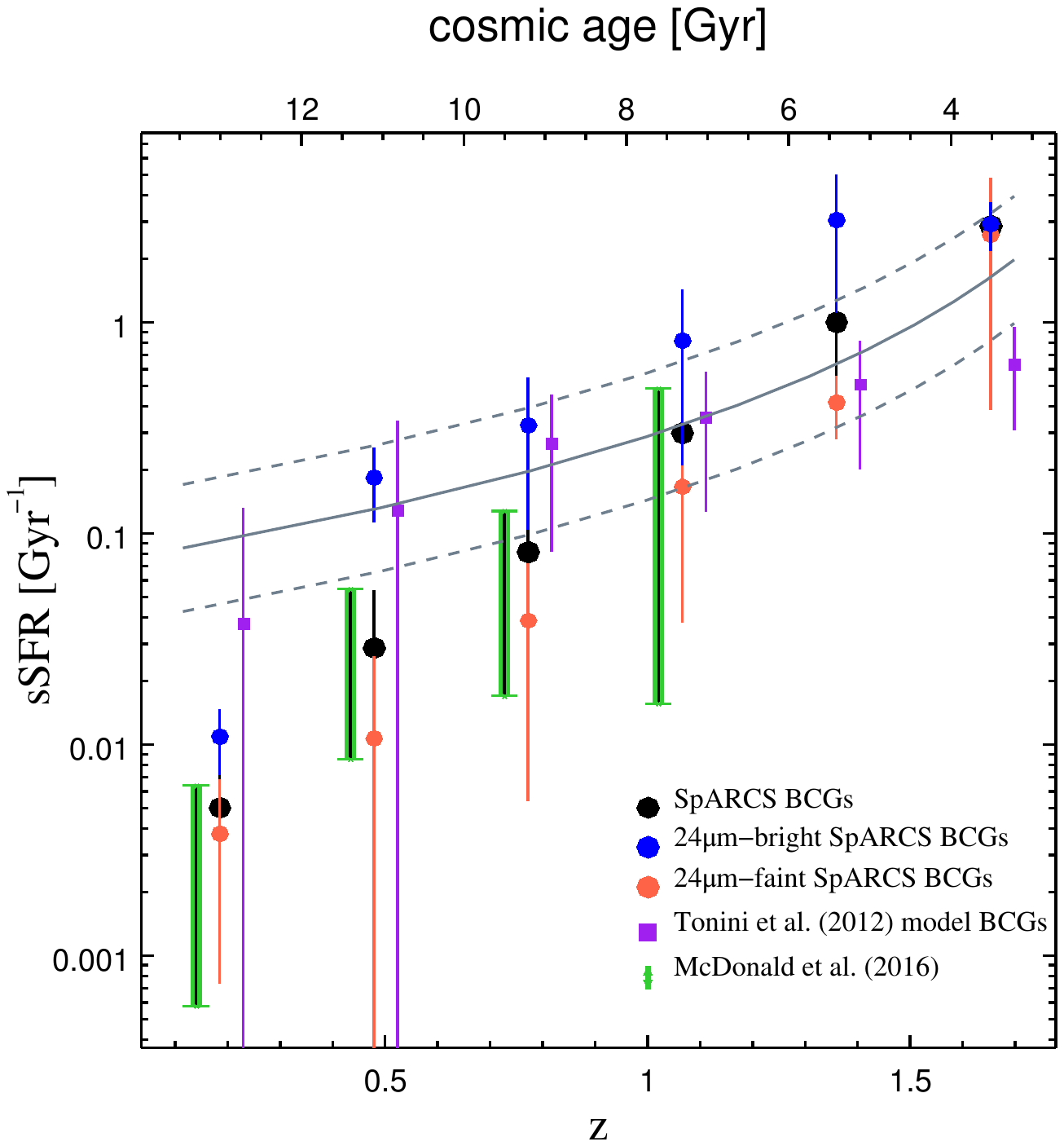}
\end{tabular}
\vspace{-1cm}
 \caption{{\it Left}: The median {\it SpARCS} BCG SFR plotted as a function of the average bin redshift, along with the average model BCG SFR from Tonini et al. (2012) and the 24${\umu}$m-inferred SFR of McDonald et al. (2016) in redshift bins matched to ours, for comparison. (The data points from the various studies have been artificially offset from one another along the redshift axis for visual clarity). The grey curve corresponds to the rest-frame $L_{IR}$ of the CE01 star-forming template that matches the adopted SWIRE MIPS 24-${\umu}$m depth (100 ${\umu}$Jy) at each redshift, roughly approximating the SWIRE $L_{IR}$ survey depth (noting that the CE01 SEDs are based on a median trend in the luminosity-luminosity correlations of only local galaxies). (We remind the reader that the BCG $L_{IR}$ values quoted in the current study are derived by directly integrating the full far-infrared SEDs, and not inferred from the 24-${\umu}$m luminosity as in our companion paper, W15.)  {\it Right}: BCG sSFR as a function of redshift, with the solid line representing Eq 13 of Elbaz et al. (2011) and tracing the position of main-sequence galaxies. The dashed curves lie a factor of 2 above and below the solid line, bounding the region of main-sequence galaxies, with quiescent and low-star-forming galaxies lying below and starburst galaxies lying above at all redshifts.}
\end{figure*}

We observe that the average bright BCG exhibits a total infrared luminosity consistent with a Luminous Infrared Galaxy (LIRG) ($L_{IR} \gid 10^{11} L_{\sun}$) between $0.5 < z < 1$ and an Ultra-luminous Infrared Galaxy (ULIRGs) ($L_{IR} \gid 10^{12} L_{\sun}$) beyond $z = 1$, whereas the faint BCGs are LIRGs between $1 < z < 1.5$ and ULIRGs beyond $z = 1.5$.  The associated star formation rates -- shown along with the average star formation rates of the model BCGs of Tonini et al. (2012) and the 24${\umu}$m-inferred star formation rates of the sample of observed star-forming BCGs of McDonald et al. (2016) (hereafter referred to as 'M16') -- are correspondingly high, with the bright BCGs producing hundreds of solar masses per year throughout the majority of the redshift range considered.  

\begin{table*}
 \centering
 \begin{minipage}{140mm}
  \hspace*{-1.7cm}\begin{tabular}{lcccccc}

    \hline              

    & $z_{avg} = 0.185$ & $z_{avg} = 0.479$ & $z_{avg} = 0.772$ & $z_{avg} = 1.066$ & $z_{avg} = 1.360$ & $z_{avg} = 1.653$  \\

    \hline

    $24\umu$m-bright              & $3.32e11 \pm 9.75e9$ & $3.06e11 \pm  1.21e10$ & $3.03e11 \pm 1.21e10$ & $2.94e11 \pm  1.40e10$ & $2.00e11 \pm 1.02e10$ & $2.42e11 \pm 1.26e10$   \\
    $24\umu$m-faint               & $2.63e11 \pm 7.55e8$ & $3.05e11 \pm 9.26e9$ & $3.49e11 \pm 1.40e10$ & $2.98e11 \pm 1.24e10$ & $1.78e11 \pm 9.37e9$ & $1.87e11 \pm 9.97e9$  \\
    All BCGs                       & $2.84e11 \pm 7.93e8$ & $3.05e11 \pm 9.82e9$ & $3.39e11 \pm 1.37e10$ & $2.95e11 \pm 1.27e11$ & $1.81e11 \pm 9.69e9$ & $1.93e11 \pm 1.08e10$  \\ 
   \hline
  \end{tabular}
  \caption{Average BCG stellar mass per data bin, in units of $M_{\odot}$, derived using the rest-frame K-band luminosity of the 0 best-fitting stacked SED, and reference $M/L_{K}$ ratios and {\it k}-corrections from Longhetti et al. (2009). }

 \end{minipage}
\end{table*}

We find that the star formation rates of the faint {\it SpARCS} BCGs are significantly lower than those of the X-ray-selected BCGs in the M16 study and those predicted by the semi-analytic model of Tonini et al. (2012) (except at the highest redshifts for the latter study), while the star formation rates of the bright {\it SpARCS} BCGs lie within the uncertainties \footnote{The error bars on the Tonini et al. (2012) mean SFR values in Figure 7a represent the standard deviation about the mean SFR. The upper-bound of the uncertainty on the M16 data points in the same figure corresponds to the mean SFR when the non-detections in their sample are assumed to be at their upper-limit SFR value, whereas the lower-bound of the uncertainty assumes a SFR value of zero for the non-detections.} of both of these studies through $z_{avg} = 1.36$.  In the two highest redshift bins, where the median star formation rates of the bright {\it SpARCS} BCGs surpass the mean Tonini et al. (2012) model BCG star formation rates, we are likely witnessing the effects of sample incompleteness; i.e., as we move towards higher redshift, the detection threshold gets brighter and we can only detect increasingly brighter objects. The result of such incompleteness is likely causing an overestimate of the true BCG star formation at the highest redshifts of our sample, where we only detect the brightest of the BCGs lying above the {\it Spitzer MIPS} 24-$\umu$m detection threshold (indicated by the solid grey line in Figure 7a).

To further confirm and characterise the star-forming nature of the BCGs implied by the far-infrared SED analysis, we consider also the sSFR, for which we derive estimates of the average BCG stellar mass in each redshift bin. Finally, we use the effective dust temperature corresponding to the peak of the infrared emission as a star-formation diagnostic, in line with Elbaz et al. (2011), who find this parameter to reliably distinguish between starburst galaxies vigorously producing new stars in spatially compact regions, and those modestly forming stars in more diffuse regions as 'main-sequence' galaxies.

\underline{BCG Stellar Mass and sSFR}

We estimate the stellar mass of the average BCG in each redshift bin using a combination of Equations (4) and (5) of Longhetti et al. (2009), 

$log_{10}(M_{gal}) = log_{10}(M/L_{K}) + 0.4K_{corr} + log_{10}(L_{o})$

where we choose the reference redshift-dependent mass-to-light ratios ($M/L_{K}$) obtained with the GRASIL spectro-photometric synthesis code and Salpeter IMF at solar metallicity (Longhetti et al. 2009), assuming the star formation history used to generate the Polletta elliptical template model we fit to the stellar emission in the SEDs (Silva et al. 1998). The chosen K correction is the reference colour K correction from Longhetti et al. (2009), which corrects to the rest-frame K band from the nearest observed IRAC band at $z_{avg} > 0.5$, as the rest-frame K band shifts longward of the observer-frame K band with increasing redshift (for our lowest redshift bin lying at $z_{avg} = 0.185$, we assume a negligible filter correction and derive the rest-frame $L_{K}$ value directly from the corresponding rest-frame stacked SED). Finally, $L_{o}$ represents the luminosity calculated from the observed IRAC flux, at the observed IRAC wavelength which redshifts the luminosity into the rest frame via the K correction term as closely as possible to the rest-frame $L_{K}$ value which we seek to estimate solar mass. This luminosity therefore is converted to solar luminosity units via division by the solar luminosity in the K band, derived from a reference value for the solar flux in the K band, $0.095$ $W m^{-2} nm^{-1}$. The stellar mass values, listed in Table 6, are all of the order of $10^{11} M_{\odot}$ and are therefore in agreement with expectations from the $M_{*}-M_{halo}$ relation (e.g., Aragon-Salamanca et al. 1998; Brough et al. 2005, 2008; Stott et al. 2008, 2010, 2012; Collins et al. 2009; Hansen et al. 2009; Lidman et al. 2012). In other words, due to the correlation between BCG stellar mass and host cluster mass, we expect the BCGs selected from similarly massive clusters at each redshift to have a similar average stellar mass. We derive the uncertainty on the stellar mass from the uncertainty on the associated $L_{o}$ value derived from the SED fit, assuming no uncertainty on the adopted $M/L_{K}$ and K correction values, as near-infrared-derived M/L ratios and K corrections (as opposed to those derived in the optical regime) are largely insensitive to the star formation history (SFH) and other spectro-photometric model parameters used to generate them (Longhetti et al. 2009). Therefore, we also do not assume a systematic uncertainty on the stellar masses derived from the bright SEDs in comparison to the faint SEDs -- due to the former exhibiting stellar emission profiles with a star-forming contribution from either of the M82 or S18 Polletta models in combination with the Polletta elliptical model -- as the M/L ratios we adopt are considered to be insensitive to the SFH.

The median specific star formation rates ($sSFR=SFR/M_{\odot}$) derived from the median BCG star formation rates and stellar masses are plotted in Figure 7b as a function of redshift (with the corresponding uncertainties propagated into the uncertainty on sSFR), along with those of the model BCGs of Tonini et al. (2012) and the sample of X-ray-detected star-forming BCGs studied in M16. From this star formation diagnostic we learn that the average {\it SpARCS} BCG transitions from a sub-main-sequence to a main-sequence star-forming galaxy at $z\sim1$, in line with a similar trend observed among the M16 BCG sample and at similar sSFR (the average stellar mass of the BCGs from M16 range from $8.0\times$ to $3.6\times$ the average SpARCS BCG stellar mass in the same redshift bin, from low to high redshift). The bright sub-sample moves into the starburst regime beyond $z\sim1$, greatly surpassing the star-forming efficiencies of the Tonini et al. (2012) BCGs, due to the combined effects of higher star formation rates and significantly smaller stellar masses. These results suggest that the average {\it SpARCS} BCG at $z \ge 1$ is forming stars at a relatively vigorous, upper-main-sequence efficiency, in agreement with the 24${\umu}$m-inferred star formation rates calculated for a smaller portion of our BCG sample in W15.

\begin{table*}
 \centering
 \begin{minipage}{140mm}
  \label{tab_one}
  \begin{tabular}{lcccccc}

   \hline    

            & $z_{avg} = 0.185$ & $z_{avg} = 0.479$ & $z_{avg} = 0.772$ & $z_{avg} = 1.066$ & $z_{avg} = 1.360$ & $z_{avg} = 1.653$  \\
   \hline
$24\umu$m-bright  & $30.57 \pm 4.18$  & $32.45 \pm 3.22$ & $30.46 \pm 3.07$  & $35.22 \pm 3.00$  & $44.71 \pm 2.78$ & $29.33 \pm 3.82$   \\
$24\umu$m-faint   & $26.26 \pm 4.17$  & $27.65 \pm 4.95$ & $25.04 \pm 2.11$  &  $38.95 \pm 7.94$ & $35.17 \pm 5.34$ & $30.74 \pm 4.12$  \\
All BCGs           & $27.05 \pm 4.12$ & $27.41 \pm 4.37$ & $25.05 \pm 2.12$  & $34.8 \pm 3.66$ & $34.79 \pm 4.70$ &  $33.85 \pm 4.37$  \\ 
   \hline
  \end{tabular}
 \caption{Effective dust temperature, in units of Kelvins, corresponding to the peak of the blackbody emission in each of the stacked far-infrared BCG SEDs.}
 \end{minipage}
\end{table*}

\underline{Effective Dust Temperature}

According to the criteria set forth in Elbaz et al. (2011), the average main-sequence galaxy should feature a significantly larger contribution from cold dust to its infrared SED than a starburst galaxy, with a relatively broad far-infrared bump suggestive of a wide range in dust temperature, and a peak centred around 90$\umu$m with an associated effective dust temperature of $\sim30$K. A typical starbust galaxy, on the other hand, would feature a narrower far-infrared bump with weaker PAH emission lines, peaking around 70-80$\umu$m, with a warmer effective dust temperature of $\sim40$K.  This would result where there is star formation occurring in bursts in relatively compact spatial regions within the galaxy, explaining the warm component, while the ambient interstellar radiation field due to stellar emission warms the dust to a cooler temperature, more diffusely throughout the galaxy, explaining the cool component.

With this in mind, we consider the effective dust temperature corresponding to the peak of the far-infrared emission in each BCG SED using Wien's law, in line with Elbaz et al. (2011), as a final step in the exploration of the star-forming nature of the BCGs in our sample (noting that we use this parameter only as a simple star-formation diagnostic in the current context, not to characterise the dust properties of the BCGs). As described in Section 5.2, we check for a contribution to the far-infrared emission from both warm and cold dust components by fitting the full SED with a combination of the star-forming, young starburst, and cold cirrus dust templates.  We estimate the effective dust temperature from the wavelength of peak emission of the star-forming fit to the SED, or the peak of the sum of the star-forming and additional cold cirrus dust component where the latter is required by the fit. The associated uncertainty on this effective dust temperature includes the uncertainty on the average redshift value as well as the spread in the peak wavelength value of the various CE01 star-forming models which are fitted to the far-infrared emission. The resulting dust temperature values are contained in Table 7.

We find that the stacked far-infrared emission of the BCGs implies an effective dust temperature in the range of $26.26 \pm 4.17$ K to $44.71 \pm 2.78$ K, with the majority of the SEDs exhibiting the broad far-infrared bump suggestive of a main-sequence galaxy. The faint SEDs at $z < 1$ present with a relatively cool average dust temperature of 26 K, while the bright BCGs in the same redshift range are warmer at an average dust temperature of 31 K. At $z > 1$, however, the effective dust temperature of all BCGs rises to greater than 30 K (within the uncertainty of the 29.33 K effective temperature of the highest-redshift bright SED), with a dust temperature characteristic of a starburst contributing to two out of three of both the bright and faint SEDs at these higher redshifts.  With respect to the previously discussed star-formation diagnostics of total infrared luminosity, star formation rate, and sSFR, we find excellent agreement with the dust temperature diagnostic: we observe the effective dust temperatures characteristic of a main-sequence galaxy among the bright BCG population at $z < 1$, and sub-main-sequence galaxies among the faint BCGs at these redshifts; and witness both populations transition to a warmer dust temperature and implicitly higher level of star formation at $z > 1$.

\section{Discussion}
\subsection{Star-forming BCGs between  $ 0.0 < z < 1.8 $}

Taken together, the separate analyses presented above come together to reveal a picture of a {\it star-forming}, as opposed to a passively evolving, BCG population throughout the $0.0 < z < 1.8$ redshift range considered in this study. This is contrary to the requirement of an early shutdown of star formation\footnote{The star formation referenced here being due specifically to the rapid cooling of the interstellar medium within the cluster galaxies merging to form the BCG at these early times, as opposed to major-merger-induced starbursts.} by $z\sim2.5$ in the semi-analytic model of DB07, after which point the BCG is predicted to assume its final identity through a relatively small number of gas-poor, minor mergers. We discover that the 24$\umu$m-faint BCGs, which outnumber the 24$\umu$m-bright BCGs by a factor of three and are therefore the most representative of our sample, exhibit an average far-infrared SED consistent with a star-forming galaxy at all redshifts. This is a rather surprising discovery, given that the mid-infrared emission alone would imply a 24$\umu$m-faint subset of the population that is completely quenched or 'off' compared to the 'on' and active 24$\umu$m-bright subset. While the average faint BCG at $z < 1$ is producing stars at a sub-main-sequence efficiency, we witness the warm dust emission from star formation in molecular clouds begin to occupy a larger fraction of the far-infrared blackbody curve of the stacked SED at $z\sim1$ (noting that the effective dust temperatures of all faint SEDs in our analysis lie within the $\sim20-60$ K range expected for an interstellar medium (ISM) heated only by star formation, Casey 2012). Now that the far-infrared component of the BCG SED has been revealed, we are able to conclude that the 24-$\umu$m 'faintness' of the majority of the BCGs in our sample at mid-infrared wavelengths is not actually a marker of intrinsic infrared-faintness. 

The notable difference in the mid-infrared flux intensity between the 24$\umu$m-bright and 24$\umu$m-faint BCGs, particularly at $z<1$, may be due in part to a difference in the {\it timing} of their most recent star formation activity. Preliminary results of a follow-up optical spectroscopic analysis of 93 SpARCS BCGs between $0 < z < 1$ (Bonaventura et al. 2017, in preparation) support a scenario in which the 24$\umu$m-bright BCGs are experiencing significant ongoing star formation throughout, while the 24$\umu$m-faint BCGs are likely to be in a low- and/or post-starburst/star-formation state -- the latter referring to a state in which star formation recently ended (or paused), within $\sim 100$ Myrs prior to the epoch of observation (e.g., Hayward et al 2014).  $H\alpha$ and/or [OII] emission-line signatures of instantaneous star formation are detected in the median-stacked optical spectra of the 24$\umu$m-bright subset of the spectroscopic sample in similar redshift bins to those used in the current study, and [OII] emission is detected amongst the 24$\umu$m-faint subset only at $z>0.7$. At $z<0.7$, the median-stacked spectra of the 24$\umu$m-faint BCGs are characterised by the Balmer stellar absorption features and relatively 'young' $4000$ $\AA$ break index of a post-starburst (or post-star-formation) galaxy (Balogh et al. 1999; Muzzin et al. 2014; Bonaventura et al. 2017, in preparation).

Various other observational studies in recent years have uncovered the ongoing star formation activity in BCGs (e.g., Vikhlinin et al. 2007; Santos et al. 2008; Hudson et al. 2010), but the majority of these studies involve BCGs residing in low-redshift, X-ray-detected clusters which preferentially host X-ray-bright cool cores (comprising $50-70\%$ of low-redshift, X-ray-detected clusters, Santos et al. 2008). As a result, the observed star formation is attributed to the cooling ICM, through direct detection or inferred from the presence of molecular gas (Johnstone et al. 1987; McNamara \& O’Connell 1992; Allen et al. 1992; Cardiel et al. 1998; Crawford et al. 1999; Edge 2001; Wilman 2006; Edwards et al. 2007; O’Dea 2010; Donahue et al. 2011; Rawle et al. 2012; McDonald et al. 2012). Until the publication of W15, M16, and the current study, it was not clear if BCGs in high-redshift clusters, either hosting or lacking a cool core, would exhibit similar star formation rates to their low-redshift counterparts.  M16, who investigate the star formation in a sample of ninety BCGs from X-ray-detected clusters out to high redshift ($0.3 < z < 1.2$), find 24-$\umu$m-inferred star formation rates similar to our bright sample in the corresponding redshift bins (see Figure 7), as well as a similar evolution in the fraction of star-forming BCGs with redshift (see W15 for details), with the highest fraction of strongly star-forming BCGs lying at $z > 1$. While BCGs in X-ray-detected versus non-X-ray-detected clusters may take different evolutionary pathways due to different environmental effects -- e.g., due to the likely presence of a reservoir of cool gas in the cores of the former and therefore a possibly unique pathway to star formation activity in these clusters -- M16 find that the star formation in the high-redshift BCGs of their sample appear to be fuelled by gas-rich galaxy mergers as opposed to cooling flows, the latter of which they find to dominate at lower redshift. 

It is interesting to consider that the optically and near-infrared selected BCGs in our sample exhibit evidence of star formation activity down to low redshift, but less than $3\%$ are candidates for radio-loud AGN (with less than $1\%$ at $z < 0.4$), which are ubiquitously associated with cool cluster cores (Sun et al. 2009). While there is not enough spatial overlap between current X-ray cluster surveys and the SWIRE fields to make a determination of what fraction of the SpARCS clusters host a cool core -- and, therefore, to what extent, if any, cluster cooling flows may be contributing to the star formation we observe -- the data and analyses considered in the present study would indicate it to be a small fraction at low redshift. It is also important to note that major gas-rich mergers are also highly likely to produce a strong radio AGN (Chiaberge et al. 2015), therefore we can also tentatively rule out major gas-rich mergers as being responsible for the bulk of the BCG star formation we observe at low redshift. At high redshift, however, we expect major merging activity to occur more frequently in cluster centres, and to disrupt cooling flows, likely contributing to the declining fraction of clusters hosting strong cool cores towards high redshift (Vikhlinin et al. 2007, Santos et al. 2008, McDonald et al. 2012). Therefore, given the dearth of strong cool cores in SpARCS clusters at low redshift, coupled with the high likelihood of major merging activity and low likelihood of cooling flows in the cores of clusters at high redshift, we expect major (and minor) 'wet' mergers to significantly contribute to the vigorous star formation in SpARCS BCGs at high redshift. The weaker star formation activity occurring in low-redshift SpARCS clusters is likely the residuals from this earlier epoch of intense merger-induced star formation (including contributions from stellar mass losses, see Section 6.2.3), as the BCG star formation and specific star formation rates continually decrease towards low redshift (Figure 7). While star formation likely continues to be sparked sporadically in BCGs at these redshifts by galaxy mergers (see Section 6.2.2), gas-rich mergers ought to be less common in cluster cores at $z<1$ than at $z>1$ due to the effects of ram-pressure stripping and strangulation on satellite galaxies merging with the central BCG.

All of this suggests a potentially fundamental difference between the growth of BCGs in the presumably non-cool-core SpARCS clusters and those residing in the cool-core environment: where low-entropy cool gas is able to condense onto the SMBH of a central BCG in the relatively relaxed cluster environment of, notably, low-redshift, X-ray-detected cool-core clusters, igniting both AGN and star formation activity, perhaps the clusters of our sample are too dynamically agitated for this process to occur. Perhaps the cool gas is consumed by star formation before it has time to reach the SMBH, or there is an insufficient quantity of gas to fuel both star formation and an AGN. Considering the results of the complementary M16 study in relation to our own, we can argue that gas-rich mergers dominate star formation in clusters at high redshift, independent of a cool core, and less-so in clusters relaxed enough to sustain a cooling flow.

\subsubsection{Model-predicted Star Formation Rates}
     
In Figure 7 we plot the star formation rates and efficiencies of the revised semi-analytic model of hierarchical structure formation of Tonini et al. (2012) (hereafter, T12), which improves upon previous such models (e.g., DB07) mainly by resolving the discrepancy between observed and previously predicted BCG K-band luminosities and near-infrared colours through the introduction of an updated spectro-photometric model: BCGs were predicted to be fainter and redder than actually observed, and this was partially responsible for perpetuating the belief that most BCGs ought to be 'red and dead' at $z < 2$. This model indicates that, not only are we expected to find significant star formation occurring within BCGs at $z < 2$, but in fact the star formation rates we measure are {\it lower} than they would be if we were able to correct for progenitor bias -- in other words, select our BCGs from the most massive clusters at each redshift, as is done in T12.  However, we point out that, while the T12 model now fully predicts the observed near-infrared colours and luminosities of the representative sample of BCGs chosen for their study, we, W15, and M16 observe a significantly steeper decrease in sSFR towards low redshift than the values predicted by the model, though at low redshifts the model is poorly constrained.

While the average {\it SpARCS} BCG in each redshift bin appears to be producing far fewer stars than predicted by the T12 model BCGs through $z_{avg} = 1.07$, the average {\it bright} {\it SpARCS} BCGs are forming stars at comparable rates and lie within the uncertainties of the T12 star formation rates through $z_{avg} = 1.07$, and actually supersede the model star formation rates at $z_{avg} = 1.36$ and beyond.  This is likely due to an increasingly incomplete BCG sample towards higher redshift where we detect only the BCGs which are brighter than the increasingly more luminous 24-$\umu$m-inferred infrared detection threshold, where the T12 model will include all the BCGs at a given epoch. Therefore at the highest redshifts we miss the less-vigorously star-forming BCGs which are included in the T12 model and bring down their average star formation rates with respect to SpARCS BCGs. Regardless, the average BCG in our sample appears to be forming stars stars at relatively vigorous, and quite unexpected, rates down to $ z\sim0.7$, and with a high-enough efficiency at $z\ge1$ to significantly increase its stellar mass on short timescales, in line with the latest semi-analytic model predictions.

\subsubsection{The Contribution to BCG Stellar-Mass Growth from Star Formation}

Integrating the average $L_{FIR}$-inferred star formation rate over the timeframe spanned by each of the six redshift bins used in our analysis and then summing the results over all bins, we find that SpARCS BCGs can generate roughly $6.5 \times 10^{11}$ $M_{\odot}$ throughout $0.0 < z < 1.8$, with $5.85 \times 10^{11}$ $M_{\odot}$ produced at $z > 1$ and $6.55 \times 10^{10}$ at $z < 1$. This roughly corresponds to the mass of the most massive BCGs residing in low redshift $\sim10^{14} M_{\odot}$ clusters (Lidman et al. 2012) and therefore represents an upper limit to the amount of stellar mass which can be added to the BCG in a $10^{14} M_{\odot}$ cluster through star formation processes, as this allows for {\it no} contribution from 'dry' merging. While low-luminosity AGN could be 'hiding' in our sample, where their mid-infrared emission is too faint to manifest in the stacked BCG SEDs, their contribution to the $L_{FIR}$-inferred star formation rate is likely to be insignificant (e.g., Mullaney et al. 2011, Kirkpatrick et al. 2015). 

However, where the average BCG sSFR lies below the star-forming main sequence at $z < 1$, it is possible that the $L_{FIR}$-inferred star formation rate is overestimating the {\it instantaneous} star formation rate of the galaxy (e.g., as would be measured through optical $H\alpha$ line emission) by up to two orders of magnitude, as demonstrated through simulations of merger-induced star formation in Hayward et al. (2014). This is expected to occur in post-starburst galaxies, where the $L_{FIR}$ is legitimately fuelled by dust heating from stars formed during a starburst within $\sim100$ Myr prior to the epoch of measurement, but which decreases much more gradually than the instantaneous star formation (see Figure 3 of Hayward et al. 2014), leading to an overestimate of the galaxy's star formation rate as measured through a standard $L_{FIR}$-SFR relation. Taking this into consideration and assuming that our $L_{FIR}$-inferred star formation rates overestimate the instantaneous BCG star formation rate by an average factor of 30 (based on inspection of Figure 3 of Hayward et al. 2014) for all of the SEDs for which a cold dust component (with peak effective wavelength of blackbody emission at 100 $\umu$m or longer) dominates the far-infrared emission (for redshift bins below $<z>=1.07$), the total contribution to BCG stellar mass from star formation reduces by a factor 5.65 ($\sim 17.7\%$) down to $1.15 \times 10^{11}$ $M_{\odot}$, $40.8\%$ of the average SpARCS BCG stellar mass.

While we cannot use our BCG sample to directly track the evolution in stellar mass of BCGs since $z\sim2$ on account of progenitor bias\footnote{As clusters grow more massive with time, we do not consider it likely that the low-redshift BCGs of our sample represent the descendents of the high-redshift BCGs, as we select BCGs from clusters of a fixed mass throughout the redshift range of the study, as opposed to the most massive clusters at each redshift.}, we can undoubtedly conclude that star formation is playing a significant role in the mass buildup of the BCG population in massive, $10^{14}$ $M_{\odot}$ clusters throughout this epoch (in addition to mass growth through stellar accretion from mergers), where DB07 predicts that BCGs grow only via a series of dissipationless, minor mergers.  In fact, Rodriguez-Gomez et al. (2015) have used Illustris data in an unprecedentedly comprehensive simulation of the stellar mass assembly of galaxies across a wide range in mass, environment, and redshift, to show that the stellar mass growth of most galaxies, whether a satellite or central galaxy, is dominated by the 'in situ' star formation resulting from mergers rather than 'ex-situ' stellar accretion from mergers, except for the most massive galaxies with stellar masses larger than a few times $10^{11} M_{\odot}$ -- a mass threshold below which the SpARCS BCG stellar masses lie.

\subsection{On the Source of the Observed Star Formation}

\subsubsection{An External Gas Supply from a Cluster Cooling Flow}

At low to intermediate redshifts, cool-core clusters ought to be common, comprising 50-70\% of all X-ray-detected clusters out to $z\sim0.4$ (Peres et al. 1998; Chen et al. 2007; Bauer et al. 2005; McDonald et al. 2011) -- whereas at high redshift, evidence has been found in support of a sharp decrease in the number density of strong cool-core clusters amongst X-ray-detected galaxy clusters (but not necessarily {\it weak} cool cores, see Santos et al. 2008). This has been attributed to the higher merger-rate in X-ray-detected clusters at high redshift (Vikhlinin et al. 2007; Samuele et al. 2011), as well as strong evidence for merger-induced star formation in their central BCGs (M16). Given that a number of studies have shown that nearly every known cooling-flow cluster hosts a radio-loud AGN within its core (Burns et al. 1990; Eilek et al. 2003, Sanderson et al. 2006; Santos et al. 2008), and that we detect candidates for radio-loud AGN amongst only $2.8\%$ of our BCG sample (20/716), we tentatively rule out cooling flows or strong cool cores as the likely dominant source of SpARCS BCG star formation.

While the sensitivity limits of the radio surveys we employ to search for radio counterparts to our optically and infrared-detected BCGs may prohibit detection of radio-loud AGN caused by a cluster cooling flow at high redshift (see Section 5.1.1), the likelihood of detectability of both the central AGN and the cluster at X-ray wavelengths increases as the central AGN transitions from the 'radio mode' low-accretion state to the 'quasar mode' high-accretion state (Hlavacek-Larrondo et al. 2013) at higher redshifts. As the coverage area of current Chandra surveys only overlaps $\sim 5\%$ of the full SWIRE survey area, we do not yet know if any SpARCS BCGs exhibit signs of quasar-mode feedback at high redshift (none have been detected in this limited X-ray coverage). An ongoing study by our group which examines the X-ray properties of the optically selected {\it SpARCS} clusters may illuminate this issue.

\subsubsection{An External Gas Supply from Merging Satellite Galaxies}

By the redshifts considered in this study -- corresponding to look-back times of up to 9.89 Gyrs  -- the amount of cool gas available for star formation in cluster satellite galaxies is expected to have decreased considerably from the previous epoch of rapid radiative cooling during initial halo collapse. It is predicted that AGN feedback nearly completely offsets such gas condensation onto a galaxy by $z\sim1$ in massive clusters (Croton et. al. 2006), and ram-pressure stripping continually removes both the cold and hot gaseous content of a satellite galaxy as it moves through the hot, dense intracluster medium (ICM) (Gunn \& Gott 1972, Larson et al. 1980, Dressler et al. 1984, Dubinksi 1998, DB07, Peng et al. 2010, Muzzin et al. 2014, Alberts et al. 2014, Boselli et al. 2014, Vijayaraghavan \& Ricker 2015). Ram-pressure rapidly halts ongoing star formation when the cold molecular component is stripped, and removes a replenishing source of fuel when the hot gaseous corona is detached from the galaxy over a longer timescale, which could eventually cool and condense onto the galaxy to form stars.  Similarly, while a hot corona remains gravitationally bound to a satellite galaxy as it moves through the ICM (up to $\sim2.4$ Gyrs after infall to the cluster for $M_{*}<10^{10} M_{\odot}$ galaxies, Vijayaraghavan \& Ricker 2015), thermal conduction from the surrounding ICM prevents its cooling and 'strangles' or 'starves' the galaxy of a fresh supply of cool gas for continued star formation -- unlike its field counterpart, which continues to accrete gas from the galactic halo (Vijayaraghavan \& Ricker 2015). These processes, combined with the predicted prevalence of high-speed galaxy-galaxy interactions in high-density cluster environments (Dubinski et al. 1998), are believed to transform blue spirals into red ellipticals in the cluster environment, and explain the characteristic quiescence of galaxies located in the cores of both low-redshift and $z \sim 1$ galaxy clusters (Muzzin et al. 2014 and references therein). It has also helped to paint a picture of 'red and dead' BCGs at $z < 2$, where they are expected to only passively accumulate stellar mass through a series of dissipationless mergers (White 1976; Ostriker \& Hausman 1977; Dubinski 1998; DB07), with the merging satellite galaxies devoid of the fuel to spark star formation. 

However, as we become increasingly aware of the star formation still ongoing in BCGs -- as illuminated by observational studies which utilise a variety of star formation diagnostics across a range of wavelengths -- a new picture of BCG formation and evolution emerges, which appears to be more in line with this distinct class of galaxies uniquely located at the gravitational centres of galaxy clusters. Significant star formation activity has been detected in cluster-core galaxies down to $z\sim1$ (e.g., Tran et al. 2010, Smail et al. 2014, Brodwin et al. 2013, Santos et al. 2015, Alberts et al. 2014), and, more surprisingly, even at $z<1$ (e.g., Rawle et al. 2012, Liu et al. 2012, W15, M16, this work). To date, star formation in BCGs has been largely attributed to a cooling ICM, given that the bulk of these studies consider BCGs in X-ray-detected clusters. Therefore, the source of the star formation in optically and/or non-X-ray-detected BCGs, such as those from SpARCS, is less obvious. While cluster galaxies exhibit similar gas depletion timescales and star formation rates to field counterparts when including recycled gas from internal stellar mass losses over the lifetime of the galaxy (Boselli et al. 2014), the threat of ram-pressure stripping on shorter timescales increases with time as galaxy clusters grow more massive, leading to relatively suppressed star formation in cluster cores at $z <\sim 1$ (Muzzin et al. 2012, 2014; Alberts et al. 2014, 2016). Nevertheless, the quantities of cold molecular gas observed to remain in gas-deficient (with respect to the field) galaxies in cluster cores at $z=0$ (Boselli et al. 2014) are non-negligible in the context of this study: the BCG acts as a reservoir for the residual gas carried by satellite galaxies that merge with it throughout its lifetime. Furthermore, given that the BCGs of our sample were selected from SpARCS clusters of the same mass at each redshift ($10^{14} M_{\odot}$), we might not expect their core galaxies to have undergone as much ram-pressure stripping as higher-mass clusters at low redshift.

Based on Alberts et al. (2014, 2016), we can assume that a typical satellite galaxy in the core of a $10^{14} M_{\odot}$ cluster at $z\sim 2$ embodies a similar molecular gas fraction to a field galaxy counterpart with the same mass and initial star formation rate. Carilli \& Walter (2013) show that the molecular gas fractions of a large sample of field galaxies evolve as $f_{gas} \sim 0.1 \times (1 + z)^2$ ($f_{gas} \equiv M_{gas}/M_{stars}$), which yields $M_{gas}=9 \times 10^{9} M_{\odot}$ for a $10^{10}$ $M_{\odot}$ satellite galaxy at this redshift, and $M_{gas}=9 \times 10^{10} M_{\odot}$ for a high-mass $10^{11}$ $M_{\odot}$ major-merger companion.  At $z=0$, we can assume a cold molecular gas mass based on stellar mass, according to an empirical relationship observed between these two quantities in Boselli et al. 2014 (referring specifically to the left panel of Figure 3 in the referenced study). According to this relationship, a $10^{10}$ $M_{\odot}$ minor-merger companion galaxy would contain an average of $\sim5.5 \times 10^{8} M_{\odot}$ of molecular gas, a factor of $\sim 3.63\times$ less than the average unperturbed, gas-rich field galaxy of the same stellar mass; and a major-merger companion would contain $M_{gas}\sim 2.5 \times 10^{9} M_{\odot}$ ($\sim 10\times$ less than a gas-rich field counterpart). According to the galaxy merger simulations discussed in the next section, a significant fraction of this gas should be available for both major and minor merger-induced star formation. 
 
We can therefore support a scenario in which a relatively significant gaseous fuel supply persists in cluster satellite galaxies merging with a SpARCS BCG, even down to low redshift, leading to a continual rejuvenation of their star formation. Depending on the epoch of infall to the cluster potential, satellite galaxies may retain both cold molecular gas and a portion of their hot coronal halo by the time they reach the cluster core and are likely to merge with the central BCG (Burke \& Collins 2013), which could ignite instantaneous and/or future star formation in and around the BCG (e.g., Tonini et al. 2011, Gabor \& Dave 2015). Residual star formation activity in satellite galaxies in the process of mass-quenching (Muzzin et al. 2012) may also persist until a merger event with the central BCG, if not completely shut down by ram-pressure stripping before such time.  Additionally, the location of the BCG at the centre of a cluster guarantees it will be fed by a cooling ICM at any time in the life of the cluster that such a cooling flow can be sustained, e.g., as the major merger rate declines and the cluster relaxes towards lower redshift. (An interesting aside, however, is that if ICM cooling is playing a role in the star formation we observe in our BCG sample, it is clearly favouring star formation over AGN activity, as we observe weak AGN activity and no tell-tale radio signature of a cooling flow and its associated AGN. In a follow-up study we will explore an environmental dependence of BCG evolution, i.e. that those in optically selected clusters may not be relaxed enough to support a steady cooling flow as in X-ray-detected clusters; and/or that a cooling flow exists but is funnelled into star formation before it reaches the SMBH (e.g., because the gas has too much angular momentum to get to the location of the SMBH before being used for star formation, or is too dense (Netzer et al. 2015)).)

We consider that the higher fraction of strongly star-forming BCGs at high redshift in our sample (see W15) may result from major mergers, as the galaxy major-merger rate is expected to be high towards $z=2$ (e.g., Vikhlinin et al. 2007); and/or a relatively large supply of cool gas remaining from the initial epoch of cluster and galaxy sub-halo collapse at $z > 2$, during which time large quantities of radiatively cooled gas contributed to both vigorous star formation and AGN activity. While the number of AGN ignited by this process is expected to peak at these redshifts, it is possible that the coincidence of luminous AGN activity in our sample with the $z=0.77-1.2$ epoch may mark the end of a phase of ICM cooling in the BCG environs. Towards lower redshift, the dwindling number of detections of both star formation and AGN activity might indicate that the gas available to fuel activity is progressively heated and exhausted, and also reflects the decline in the major merger rate as the host cluster relaxes.


\underline{Merger-induced Star Formation}

While DB07 predicts that BCGs passively gain stellar mass through a series of gas-poor minor mergers since z=2 (with 80\% of the final BCG stellar mass having been formed by $z=3$), the semi-analytic model of Laporte et al. (2013), and the simulations of galaxy and black hole growth presented in Croton et al. (2006), predict that {\it major} mergers, in addition to minor, play a significant role in the mass growth of BCGs through to late times. The cosmological simulation of BCG growth presented in Dubinski et al. (1998) produces major mergers as late as $z \sim 0.4$, marking a period of intense merging activity and tidal stripping in the environs of the model BCG.  In fact, through a study of nineteen high-redshift galaxy clusters between $z=0.84$ and $z=1.46$, half of which are drawn from the same {\it SpARCS} survey as our clusters, Lidman et al. (2013) observe a major merger rate of $0.38 \pm 0.14$ $Gyr^{-1}$ among BCGs at $z\sim 1$, and that major mergers contribute $\sim 50\%$ of BCG stellar mass between $z \sim 1$ and the present.

It is outside the scope of the present study to quantify the fraction of BCG star formation which may be attributed to major and minor mergers. However, we can make an approximate, 'order of magnitude' calculation to understand if the previously mentioned BCG major merger rate of Lidman et al. (2013), and the galaxy gas mass fractions calculated according to the Carilli \& Walter (2013) relation come together to yield a BCG stellar-mass growth matching that implied by our SED analysis. To do this, we consider the fraction of gas available for conversion into stars which is predicted by the galaxy-galaxy merger simulations presented in Mihos \& Hernquist (1994,1996). This study shows that both major and minor mergers -- which they define as those having a mass ratio of 1:1 and 1:10, respectively -- contribute mass to the combined, or more massive, system, through star formation processes, with major mergers resulting in $65-85\%$ of the total gas content of the combined galaxies being converted into stars through bursts of star formation over a timescale of 50-150 Myr, independent of galaxy morphology or orbital geometry (Cox et al. 2005 find a smaller percentage, 50\%). In the case of a minor merger, the majority of the gas of the combined system collapses onto the core of the more massive galaxy, triggering a starburst, with 50\% of the gas being consumed over 60 Myr (the strength of the starburst in this scenario is shown to be dependent on the morphology of the interacting galaxies, with bulge/disk galaxies producing a weaker starburst).  The 'collisional starburst' model of Somerville et al. (2001) -- which predicts that the fraction of the combined cold gas from two galaxies in a merger that is turned into stars is governed by the relation $e_{burst} = 0.56 \times (m_{sat}/m_{central})^{0.7}$ -- is found by Croton et al. (2006) to agree with the simulations of Mihos \& Hernquist (1994,1996) in their (the former's) implementation of this starburst model in their simulations of galaxy and supermassive black hole.

Based on several simplifying assumptions, we estimate that a SpARCS BCG can gain nearly one quarter of its stellar mass (23.34\%) at $z < 1$ through major-merger-induced star formation alone, purely from the gas deposited by massive merging galaxies (i.e., not including the accumulated stellar mass from previously formed stars in the merging companion galaxy). This estimate does not include any gas which may already be present within and around the BCG prior to merging (which could be significant in the event of central ICM cooling), or the additional star formation caused by a number of gas-rich minor majors. The assumptions we make in this calculation are as follows: 1) a constant\footnote{As noted in Lidman et al. (2013), the BCG major merger rate is theoretically expected to decrease towards lower redshift, however the rate they measure at low redshift is comparable to the one they measure at z=1.} major merger rate of $0.38$ $Gyr^{-1}$ between $z=1$ and $z=0$, as in Lidman et al. (2013) (which is consistent with the average merger rate measured by Burke \& Collins (2013), and references therein, over a similar redshift range);  2) the stellar mass of a merging companion galaxy equal to 62.5\% of the average BCG stellar mass (Lidman et al. 2013) at $z=1$ ($1.84 \times 10^{11} M_{\odot}$ and $2.95 \times 10^{11} M_{\odot}$, respectively), which is similar to the average BCG stellar mass at all redshifts considered (see Table 6);  3) the gas mass fraction of a $1.84 \times 10^{11} M_{\odot}$ merging companion galaxy calculated according to the aforementioned Carilli \& Walter (2013) relation for field galaxies between $z=1$ and $z=0$, reduced by the corresponding redshift-dependent ratio between the average SFR of cluster-core and field galaxies from Figure 5 of Alberts et al. (2014) (assuming a linear correspondence between average SFR and molecular gas mass fraction, Kennicutt \& Evans 2012); and 4) 85\% of the total accumulated gas from the merging companion galaxy is available to form stars.

If, instead, we consider only the 10 minor, gas-poor mergers predicted by DB07 to constitute the bulk of the stellar mass growth of BCGs at late times, assuming a minor companion stellar mass of $10^{10} h^{-1} M_{\odot}$ (h=0.73, DB07) and associated redshift-dependent gas fractions as calculated above, we arrive at a more modest percentage of a $\sim 5\%$ contribution from star formation to the BCG stellar mass growth at $z<1$ (again, not including the gas mass of the BCG in the estimation).

{\subsubsection{An Internal Gas Supply from Stellar Mass Loss}

An additional and likely significant source of fuel for the infrared-inferred SpARCS BCG star formation, particularly at low redshift, is recycled stellar ejecta from the internal evolved stellar population, in the form of gas mass released by Asymptotic Giant Branch (AGB) stars, Type Ia supernovae, and massive stars (Segers et al. 2015 and references therein). Segers et al. (2015) employ the cosmological, hydrodynamical simulations from the EAGLE (Evolution and Assembly of GaLaxies and their Environments) and OWLS (OverWhelmingly Large Simulations) projects to trace the fractional contribution of recycled stellar mass loss to the cosmic SFR density and cosmic stellar mass density as a function of redshift, to find an increasing contribution with decreasing redshift, with the contribution to the SFR density nearly tripling between z=2 and z=0 from 12\% to 35\%, and that to the stellar mass density nearly doubling over the same redshift range, from 10\% to 19\% (as approximated from Figure 3 of Segers et al. 2015). They also measure the fractional contribution to the SFR and stellar mass of satellite and central galaxies at $z = 0$ from recycled gas, to find that an average of $\sim 30\%$ of the SFR and $\sim 25\%$ of the stellar mass of a $10^{11.45} M_{\odot}$ central galaxy (the average SpARCS BCG stellar mass) can be attributed to this internal fuel source. The contribution to the SFR of satellite galaxies is higher, with a median contribution of $\sim 55\%$ for a $10^{10.5} M_{\odot}$ satellite galaxy, $1.3\times$ higher than that of a similarly massive central galaxy (42.5\%), suggesting yet an additional source of star formation for the BCG: the sustained star formation of merging satellite galaxies from their internal stellar mass losses. In fact, as noted in Segers et al. (2015), Kennicutt et al. (1994) observe that recycling-fuelled star formation can extend the lifetime of a gaseous disc by up to a factor of 4, allowing for SFRs to be sustained for periods comparable to the Hubble time. The difference in the fractional contribution to SFR from stellar mass loss to the average central versus satellite galaxy is attributed by Segers et al. (2015) to the difference in mass between the two galaxy populations, and the associated efficiencies of stellar feedback (for lower-mass galaxies) and AGN feedback (for higher-mass galaxies) in regulating the SFR: in the inefficient feedback regime ($10^{10.5} M_{\odot}$), the recycling-fuelled SFR fraction of satellite galaxies with low gas fractions can reach as high as $\sim90\%$.

Therefore, while the simulations of Segers et al. (2015) suggest that ongoing star formation in massive central galaxies is largely attributed to unprocessed gas -- i.e., from an external source, such as a cooling flow or gas stripped from infalling satellite galaxies -- the average contribution from internal stellar mass loss is non-negligible at $\sim 30\%$ (at $z=0$), and becomes more significant when considering the contribution of stellar mass loss from individual merging satellite galaxies over the lifetime of the BCG.}

\subsection{Hosts to Radiatively Weak AGN}

An interesting result of the current study is that only 12\% of the BCG sample are classified as continuum-dominated by the IRAC colour-colour diagnostic, with bona fide luminous AGN comprising a mere 1.4\% (noting the potential for contamination from high-redshift star-forming galaxies to this selection).  While AGN residing in host-dominated systems can 'hide' in all four regions of the IRAC colour-colour space (see Section 5.1.1), implying that we could potentially be missing many AGN in composite systems in our sample, their presence in any significant number would be expected to manifest in the stacked infrared SED longward of the IRAC bands, which we do not find: the average SED of both the bright and faint BCG populations is best described by a purely star-forming template, with no measurable AGN contribution, throughout the 8-1000 $\umu$m microns. The Mid-infrared Radio Correlation adds potentially 16 radio-loud AGN candidates, but reveals more relevantly that the mid-infrared-selected AGN at $z < 1$ are largely radio-quiet, meaning they likely lack jets associated with a black-hole accretion disk, and that the radio mode of AGN feedback is not likely to be occurring in these systems. 

Take together, these independent sets of observations strongly suggest that any AGN in our sample which are not classified as continuum-dominated based on their IRAC colours, are likely to be weak emitters. This echoes the findings of Fanidakis et al. (2010), that the most massive black holes are quietly accreting, and also that the space density of weak AGN peaks at $z < 2$ -- we remind the reader that 76/87 of the mid-infrared-identified AGN are detected between $z=0.77$ and $z=1.3$. In a future study we seek to gain further evidence in support of this claim, and consider the implications of a scenario in which the cool gas within the environs of the BCGs in our sample may be preferentially turning into stars before it can fall into the central supermassive black hole and ignite a luminous AGN, in the absence of observable radiative and mechanical feedback.

\section{Summary}

Residing at the bottom of the gravitational potential wells of the largest collapsed structures in the Universe, we expect BCGs to be subjected to the combined effects of many different physical processes at any given time, such as galaxy merging, cluster cooling flows, AGN feedback, and star formation, which presents a challenge to pinpointing the dominant mechanism(s) driving their formation and evolution. In an attempt to disentangle this complexity and develop a clearer, more complete picture of BCG evolution since $z\sim2$, we model the stacked infrared SEDs of the largest sample of optically selected BCGs in multiple flux and redshift bins out to $z=1.8$. Given that the BCGs in our sample are $z < 2$ cluster galaxies and therefore expected to be passively evolving ellipticals -- but, at the same time, likely hosts to star formation and AGN due to the in-fall of cluster galaxies and intracluster gas towards the centre of the cluster over time -- we fit the BCG SEDs with a combination of star-forming, elliptical, quiescent, and AGN model templates selected from various template libraries in the literature, allowing the fit to flexibly pick and choose, so to speak, which components to keep and reject and allow the true nature of the BCGs to reveal themselves.

In addition to the {\it Spitzer} IRAC colour-colour diagnostic of Sajina et al. (2005), we use our SED-derived estimates of BCG total infrared luminosity, star formation rate and efficiency, stellar mass, and effective dust temperature to determine the infrared character of the average BCG in each redshift bin. We find the following:

\begin{enumerate}

\item After splitting the sample of {\it SpARCS} BCGs into two flux bins, those which are and are not detected in the {\it Spitzer} MIPS 24$\umu$m band, and further subdividing each of those bins into six redshift bins between z=0 and z=1.8, we are surprised to discover that the stacked far-infrared SEDs of both subsamples match that of a star-forming galaxy with an effective dust temperature between 25 and 45 K, as opposed to a 'red and dead' elliptical galaxy or an obvious AGN. We witness the $L_{FIR}$-inferred median star formation rates of the 24$\umu$m-faint BCGs transition from sub-main-sequence to main-sequence levels beyond $z=1$, and the 24$\umu$m-bright BCGs produce stars with the efficiency of an upper-main-sequence/starbust galaxy throughout most of the redshift range considered.

\item The two BCG sub-populations are remarkably similar in their far-infrared properties, where the 24$\umu$m-faint systems are not in fact 'off' or inactive in comparison to the 'on' 24$\umu$m-bright BCGs, as the mid-infrared portion of the SED, alone, might suggest.  This allows us to conclude that the faint BCGs, which outnumber the bright BCGs by a factor of three in our sample, are not intrinsically infrared-faint, as expected. We attribute the difference in their mid-infrared flux intensity to a difference in the {\it timing} of their most recent star formation activity, with the 24$\umu$m-bright BCGs being 'caught in the act' of instantaneous star formation, and the 24$\umu$m-faint BCGs observed in a post-starburst state, where star formation could have very recently ended (or paused).

\item All of the physical parameters we derive from the stacked luminosity SEDs are generally consistent in their classification of the average bright and faint BCG in each redshift bin as either a sub-main-sequence, main-sequence, or starburst star-forming candidate, with the sSFR analysis suggesting a starbursting bright population at intermediate-to-high redshifts, and a faint population which transitions from a low to a relatively high level of star formation activity with increasing redshift. 

\item The median star formation rates we infer from the stacked infrared SEDs suggest that a potentially significant fraction of BCG stellar mass growth since $z\sim2$ can be attributed to star formation processes, yielding a total contribution of $\sim40.8\%$ of the average SpARCS BCG stellar mass over this redshift range (with the majority of the mass growth occurring at $z>1$). This result conflicts with the prediction of DB07 that BCGs only passively evolve through a series of gas-poor minor mergers since $z=2$, but is in better agreement with the updated semi-analytic model of Tonini et al. (2012), which also finds a star-forming BCG through to late times. 

\end{enumerate}

Our final conclusion is that the optically and near-IR-selected BCGs in our sample at $0.0 < z < 1.8$ are not 'red and dead', but what we refer to as {\it red but not dead}: while the bulk of their stellar population appears to have been in place since $z>2$, as predicted, they retain a pulse of star formation activity -- not enough to shift the weight of their stellar population towards young, blue stars, but enough to substantially build their stellar mass on short timescales down to $z \sim 0.7$, and more gently subsequent to this time. This is in agreement with the model BCG data of Tonini et al. (2012), where BCGs with average star formation rates of $50-250$ $M_{\odot} yr^{-1}$ are produced at $z < 2$ while maintaining 'red' optical colours, with 50\% of their stellar population being as old as the Universe at each epoch sampled. We find that, despite the inherent complexity in studying BCGs on account of their unique environment in the Universe, and the various limitations and biases plaguing BCG data sets, the set of distinctive properties the environment appears to bestow upon them illuminates the overarching physical parameters guiding their formation and evolution -- and, by this very distinction from other classes of galaxies, actually informs upon galaxy formation and evolution in general.

\section{Acknowledgements}

The authors thank Chiara Tonini for providing the BCG model data referenced from Tonini et al. (2012).

GW acknowledges financial support for this work from NSF grant AST-1517863 and from NASA through programs GO-13306, GO-13677, GO-13747 \& GO-13845/14327 from the Space Telescope Science Institute, which is operated by AURA, Inc., under NASA contract NAS 5-26555.

\label{lastpage}


\begin{thebibliography}{99}
\bibitem[\protect\citeauthoryear{Alberts et al.}{2014}]{b1} Alberts S. et al., 2014, MNRAS, 437, 437
\bibitem[\protect\citeauthoryear{Alberts et al.}{2016}]{b2} Alberts S. et al., 2016, ApJ, 825, 72
\bibitem[\protect\citeauthoryear{Aragon-Salamanca et al.}{1993}]{b3} Aragon-Salamanca A., Ellis R. S., Couch W. J., Carter D., 1993, MNRAS, 262, 764
\bibitem[\protect\citeauthoryear{Bai et al.}{2014}]{b4} Bai L. et al., 2014, ApJ, 789, 134
\bibitem[\protect\citeauthoryear{Bauer et al.}{2005}]{b5} Bauer F. E., Fabian A. C., Sanders J. S., Allen S. W., Johnstone R. M., 2005, MNRAS, 359, 1481
\bibitem[\protect\citeauthoryear{Bell et al.}{2005}]{b6} Bell E.F., 2005, ApJ, 625, 23
\bibitem[\protect\citeauthoryear{Blanton et al.}{2004}]{b7} Blanton E. L., 2004, in Proceedings of The Riddle of Cooling Flows in Galaxies and Clusters of Galaxies, ed. T. Reiprich, J. Kempner, \& N. Soker (arXiv:astro-ph/0402342)
\bibitem[\protect\citeauthoryear{Boselli et al.}{2014}]{b8} Boselli A. et al., 2014, A\&A, 564, A67
\bibitem[\protect\citeauthoryear{Bower et al.}{1992}]{b9} Bower R. G., Lucey J. R., Ellis R. S., 1992, MNRAS, 254, 601
\bibitem[\protect\citeauthoryear{Brodwin et al.}{2013}]{b10} Brodwin M. et al., 2013, ApJ, 779, 138
\bibitem[\protect\citeauthoryear{Burke et al.}{2000}]{b11} Burke D. J., Collins C. A., Mann R.G., 2000, ApJ, 532, 105
\bibitem[\protect\citeauthoryear{Burke \& Collins}{2013}]{b12} Burke C., Collins C. A., 2013, MNRAS, 434 2856
\bibitem[\protect\citeauthoryear{Burns}{1990}]{b13} Burns J. O. 1990, AJ, 99, 14
\bibitem[\protect\citeauthoryear{Casey}{2012}]{b14}Casey C., 2012 MNRAS, 425,309
\bibitem[\protect\citeauthoryear{Chary \& Elbaz}{2001}]{b15} Chary R., Elbaz D., 2001, ApJ, 556, 562
\bibitem[\protect\citeauthoryear{Chen et al.}{2007}]{b16} Chen Y., Reiprich T. H., Böhringer H., Ikebe Y., Zhang Y.-Y., 2007, A\&A, 466, 805
\bibitem[\protect\citeauthoryear{Cox et al.}{2005}]{b17} Cox T. J., Jonsson P., Primack J. R., Somerville R. S., 2005, MNRAS, 384, 386
\bibitem[\protect\citeauthoryear{Croton et al.}{2005}]{b18} Croton D.J., Springel V., White S. D. M., De Lucia G., Frenk C. S., Gao L., Jenkins A., Kauffmann G., Navarro J. F., 
\bibitem[\protect\citeauthoryear{Dale et al.}{2014}]{b19} Dale D. A., Helou G., Magdis G. E., Armus L., Díaz-Santos T., Shi Y., 2014, ApJ, 784, 83
\bibitem[\protect\citeauthoryear{De Lucia \& Blaizot}{2007}]{b20} De Lucia G., Blaizot J., 2007, MNRAS, 375, 2 
\bibitem[\protect\citeauthoryear{Donahue et al.}{2009}]{b21} Donahue M., 2009, in The Monster's Fiery Breath: Feedback in Galaxies, Groups, and Clusters. AIP Conference Proceedings, 1201, pp. 177 
\bibitem[\protect\citeauthoryear{Donahue et al.}{2010}]{b22} Donahue et al., 2010, ApJ, 715, 881
\bibitem[\protect\citeauthoryear{Dressler}{1984}]{b23} Dressler A., 1984, ARA\&A
\bibitem[\protect\citeauthoryear{Dubinski}{1998}]{b24} Dubinski J., 1998, ApJ, 502, 141
\bibitem[\protect\citeauthoryear{Dutson et al.}{2014}]{b25} Dutson K. L., Edge A. C., Hinton J. A., Hogan M. T., Gurwell M. A., Alston W. N., 2014, MNRAS, 442, 2048
\bibitem[\protect\citeauthoryear{Eilek}{2003}]{b26} Eilek J. A., 2003, Proceedings of The Riddle of Cooling Flows in Galaxies and Clusters of Galaxies, ed. T. Reiprich, J. Kempner, \& N. Soker (arXiv:astro-ph/0310011)
\bibitem[\protect\citeauthoryear{Elbaz et al.}{2011}]{b27} Elbaz D. et al., 2011, A\&A, 533, 26
\bibitem[\protect\citeauthoryear{Fabian et al.}{2001}]{b28} Fabian A.C., 2001, in Particles and Fields in Radio Galaxies Conference, ASP Conf. Proc., Vol. 250. Astron. Soc. of the Pac., San Francisco, p.471
\bibitem[\protect\citeauthoryear{Gabor \& Dave}{2015}]{b29} Gabor J.M. \& Dave R., 2015, MNRAS, 447, 374
\bibitem[\protect\citeauthoryear{Gomes et al.}{2015}]{b30} Gomes J.M. et al., 2015, A\&A, 586, 22
\bibitem[\protect\citeauthoryear{Hashimoto, Henry \& Boehringer}{2014}]{b31} Hashimoto Y., Patrick J.P., Boehringer H., 2014, MNRAS 440, 588
\bibitem[\protect\citeauthoryear{Goulding et al.}{2012}]{b32} Goulding A. D. et al., 2012, ApJ, 755, 5
\bibitem[\protect\citeauthoryear{Hicks, Mushotsky \& Donahue}{2010}]{b33} Hicks A. K., Mushotzky R., Donahue M., 2010, ApJ, 719, 1844
\bibitem[\protect\citeauthoryear{Hoeting et al.}{1999}]{b34} Hoeting J. A., 1999, Statistical Science, Vol. 14, No. 4, 382–417
\bibitem[\protect\citeauthoryear{Idzi}{2007}]{b35} Idzi R., 2007, PhD Thesis, Johns Hopkins University
\bibitem[\protect\citeauthoryear{Katayama, Hayashida \& Takahara}{2014}]{b36} Katayama H., Hayashida K., Takahara F., 2014, ApJ, 585, 687
\bibitem[\protect\citeauthoryear{Kennicutt}{1983}]{b37} Kennicutt R. C., Jr., 1983, ApJ, 272, 54
\bibitem[\protect\citeauthoryear{Kirkpatrick et al.}{2012}]{b38} Kirkpatrick A., 2012, ApJ, 759, 139
\bibitem[\protect\citeauthoryear{Kirkpatrick et al.}{2013}]{b39} Kirkpatrick A., 2013, ApJ, 763, 123
\bibitem[\protect\citeauthoryear{Laporte et al.}{2013}]{b40} Laporte C. F. P., White S. D. M., Naab T., Gao L., 2013, MNRAS, 435, 901
\bibitem[\protect\citeauthoryear{Lidman et al.}{2012}]{b41} Lidman C. et al., 2012, MNRAS, 427, 550
\bibitem[\protect\citeauthoryear{Lidman et al.}{2013}]{b42} Lidman C. et al., 2013, MNRAS, 433, 825
\bibitem[\protect\citeauthoryear{Liu, Mao, Meng}{2012}]{b43} Liu F. S., Mao S., Meng X. M., 2012, MNRAS, 423, 422
\bibitem[\protect\citeauthoryear{Lonsdale et al.}{2004}]{b44} Lonsdale C. et al., 2004, ApJ, 154, 54
\bibitem[\protect\citeauthoryear{McDonald et al.}{2011}]{b45} McDonald M. et al., ApJ, 742L, 35
\bibitem[\protect\citeauthoryear{McDonald et al.}{2016}]{b46} McDonald M. et al., ApJ, 817, 86
\bibitem[\protect\citeauthoryear{McNamara et al.}{2001}]{b47} McNamara B. R. et al. 2001, in Clusters of galaxies and the high redshift universe observed in X-rays, Recent results of XMM-Newton and Chandra. Proceedings of XXI Moriond conference: Galaxy Clusters and the High Redshift.
\bibitem[\protect\citeauthoryear{Mendez et al.}{2013}]{b48} Mendez A.J. et al., 2013, ApJ, 770, 40
\bibitem[\protect\citeauthoryear{Mihos \& Hernquist}{1994}]{b49} Mihos J.C, Hernquist L., 1994, ApJ, 425, 13
\bibitem[\protect\citeauthoryear{Mihos \& Hernquist}{1996}]{b50} Mihos J.C, Hernquist L., 1996, ApJ, 464, 641
\bibitem[\protect\citeauthoryear{Mullaney et al.}{2011}]{b51} Mullaney J. R., Alexander D. M., Goulding A. D., Hickox, R. C., 2011, MNRAS, 414, 1082
\bibitem[\protect\citeauthoryear{Muzzin et al.}{2008}]{b52} Muzzin A. et al., 2008, ApJ, 686, 966
\bibitem[\protect\citeauthoryear{Muzzin et al.}{2009}]{b53} Muzzin A. et al., 2009, ApJ, 698, 1934
\bibitem[\protect\citeauthoryear{Muzzin et al.}{2010}]{b54} Muzzin A., van Dokkum P., Kriek M., Labbé I., Cury I., Marchesini D., Franx M., 2010, ApJ, 725, 742
\bibitem[\protect\citeauthoryear{Muzzin et al.}{2014}]{b55} Muzzin et al., 2014, ApJ, 796, 65
\bibitem[\protect\citeauthoryear{Nguyen et al.}{2010}]{b56} Nguyen et al., 2010, A\&A, 518, 5
\bibitem[\protect\citeauthoryear{Ostriker \& Hausman}{1977}]{b57} Ostriker J. P., Hausman M. A., ApJ, 217, 125
\bibitem[\protect\citeauthoryear{Peres et al.}{1998}]{b58} Peres C.B., Fabian A. C. , Edge A. C., Allen S. W., Johnstone R. M., White D. A., 1998, MNRAS, 298, 416
\bibitem[\protect\citeauthoryear{Polletta et al.}{2007}]{b59} Polletta M. et al., 2007, ApJ, 663, 81
\bibitem[\protect\citeauthoryear{Rowan-Robinson}{2008}]{b60} Rowan-Robinson M., et al., 2008, MNRAS, 386, 697
\bibitem[\protect\citeauthoryear{Rowan-Robinson}{2008}]{b61} Rowan-Robinson M., et al., 2010, MNRAS, 409, 2
\bibitem[\protect\citeauthoryear{Rawle et al.}{2012}]{b62} Rawle et al., 2012, ApJ, 747, 29
\bibitem[\protect\citeauthoryear{Sajina, Lacy \& Scott}{2005}]{b63} Sajina A., Lacy M., Scott, D., 2005, ApJ, 621, 256
\bibitem[\protect\citeauthoryear{Samuele}{2011}]{b64} Samuele R., McNamara B. R., Vikhlinin A., Mullis C. R., 2011, ApJ, 731, 31
\bibitem[\protect\citeauthoryear{Sanderson, Ponman, \& O’Sullivan}{2006}]{b65} Sanderson A. J. R., Ponman T. J., O’Sullivan E., 2006, MNRAS, 372, 1496
\bibitem[\protect\citeauthoryear{Santos et al.}{2008}]{b66} Santos J. S. , Rosati P., Tozzi P., Böhringer H., Ettori S., Bignamini A., A\&A, 483, 35
\bibitem[\protect\citeauthoryear{Santos et al.}{2015}]{b67} Santos J. S. et al., 2015, MNRAS, 447, L65
\bibitem[\protect\citeauthoryear{Shi, Helou \& Armus}{2013}]{b68} Shi Y., Helou G., Armus L., 2013, ApJ, 777, 6
\bibitem[\protect\citeauthoryear{Silva et al.}{1998}]{b69} Silva L., Granato G.L., Bressan A., Danese L., 1998, ApJ, 509, 103
\bibitem[\protect\citeauthoryear{Smail et al.}{2014}]{b70} Smail I. et al., 2014, ApJ, 782, 19
\bibitem[\protect\citeauthoryear{Somerville, Primack \& Faber}{2001}]{b71} Somerville R. S., Primack J. R., Faber S. M., 2001, MNRAS, 320, 504
\bibitem[\protect\citeauthoryear{Stott et al.}{2012}]{b72} Stott J.P. et al., 2012, MNRAS, 422, 2213
\bibitem[\protect\citeauthoryear{Stott et al.}{2010}]{b73} Stott J.P. et al., 2010, ApJ, 718, 23
\bibitem[\protect\citeauthoryear{Stott et al.}{2008}]{b74} Stott J.P. et al., 2008, MNRAS, 384, 1502
\bibitem[\protect\citeauthoryear{Surace et al.}{2005}]{b75} Surace et al., 2005, The SWIRE Data Release 2: Image Atlases and Source Catalogs for ELAIS-N1, ELAIS-N2, XMM-LSS, and the Lockman Hole, \url{http://swire.ipac.caltech.edu/swire/astronomers/data_access.html}
\bibitem[\protect\citeauthoryear{Sun et al.}{2009}]{b76} Sun et al., 2009, ApJ, 704, 1586S
\bibitem[\protect\citeauthoryear{Tacchella}{2015}]{b77} Tacchella et al., 2015, Science, 348, 314
\bibitem[\protect\citeauthoryear{Tonini et al.}{2012}]{b78} Tonini C., Bernyk M., Croton D., Maraston C., Thomas D., 2012, ApJ, 759, 43
\bibitem[\protect\citeauthoryear{Tran et al.}{2010}]{b79} Tran K.-V. H. et al., 2010, ApJ, 719, L126
\bibitem[\protect\citeauthoryear{van Dokkum et al.}{1998}]{b80} van Dokkum P.G., Franx M., Kelson D.D., Illingworth G., 1998 ApJ, 504, 17 
\bibitem[\protect\citeauthoryear{Vikhlinin et al.}{2007}]{b81} Vikhlinin A., Burenin R., Forman W. R., Jones C., Hornstrup A., Murray S. S., Quintana H., 2007, in Heating versus Cooling in Galaxies and Clusters of Galaxies, Eso Astrophysics Symposia. Springer-Verlag Berlin Heidelberg, 2007, p. 48
\bibitem[\protect\citeauthoryear{Vijayaraghavan \& Ricker}{2015}]{b82} Vijayaraghavan R. \& Ricker P.M., 2015, MNRAS, 449, 3 
\bibitem[\protect\citeauthoryear{Webb et al.}{2015}]{b83} Webb T.M.R., et al., 2015, ApJ, 814, 96
\bibitem[\protect\citeauthoryear{Whiley et al.}{2008}]{b84} Whiley et al., 2008, MNRAS, 387, 1253
\bibitem[\protect\citeauthoryear{White et al.}{1976}]{b85} White S.D.M., 1976, MNRAS, 174, 19
\bibitem[\protect\citeauthoryear{Wilson et al.}{2009}]{b86} Wilson G. et al., 2009, ApJ, 698, 1943
\bibitem[\protect\citeauthoryear{Yoshida et al.}{2005}]{b87} Yoshida N. et al., 2005, MNRAS, 365, 11 
\bibitem[\protect\citeauthoryear{Zhao, Aragón-Salamanca \& Conselice}{2015}]{b88} Zhao, D., Aragón-Salamanca A., Conselice, C. J., 2015, MNRAS, 448, 2530
\end{thebibliography}
\end{document}